\documentclass{jfm}

\usepackage{graphicx}
\usepackage{epstopdf}
\usepackage{amsmath,rotating}
\usepackage{dashrule}
\usepackage{psfrag}
\usepackage{latexsym}
\usepackage{mathrsfs}
\usepackage{pifont}
\usepackage{amsmath}
\usepackage{subfigure}
\usepackage{mathtools}
\usepackage{mathcomp}
\usepackage{mathdots}
\usepackage{rotating}
\usepackage{array}
\usepackage{color}
\usepackage{algorithm}
\usepackage{algorithmic}
\usepackage[normalem]{ulem}

\usepackage{gensymb}
\usepackage{amssymb}
\usepackage[dvipsnames]{xcolor}
\usepackage{tikz}
\usepackage{natbib}
\usetikzlibrary{arrows,shapes,trees,calc,positioning}
\usetikzlibrary{matrix}

\newcommand{\floor}[1]{\left\lfloor #1 \right\rfloor}

\newcommand{\enma}[1]   {\ensuremath{#1}}

\newcommand{\beq}{\begin{equation}}
\newcommand{\eeq}{\end{equation}}
\newcommand{\bseq}{\begin{subequations}}
\newcommand{\eseq}{\end{subequations}}
\newcommand{\beqn}{\begin{eqnarray}}
\newcommand{\eeqn}{\end{eqnarray}}
\newcommand{\ba}{\begin{array}}
\newcommand{\ea}{\end{array}}
\newcommand{\bct}{\begin{center}}
\newcommand{\ect}{\end{center}}
\newcommand{\btmz}{\begin{itemize}}
\newcommand{\etmz}{\end{itemize}}
\newcommand{\benum}{\begin{enumerate}}
\newcommand{\eenum}{\end{enumerate}}

\newcommand{\cH}{\enma{\mathcal H}}

\newcommand{\diag}      {\enma{\mathrm{diag}}}

\newcommand{\trace}     {\enma{\mathrm{trace}}}

\newcommand{\bv}{{\bf v}}

\newcommand{\matbegin}{
        \left[
}
\newcommand{\matend}{
        \right]
}

\newcommand{\tbt}[4]{
  \matbegin \begin{array}{cc}
       #1 & #2 \\ #3 & #4
       \end{array} \matend }

\newcommand{\be}{\begin{equation}}
\newcommand{\ee}{\end{equation}}

\newcommand{\cplxs}{ C\kern -.35em \rule{0.03 em}{.7 ex}~   }

\def\complex{\hbox{C\kern -.45em \rule{0.03 em}{1.5 ex}}~}

\newcommand{\bi}{\begin{itemize}}
\newcommand{\ei}{\end{itemize}}

\newcommand{\bu}{{\bf u}}

\newcommand{\bbZ}{\mathbb{Z}}

\newcommand{\bbR}{\mathbb{R}}

\newcommand{\btab}{\begin{tabular}}
\newcommand{\etab}{\end{tabular}}
\newcommand{\bd}{{\bf d}}

\newcommand{\bvarphi}{\mbox{\boldmath$\varphi$}}

\newcommand{\bpsi}{\mbox{\boldmath$\psi$}}

\newcommand{\mrd}{\mathrm{d}}
\newcommand{\mre}{\mathrm{e}}
\newcommand{\mri}{\mathrm{i}}

\newcommand{\ds}{\displaystyle}

\newcommand{\eps}{{\epsilon}}

\newcommand{\bA}{\mathbf{A}}
\newcommand{\bB}{\mathbf{B}}
\newcommand{\bC}{\mathbf{C}}

\newcommand{\bL}{\mathbf{L}}

\newcommand{\DefinedAs}[0]{\mathrel{\mathop:}=}
\newcommand{\vsp}{\vspace*{0.15cm}}
\definecolor{bgblue}{rgb}{0.04,0.19,0.53}
\definecolor{dblue1}{rgb}{0,0.3,0.7}
\definecolor{dred}{rgb}{0.4,0.2,0}

\definecolor{bgblue}{rgb}{0.04,0.19,0.53}
\definecolor{dblue1}{rgb}{0,0.3,0.7}
\definecolor{dred}{rgb}{0.4,0.2,0}

\shorttitle{Turbulent channel flow over riblets}
\shortauthor{W. Ran, A. Zare, and M. Jovanovi\'{c}}

\title{Model-based design of riblets \\[0.1cm] for turbulent drag reduction}

\author{Wei Ran\aff{1},
  Armin Zare\aff{2},
 \and Mihailo R.\ Jovanovi\'{c}\aff{3}
 \corresp{\email{mihailo@usc.edu}}
 }

\affiliation{\aff{1} Department of Aerospace and Mechanical Engineering, \\
University of Southern California, Los Angeles, CA 90089, USA
\aff{2} Department of Mechanical Engineering, \\
University of Texas at Dallas, Richardson, Texas 75080, USA
\aff{3} Ming Hsieh Department of Electrical and Computer Engineering, \\
University of Southern California, Los Angeles, CA 90089, USA}

\begin{document}

\maketitle

\begin{abstract}
Both experiments and direct numerical simulations have been used to demonstrate that riblets can reduce turbulent drag by as much as $10\%$, but their systematic design remains an open challenge. In this paper, we develop a model-based framework to quantify the effect of streamwise-aligned spanwise-periodic riblets on kinetic energy and skin-friction drag in turbulent channel flow. We model the effect of riblets as a volume penalization in the Navier-Stokes equations and use the statistical response of the eddy-viscosity-enhanced linearized equations to quantify the effect of background turbulence on the mean velocity and skin-friction drag. For triangular riblets, our simulation-free approach reliably predicts drag-reducing trends as well as mechanisms that lead to performance deterioration for large riblets. We investigate the effect of height and spacing on drag reduction and demonstrate a correlation between energy suppression and drag-reduction for appropriately sized riblets. We also analyze the effect of riblets on drag reduction mechanisms and turbulent flow structures including very large scale motions. Our results demonstrate the utility of our approach in capturing the effect of riblets on turbulent flows using models that are tractable for analysis and optimization.
\end{abstract}


	\vspace*{-8ex}
\section{Introduction}
\label{sec.intro}

Surface roughness typically increases skin-friction drag and degrades performance of engineering systems that involve the motion of rigid bodies in turbulent flows, e.g., ships and submarines with biofouled hulls~\citep{schbenholher11}. Using both experiments and numerical simulations,~\cite{yusuta17} reported an increase in skin-friction drag by about $41\%$ per year because of marine fouling growth. In contrast, carefully designed surface corrugations can reduce skin-friction drag by as much as $10\%$~\citep{becbruhaghoehop97,gad00}. Patterned surface modifications have been used to reduce drag in a number of engineering applications~\citep{cousav89}. Success stories include the $2\%$ drag reduction by spanwise-periodic riblets in commercial aircrafts~\citep{szo91}, and the $7\%$ drag reduction by shark-skin-inspired design of swimsuits for olympic swimmers~\citep{bendawblaell02,molteropppen04}.

\subsection{Previous studies of drag reduction by riblets}

Given the potential economic benefits of riblets, many experimental and numerical studies have been dedicated to examining the dependence of skin-friction drag on design parameters~\citep{wal82,wallin84,chomoikim93,becbruhaghoehop97,becbruhag00,garjim11a}. These efforts provided a broad range of guidelines for characterizing the drag-reducing trends of riblets based on their size and shape (blade-like, triangular, T-shaped, etc.). In particular, turbulent drag-reduction as a function of various metrics of size (e.g., rib spacing or groove area) appears to follow a consistent trend over a host of riblet shapes, e.g., see~\citet[figure~15]{becbruhaghoehop97}. For example, for small riblets (i.e., in the so-called ``viscous'' regime), drag reduction is proportional to the riblet size. This linear trend gradually saturates at an optimal size before eventually degrading and leading to a drag increase for large riblets. Furthermore, for various riblet shapes, the optimal riblet spacing (in inner units) satisfies $s^+ \in [10,20]$~\citep{becbruhaghoehop97}.~\citet{garjim11a,garjim11b} discovered that the cross-sectional area $A_g$ of the grooves provides the best predictor of drag-reducing trends over various shapes and identified $l_g^+ = \sqrt{A_g^+} \approx10.7\pm1$ as the optimal size of riblets.

Beyond parametric studies, a considerable effort was made to uncover mechanisms responsible for drag reduction. The presence of small-size riblets results in the suppression of the cross-flows introduced by near-wall turbulence, which weakens the near-wall quasi-streamwise vortices and pushes them away from the wall. This limits the transfer of mean momentum toward the wall and creates a zone of suppressed turbulence within the grooves, thereby reducing skin-friction drag~\citep{chomoikim93,sirkar97,leelee01}.

Various notions of protrusion height have been proposed to quantify the effect of riblets on near-wall turbulence.~\cite{becbar89} defined the protrusion height as the offset between the virtual origin for the mean flow and a measure of the average wall location. In contrast,~\cite{lucmanpoz91} proposed to use the difference between the virtual origin for the streamwise and spanwise flows. For blade-like and scalloped riblets, the latter approach provides a good indicator of the shift in mean velocity and it offers a better surrogate for predicting drag reduction in the viscous regime.~\citet{gargomfai19,ibrgomgar19} proposed to quantify the shift in turbulence arising from quasi-streamwise vortices as a function of the wall-normal and spanwise slip lengths. This method can be used to predict the shift in the mean velocity, which is typically difficult to quantify in flows over complex surfaces. On the other hand, by examining 2D/3D roughness,~\cite{orlleo06} demonstrated a linear relation between the roughness function, i.e., shift of mean velocity in the logarithmic region, and the rms of wall-normal velocity at the tip of roughness elements.

Both experiments and simulations have been used to demonstrate that the drag-reducing performance of riblets eventually saturates and degrades with increase in their size.~\cite{goltua98} suggested that the creation of small secondary streamwise vortices around the tips of riblets by the unsteady cross-flow degrades performance.~\citet{chomoikim93,suzkas94,leelee01} related this phenomenon to the lodging of streamwise vortices into the grooves, which breaks down the viscous regime near the wall. More recently, the numerical study of~\cite{garjim11b} suggested that the breakdown of the viscous regime is accompanied by the emergence of spanwise rollers of typical streamwise length $\lambda_x^+\sim150$ that develop from a two-dimensional Kelvin-Helmholtz (\mbox{K-H}) instability. The emergence of these coherent flow structures was also connected to an increase in the Reynolds shear stress in the vicinity of the corrugated surface.

While these studies offer valuable insights into drag reduction mechanisms, their reliance on costly experiments and simulations has hindered the {optimal} design of riblet-mounted surfaces. This motivates the development of low-complexity models that capture the essential physics of turbulent flows over riblets and are well-suited for analysis, design, and optimization. Previously proposed notions of protrusion height~\citep{becbar89,lucmanpoz91}, spanwise slip length~\citep{gargomfai19,ibrgomgar19}, and the roughness function~\citep{orlleo06} provide surrogate measures for the performance of specific riblet geometries, but are typically constrained to the viscous regime. The self-regular model for wall turbulence regeneration proposed by~\cite{banhel14} accounts for the spatio-temporal evolution of flow structures over patterned surfaces and matches experimental results for transitional and turbulent flows at low Reynolds numbers. More recently, the receptivity of channel flow over riblets was studied using the $\cH_2$ norm of the linearized dynamics~\citep*{kasdunpap12} and the resolvent analysis~\citep{chaluh19}. While the former study used a change of coordinates to translate spatially-periodic geometry into spatially-periodic differential operators, the latter utilized a volume penalization technique to represent the effect of riblets as a feedback term in the dynamics. Moreover, \cite{chaluh19} showed that the dependence of the resolvent gain on the spacing of riblets closely follows previously reported drag-reducing trends in turbulent flows.

Prior model-based efforts have shown promise in predicting the energetics of turbulent flows in the presence of riblets. However, apart from~\cite{chaluh19}, such studies do not account for the interactions among harmonics of flow fluctuations that are induced by spatially-periodic geometry. Furthermore, in the absence of a systematic framework to quantify the influence of background turbulence on the mean velocity, prior studies cannot provide accurate predictions of skin-friction drag in the presence of riblets. In this paper, we account for dynamical interactions and utilize turbulence modeling in conjunction with the eddy-viscosity-enhanced linearized NS equations to quantify the effect of background turbulence on skin-friction drag in turbulent channel flow over riblets.

\vspace*{-2ex}
\subsection{Preview of modeling framework and main results}

The linearized NS equations have been used to capture structural and statistical features of transitional~\citep{butfar92,tretrereddri93,farioa93,bamdah01,mj-phd04,jovbamJFM05,ranzarhacjovPRF19b} and turbulent~\citep*{mcksha10,hwacosJFM10b,zarjovgeoJFM17,zargeojovARC20} wall-bounded shear flows. In these studies, the effect of disturbances was modeled as an additive source of deterministic or stochastic excitation in the NS equations. Such an input-output approach~\citep{jovARFM21} has also been used for the model-based design of sensor-free control strategies for suppressing turbulence via streamwise traveling waves~\citep*{moajovJFM10,liemoajovJFM10} and transverse wall oscillations~\citep{jovPOF08,moajovJFM12}, as well as feedback control strategies~\citep{kimbew07} including opposition control~\citep*{luhshamck14,toeluhmck19}.

A challenging aspect of control design for turbulent flows is to quantify the effect of background turbulence on the mean velocity around which we study the dynamics of fluctuations. This effect is often captured by turbulent eddy viscosity models that are prescribed for specific flow configurations and do not account for the influence of control. To capture the influence of background turbulence,~\cite{moajovJFM12} developed a framework to determine the turbulent viscosity of channel flow in the presence of control from the statistics of the eddy-viscosity-enhanced linearized NS equations. This study showed that, by accounting for the influence of fluctuation dynamics on the turbulence model, reliable predictions of the mean velocity and the skin-friction drag can be obtained.

In this paper, we extend the framework developed in~\cite{moajovJFM12} to quantify the effect of riblets on turbulent channel flow. Following~\citet{chaluh19} we use a volume penalization technique~\citep{khaangparcal00} to approximate the effect of {the} spatially-periodic surface on {the} turbulent flow. This method introduces a static feedback term that captures the shape of riblets via a resistive function in the momentum equation. Additionally, we augment kinematic viscosity with turbulent eddy viscosity and examine the dynamics of flow fluctuations around the steady-state solution of the modified governing equations. The spatially-periodic nature of the mean flow introduces interactions between different harmonics of the mean and fluctuating velocity fields, which complicates frequency response analysis relative to the flow over smooth walls. We utilize the second-order statistics of velocity fluctuations to determine the turbulent viscosity for the flow over riblets and compute their effect on \mbox{skin-friction drag.}

We use our simulation-free approach to examine the effect of triangular riblets {on} turbulent channel flow. For various shapes and sizes of {triangular} riblets, our results are in close agreement with experimental and numerical studies~\citep{becbruhaghoehop97,garjim11b}. We also study the kinetic energy of velocity fluctuations and observe a strong correlation between energy suppression and drag-reduction trends for certain sizes of riblets. In addition, we use our model to examine dominant flow structures and mechanisms for drag reduction. The close agreement between ours and prior experimental and DNS results demonstrates {the utility of our model-based approach.}

\vspace*{-2ex}
\subsection{Paper outline}

The rest of our presentation is organized as follows. In \S~\ref{sec.probform}, we formulate the problem and provide an overview of the volume penalization technique that is used to account for the presence of riblets. In \S~\ref{sec.fluctuations}, we utilize the linearized eddy-viscosity-enhanced NS equations to study the dynamics of velocity fluctuations around the turbulent base flow induced by riblets. The second-order statistics computed from this linearized model are used to modify turbulent viscosity and refine predictions of the mean velocity and skin-friction drag in turbulent channel flow over riblets. In \S~\ref{sec.result}, we demonstrate the merits of our framework and its ability to capture the drag-reducing trends of triangular riblets. In \S~\ref{sec.flowstructures}, we show that our framework captures the effect of riblets on the typical flow structures and uncovers mechanisms for drag reduction. Finally, in \S~\ref{sec.conclution}, we provide {a} summary of our results and {an} outlook for future research directions.

	\vspace*{-2ex}
\section{Problem formulation}
\label{sec.probform}

\begin{figure}
        \begin{center}
        \begin{tabular}{cccc}
        \hspace{-.15cm}
        \subfigure[]{\label{fig.channel}}
        &&\hspace{.2cm}
        \subfigure[]{\label{fig.ribletschannel}}
        &
        \\[-.1cm]
        &
        \begin{tabular}{c}
                \includegraphics[width=.44\textwidth]{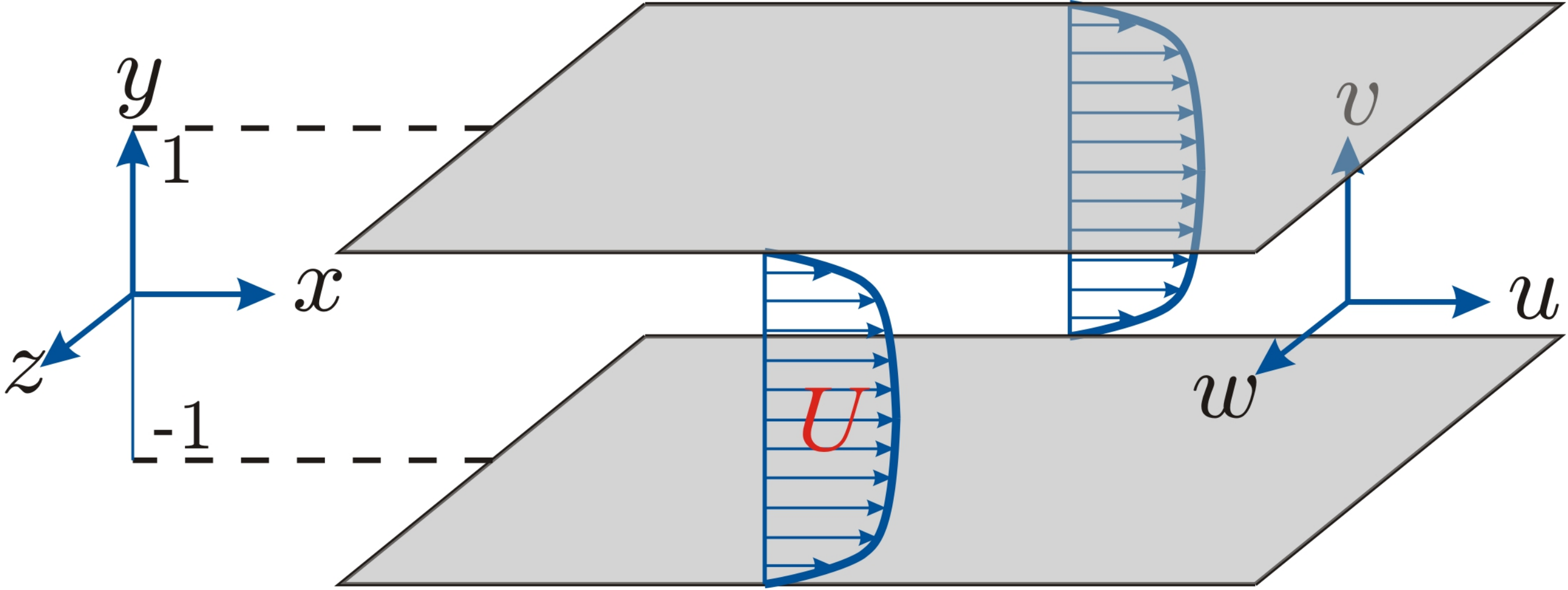}
        \end{tabular}
        &&
        \begin{tabular}{c}
                \includegraphics[width=.44\textwidth]{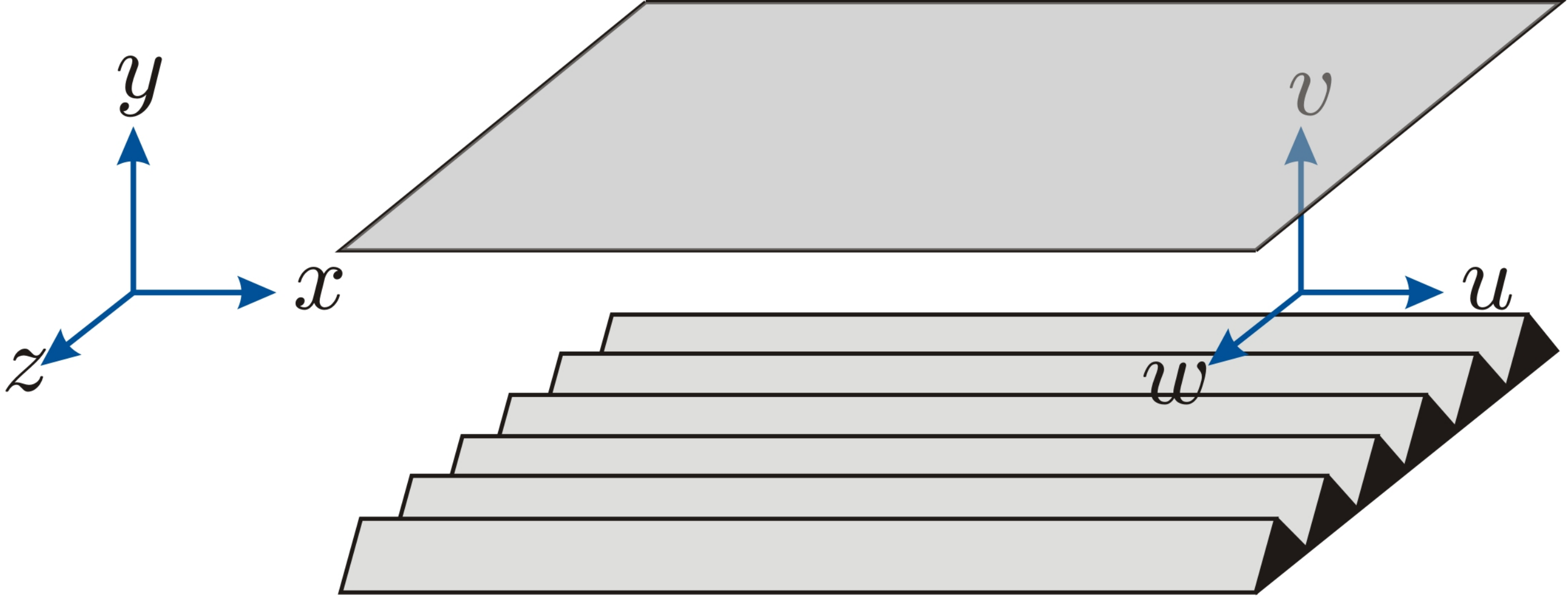}
        \end{tabular}
        \end{tabular}
        \end{center}
        \caption{(a) Fully-developed pressure-driven turbulent channel flow. (b) Channel flow with spanwise-periodic riblets on the lower wall.}
        \label{fig.channelflow}
\end{figure}

The pressure-driven channel flow of incompressible Newtonian fluid, with geometry shown in figure~\ref{fig.channel}, is governed by the Navier-Stokes and continuity equations
\begin{align}
	\label{eq.NSeqn}
	\ba{rcl}
    		\partial_t\bu
    		& \!\!=\!\! &
    		\;-\;
    		\left( \bu \cdot \nabla \right)\, \bu
     		\;-\;
    		\nabla P
    		 \;+\;
    		\dfrac{1}{Re_\tau}\, \Delta \bu,
    		\\[.15cm]
    		0
    		& \!\!=\!\! &
    		\nabla \cdot \bu,
	\ea
\end{align}
where $\bu$ is the velocity vector, $P$ is the pressure, $\nabla$ is the gradient operator, $\Delta=\nabla\cdot\nabla$ is the Laplacian, $(x,y,z)$ are the streamwise, wall-normal, and spanwise directions, and $t$ is time. The friction Reynolds number $Re_\tau=u_\tau \delta/\nu$ is defined in terms of the channel's half-height $\delta$ and the friction velocity $u_\tau=\sqrt{\tau_w/\rho}$, where $\tau_w$ is the wall-shear stress (averaged over horizontal directions and time), $\rho$ is the fluid density, and $\nu$ is the kinematic viscosity. In~\eqref{eq.NSeqn} and throughout this paper, spatial coordinates are nondimensionalized by $\delta$, velocity by $u_\tau$, time by $\delta/u_\tau$, and pressure by $\rho u_\tau^2$. We also assume that the bulk flux, which is obtained by integrating the streamwise velocity over spatial dimensions and time, remains constant via adjustment of the uniform streamwise pressure gradient $\partial_x P$.

When the lower channel wall is corrugated with a spanwise-periodic surface $r (z)$ that is aligned with the flow, as shown in Fig.~\ref{fig.ribletschannel}, boundary conditions on $\bu$ are given by the no-slip and no penetration conditions,
\begin{align}
\label{eq.BCs}
    	\bu(x,y=1,z,t)
	\,=\,
    	\bu(x,y=-1+r(z),z,t)
	\,=\,
	0.
\end{align}
Solving the NS equations~\eqref{eq.NSeqn} subject to these boundary conditions requires a stretched mesh that conforms to the geometry dictated by a shape function $r(z)$. This approach is computationally inefficient because it requires a large number of discretization points to resolve the grid in the vicinity of the wall. This motivates the development of low-complexity models for analysis, optimization, and design. The key challenge is to capture the effect of riblets on the turbulent flow so that skin-friction drag is accurately predicted.

As skin-friction drag depends on the gradient of the turbulent mean velocity at the wall, a natural first step is to determine an approximation to the mean velocity in the presence of riblets. To this end, we adopt the Reynolds decomposition to split the velocity and pressure fields into their time-averaged mean and fluctuating parts as
\begin{align}
\label{eq.decomp}
        \ba{rclrclrcl}
        \bu
        &\!\!=\!\!&
        \bar{\bu} \;+\; \bv,
        &\quad
        \langle\bu\rangle
        &\!\!=\!\!&
        \bar{\bu},
        &\quad
        \langle \bv\rangle
        &\!\!=\!\!&
        0,
        \\[0.15cm]
        P
        &\!\!=\!\!&
        \bar{P} \;+\; p,
        &\quad
        \langle P\rangle
        &\!\!=\!\!&
        \bar{P},
        &\quad
        \langle p\rangle
        &\!\!=\!\!&
        0.
        \ea
\end{align}
Here, $\bar{\bu}=[\,U\,~V\,~W\,]^T$ is the vector of mean velocity components, $\bv=[\,u\,~v\,~w\,]^T$ is the vector of velocity fluctuations, $p$ is the fluctuating pressure field around the mean $\bar{P}$, and $\langle\cdot\rangle$ denotes the expected value,
\begin{align}
	\langle \bu(x,y,z,t) \rangle
	\;=\;
	\lim_{T \, \to \, \infty}\dfrac{1}{T} \int^T_0 \bu(x,y,z,t+\tau)\, \mrd \tau.
\end{align}
Substituting the decomposition~\eqref{eq.decomp} into the NS equations~\eqref{eq.NSeqn} and taking the expectation yields the Reynolds-averaged NS equations
\begin{align}
\ba{rcl}
    		\partial_t\bar{\bu}
    		&\!\!=\!\!&
    		-\,
    		\left( \bar{\bu} \cdot \nabla \right) \bar{\bu}
     		\,-\,
    		\nabla \bar{P}
    		 \,+\,
    		\dfrac{1}{Re_\tau} \Delta \bar{\bu}
		\,-\,\nabla\cdot\langle\bv\bv^T\rangle,
    		\\[.15cm]		
    		0
    		&\!\!=\!\!&
    		\nabla \cdot \bar{\bu}.
    \ea
    \label{eq.NSmean}
\end{align}
The Reynolds stress tensor $\langle\bv\bv^T\rangle$ quantifies the transport of momentum arising from turbulent fluctuations~\citep{mcc91}, and its value significantly affects the solution of equations~\eqref{eq.NSmean}. The difficulty in obtaining the fluctuation correlations stems from closure problem. We overcome this challenge by utilizing the turbulent viscosity hypothesis~\citep{mcc91}, which considers the turbulent momentum to be transported in the direction of the mean rate of strain
\begin{align}
\label{eq.eddyvis}
        \langle\bv\bv^T\rangle \;-\; \dfrac{1}{3} \, \trace \, ( \langle\bv\bv^T\rangle ) \, I
        \;=\;
        -
        \dfrac{\nu_T}{Re_\tau}\left( \nabla\bar{\bu} \;+\; (\nabla\bar{\bu})^T \right),
\end{align}
where $\nu_T(y)$ is the turbulent eddy viscosity normalized by molecular viscosity $\nu$, and $I$ is the identity operator. As we discuss  {in} what follows, our choice of turbulence model is motivated by~\cite{moajovJFM12}  {that} demonstrated  {its utility in capturing} the effect of transverse wall oscillations on turbulent drag and kinetic energy in \mbox{channel flow.}

	\vspace*{-2ex}
\subsection{Modeling surface corrugation}
\label{sec.ribletmodel}

To account for the effect of riblets, we use the volume penalization technique proposed by~\cite{khaangparcal00}. In this method, the solid obstruction of the flow is modeled as a spatially-varying permeability function $K$ that enters the governing equations through an additive body forcing term. This modulation  {along with the turbulent viscosity hypothesis~\eqref{eq.eddyvis}} brings the mean flow equations~\eqref{eq.NSmean} into the following form,
\begin{align}
	\label{eq.NSvisriblet}
	\ba{rcl}
    		\partial_t\bar{\bu}
    		& \!\!=\!\! &
    		-\,
    		\left( \bar{\bu} \cdot \nabla \right) \bar{\bu}
     		\;-\;
    		\nabla \bar{P} \;-\; {K}^{-1}\bar{\bu}
            	\, + \,
    		\dfrac{1}{Re_\tau}
		\,
		\nabla\cdot\left((1\,+\,\nu_T)( \nabla\bar{\bu} \;+\; (\nabla\bar{\bu})^T )\right),
    		\\[.15cm]
    		0
    		& \!\!=\!\! &
    		\nabla \cdot \bar{\bu}.
	\ea
\end{align}

\begin{figure}
\begin{center}
\begin{tabular}{cccc}
\hspace{-.3cm}\subfigure[]{\label{fig.invKomz30}}
&&\hspace{-.8cm}
\subfigure[]{\label{fig.Amyomz30}}
&
\\[-.4cm]
\hspace{-.3cm}
\begin{tabular}{c}
\vspace{.4cm}
\rotatebox{90}{\normalsize $y$}
\end{tabular}
\hspace{-.4cm}
&
\begin{tabular}{c}
\includegraphics[width=.45\textwidth]{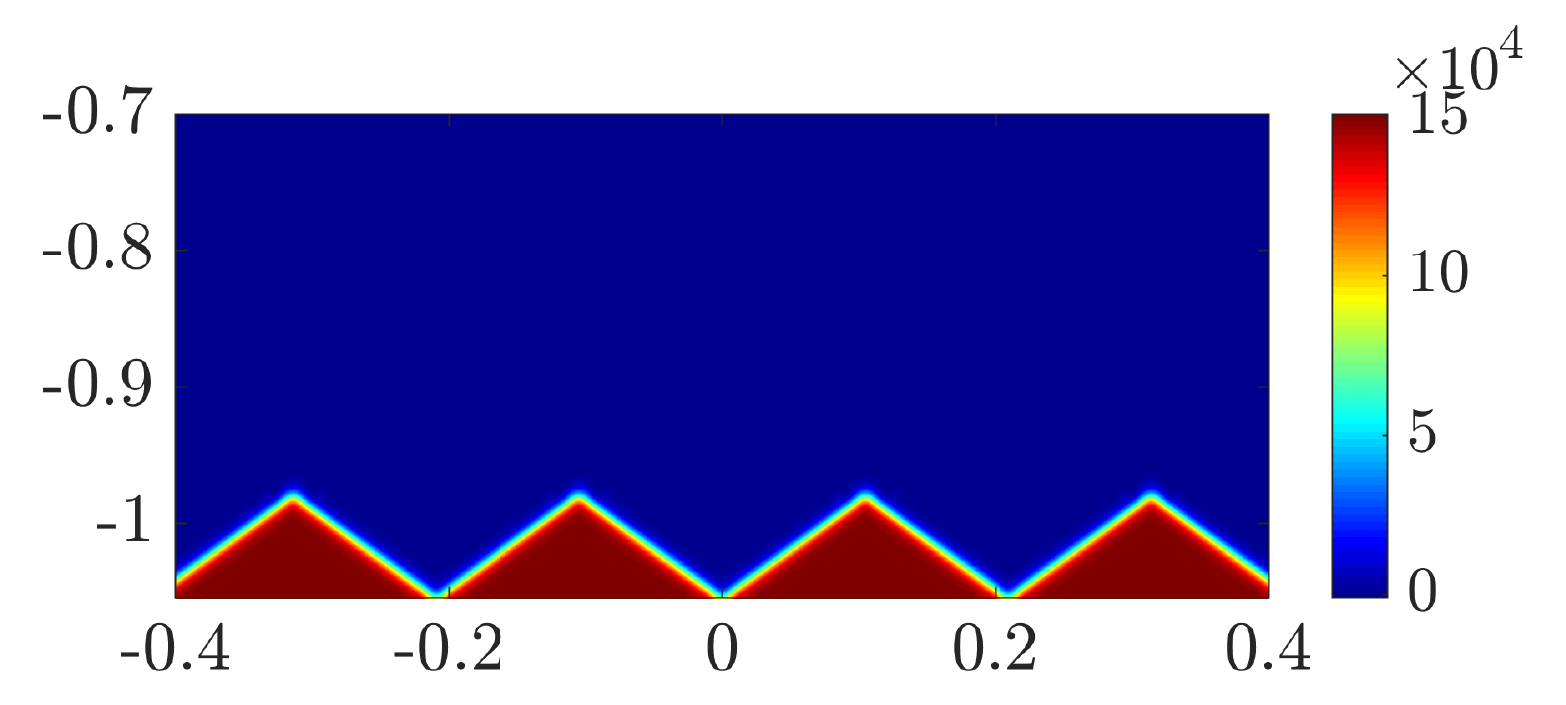}
\\[-.1cm]
\hspace{-.5cm}{\normalsize $z$}
\end{tabular}
&\hspace{-.4cm}
\begin{tabular}{c}
\vspace{.4cm}
\rotatebox{90}{\normalsize $a_m(y)$}
\end{tabular}
&\hspace{-.4cm}
\begin{tabular}{c}
\includegraphics[width=.45\textwidth]{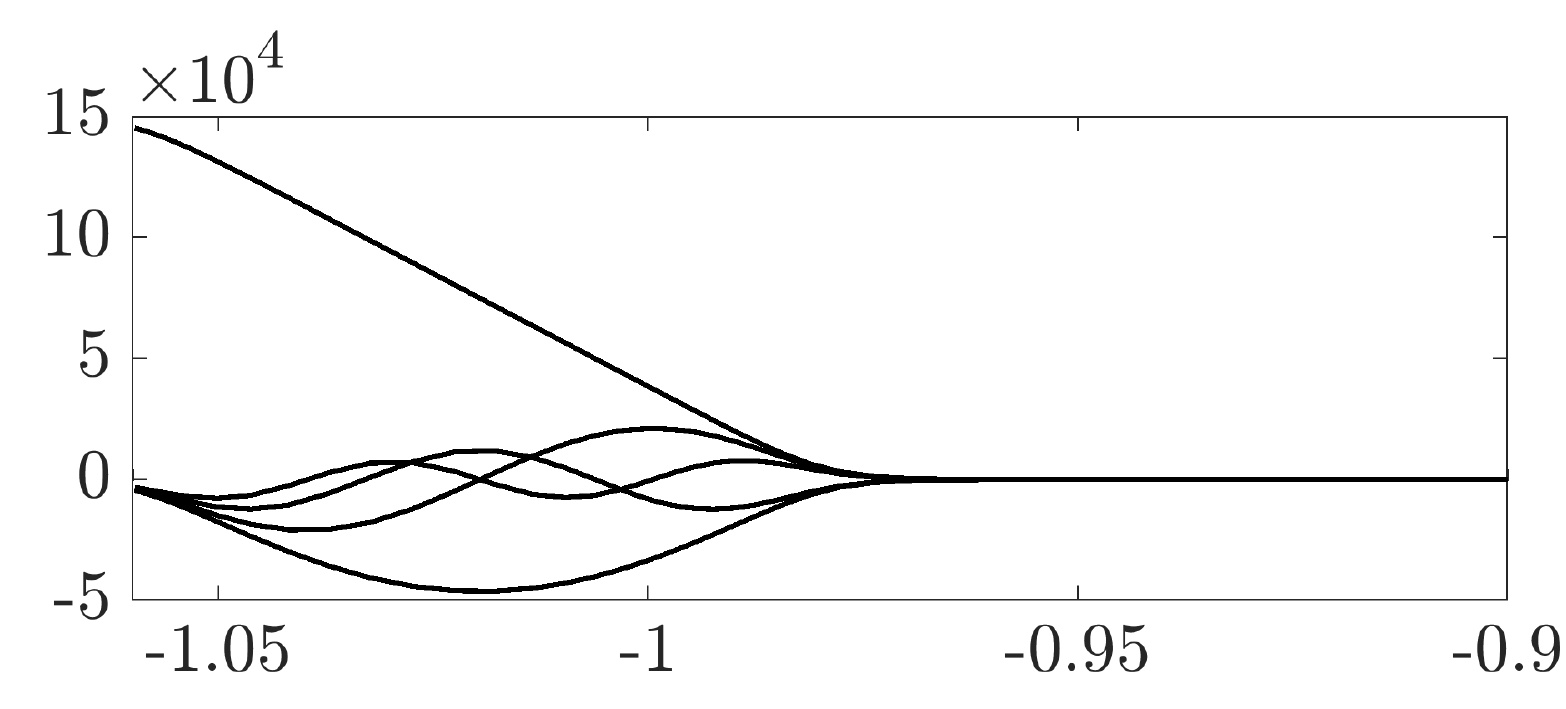}
\\[-.1cm]
{\normalsize $y$}
\end{tabular}
\end{tabular}
\end{center}
\caption{(a) Resistance function $K^{-1}(y,z)$ given by equation~\eqref{eq.invKform} with $h=0.0804$, $R = 1.5\times 10^5$, $s_f = 141$, $\omega_z=30$, and $r_p = 0.7434$. (b) The first five Fourier coefficients $a_m(y)$ with successively decreasing amplitudes corresponding to riblets shown in (a).}
\label{fig.invK}
\end{figure}

The permeability function $K$ takes on two values: within the fluid, $K \to \infty$ yields back the original mean flow equations~\eqref{eq.NSmean}; and within the riblets, $K \to 0$ forces the velocity field to zero. Following~\cite{chaluh19}, we account for streamwise-constant, spanwise-periodic corrugation by considering the harmonic resistance
\begin{align}
\label{eq.invK}
	K^{-1}(y,z)
	\;=\;
	\sum_{m \, \in \, \bbZ} a_m(y) \exp({\mri m \omega_z z}).
\end{align}
Here, $\omega_z$ is the fundamental spatial frequency of the riblets and $a_m(y)$ are the Fourier series coefficients of $K^{-1}(y,z)$. The dependence of the resistance function $K^{-1}(y,z)$ on $y$ and $z$ follows from the geometry of riblets. Figure~\ref{fig.invK} shows a resistance function and corresponding dominant coefficients for triangular riblets.

\begin{figure}
\begin{center}
\begin{tabular}{cc}
\begin{tabular}{c}
\vspace{.4cm}
\rotatebox{90}{\normalsize $K^{-1}(y,\pi/\omega_z)$}
\end{tabular}
&\hspace{-.5cm}
\begin{tabular}{c}
\includegraphics[width=.5\textwidth]{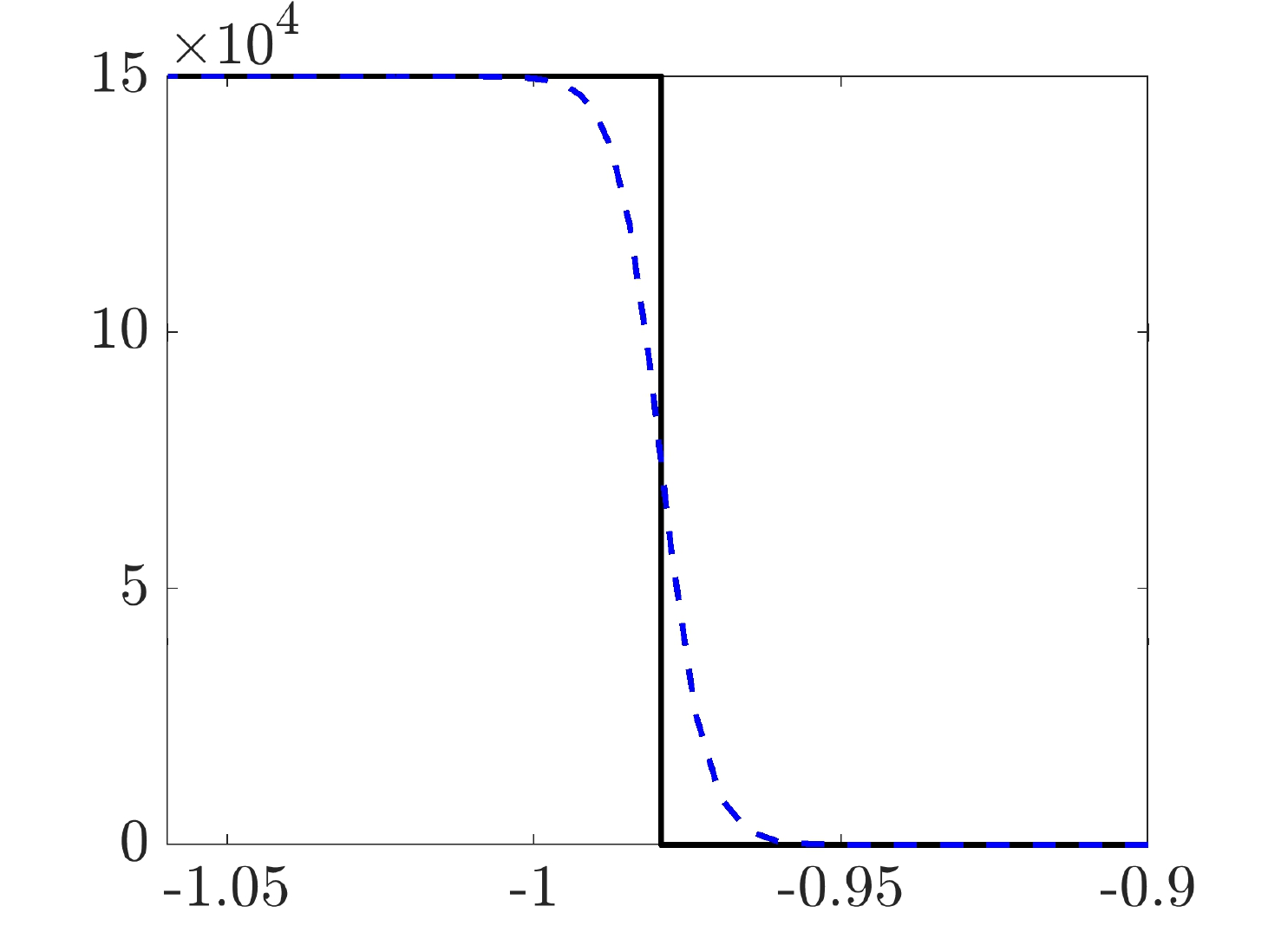}
\\[-.1cm]
{\normalsize $y$}
\end{tabular}
\end{tabular}
\end{center}
\caption{The wall-normal dependence of the resistance function {$K^{-1}(y,z)$} at the tip {($z=\pi/30$) of the triangular riblet given in figure~\ref{fig.invKomz30}}. The dashed curve results from equation~\eqref{eq.invKform} and it represents a smooth hyperbolic approximation to the step function (solid line). Here, {$h=0.0804$, $R = 1.5\times 10^5$, $s_f = 141$, $\omega_z=30$, $r_p = 0.7434$}, and the function $r(z)$ represents triangular riblets.}
\label{fig.invKy}
\end{figure}

In practice, we construct a resistance function $K^{-1}(y,z)$, e.g., the one shown in figure~\ref{fig.invKomz30}, and then compute $a_m(y)$ using the Fourier transform in $z$. Ideally, at any spanwise location $z$, the resistance should emulate a wall-normal step function at the interface of the solid riblet surface and the fluid; see figure~\ref{fig.invKy}. However, in favor of wall-normal differentiability, we use the hyperbolic approximation
\begin{align}
\label{eq.invKform}
    K^{-1}(y, z)
    \;=\;
    \dfrac{R}{2} \left( 1 \,-\, \tanh \left( s_f ( y\,+\,1\,-\,r(z) ) \right) \right)
\end{align}
where $-1+r(z)$ indicates the location of the lower corrugated wall (cf.\ Eq.~\eqref{eq.BCs}), $s_f$ is a smoothness factor that modifies the slope of the hyperbolic curve, and $R$ is a resistance rate that controls the accuracy of the solution in the solid region. While larger values of $s_f$ yield a better approximation of the step function, they require the use of a larger number of harmonics to maintain the smoothness of the resistance field. Herein, we choose $s_f$ to be inversely proportional to the height of the riblets $h$, i.e., $s_f = 3.6\pi/h$. On the other hand, while large values of the resistance rate $R$ induce a smaller velocity field within the riblets, they may trigger spurious negative solutions. In view of this fundamental trade-off, we relax the non-negativity constraint on $\bar{\bu}$ and choose $R$ to guarantee that the solution to~\eqref{eq.NSvisriblet} is larger than $-1\times10^{-6}$. In particular, for turbulent channel flow with $Re_\tau=186$ over the triangular lower-wall riblets with frequency $\omega_z=30$ and height to spacing ratio $h/s=0.38$, our computational experiments show that $R = 1.5 \times 10^{5}$, $s_f=141$, and $25$ spanwise harmonics ($m=-12, \dots, 12$) yield small negative mean velocity while preserving the smoothness of the resistance field. For triangular riblets with $\omega_z=30$ and height $h=0.0804$, figure~\ref{fig.invKomz30} shows the resistance field $K^{-1}$ resulting from Eq.~\eqref{eq.invKform} with,
\begin{align}
\label{eq.rz}
	r(z)
	\;=\;
	-h\,r_p \; + \; \dfrac{h\,\omega_z}{\pi}
	\left|
	z \; - \;
	\dfrac{2\pi}{\omega_z}
	\Big(1 \, + \, \floor{\dfrac{z \, \omega_z}{2 \pi}  \, - \, \dfrac{1}{2}}\Big)
	\right|.
\end{align}
Here, $| \cdot |$ is the absolute value, $\floor \cdot$ is the floor function, and $r_p$ denotes the proportion of the riblet height in the extended channel, i.e., below $y=-1$. In this study, we tune $r_p$, and thereby adjust the wall-normal position of riblets, so that the mean velocity profile resulting from~\eqref{eq.NSvisriblet} has the same bulk as the channel flow with smooth walls.  {While the choice of $r(z)$ in equation~\eqref{eq.rz} corresponds to triangular riblets, which are used as a case study throughout this paper, the shape function $r(z)$ can be selected to account for an arbitrary spanwise-periodic surface corrugation.}

For a given smoothness factor $s_f$, we start from an initial choice of $r_p$ and resistance rate $R$ and iterate steps (i)-(iii) below to identify $r_p$ and the largest $R$ that ensure that the mean velocity is greater than $-1\times10^{-6}$ and that it satisfies the constant bulk flux condition.
\begin{enumerate}
\vsp
\item{Determine the shape function $r(z)$ to capture the desired geometry of riblets.}
\vspace{.1cm}
\item{Use the shape function $r(z)$ to construct the resistance function $K^{-1}(y,z)$ using the hyperbolic approximation~\eqref{eq.invKform} and derive the Fourier series coefficients $a_m(y)$.}
\vspace{.1cm}
\item{Solve~\eqref{eq.NSvisriblet} for $\bar{\bu}$ and check to see if it has the same bulk as the turbulent channel flow with smooth walls.}
\end{enumerate}

	\vspace*{-3ex}
\subsection{The turbulent mean velocity}
\label{sec.baseflow}

We approach the problem of quantifying the influence of riblets on skin-friction drag by developing robust models that approximate the turbulent viscosity $\nu_T$ in equations~\eqref{eq.NSvisriblet}. Several studies have offered expressions for $\nu_T$ that yield the turbulent mean velocity in the flow over smooth walls~\citep{mal56,ces58,reytie67}. The following turbulent viscosity model for channel flow was developed by~\cite{reytie67} as an extension of the model introduced by~\cite{ces58} for pipe flow:
\begin{align}
    \nu_{Ts}(y)
    \; = \;
    \dfrac{1}{2} \left( \left( 1\,+ \left(\dfrac{c_2}{3} Re_\tau(1 \,- y^2)(1 \,+\, 2y^2)\,(1 \,- \mre^{-(1- |y|)Re_\tau/c_1})\right)^2\right)^{1/2}-\,1\right).
    \label{eq.nuT0}
\end{align}
In this expression, parameters $c_1$ and $c_2$ are selected to minimize the least squares deviation between the mean streamwise velocity obtained in experiments and simulations and the steady-state solution to Eq.~\eqref{eq.NSvisriblet} without riblets using the averaged wall-shear stress $\tau_w=1$ and $\nu_T$ given by Eq.~\eqref{eq.nuT0}. For turbulent channel flow with $Re_\tau=186$, the optimal parameters $c_1=46.2$ and $c_2=0.61$ {provide the best fit to the mean velocity in a turbulent channel flow resulting from DNS~\citep{deljim03,deljimzanmos04}. For the turbulent channel flow with $Re_\tau=547$ discussed in \S~\ref{sec.dragreduction} and \S~\ref{sec.VLSM} these parameters are $c_1=29.4$ and $c_2=0.45$.} Even though the turbulent viscosity model given by Eq.~\eqref{eq.nuT0} does not hold in the presence of riblets, we use $\nu_{Ts}$ as a starting point for determining the mean flow in the presence of riblets. Furthermore, in the vicinity of the solid wall the flow is dominated by viscosity and, for small-size riblets, the flow in the grooved region can be assumed to be laminar. Thus, we consider small-size riblets and set $\nu_T = 0$ for $y \leq -1$.

As shown in \S~\ref{sec.ribletmodel}, a harmonic resistance function $K^{-1}(y, z)$ is used to model a spatially periodic surface corrugation; cf.~\eqref{eq.invK}. The corresponding base flow, i.e., the solution to the steady-state mean flow equations~\eqref{eq.NSvisriblet}, can be also decomposed into the Fourier series
\begin{align}
\label{eq.baseflow}
	\bar{\bu}(y,z)
	\;=\;
	\sum_{m \, \in \, \bbZ}
	\bar{\bu}_m(y) \exp({\mri m \omega_z z}).
\end{align}
The steady-state solution to the nonlinear mean flow equations~\eqref{eq.NSvisriblet} is obtained via Newton's method and it only contains a streamwise velocity component, $\bar{\bu} = [\,\bar{U}(y,z) \,~0\,~0\,]^T$. Since the spanwise and wall-normal base flow components are zero, the nonlinear terms in mean flow equation~\eqref{eq.NSvisriblet} vanish and the equation for $\bar{U}(y,z)$ is linear,
\begin{align}
\label{eq.baseflow-U}
	\left(1 \,+\, \nu_T\right) \Delta \bar{U} \;+\; \nu'_T\,\bar{U}' \;-\; K^{-1}\bar{U}
	\;=\;
	Re_\tau \bar{P}_{x}.
\end{align}
Here $\bar{U}'$ denotes the wall-normal derivative of $\bar{U}$ and $\bar{P}_{x}$ is the mean pressure gradient. Inclusion of the harmonics of $K^{-1}$ yields the equation for the $m$th harmonic $\bar{U}_m$,
\begin{align*}
	\underbrace{\left[\left(1\,+\,\nu_T\right) \left(\partial_y^2 \,-\, m^2\omega_z^2\right) \,+\, \nu'_T\partial_y \,-\, a_0\right]}_{\bL_{m,0}}\bar{U}_m
	\;+\,
	\underbrace{\ds{\sum_{n \, \in \, \bbZ \setminus \! \{ 0 \}}} a_n}_{\bL_{m,n}} \bar{U}_{m-n}
	\;=\;
	\left\{
	\ba{ll}\!\!
        Re_\tau\, \bar{P}_{x}, & m=0
        \\[0.2cm]
        \!\!
       	 0, & m\neq 0
        \ea
        \right.
\end{align*}
which amounts to the following bi-infinite matrix form: 
\begin{align}
\label{eq.baseflow-U-matrix}
	\left[
	\ba{ccccc}
         \ddots \!\!&\!\! \vdots \!\!&\!\! \vdots \!\!&\!\! \vdots \!\!&\!\! \rotatebox{90}{$\ddots$}
         \\[0.1cm]
         \cdots \!\!&\!\! \bL_{m-1,0} \!\!&\!\! \bL_{m-1,+1} \!\!&\!\! \bL_{m-1,+2} \!\!&\!\! \cdots
         \\[0.15cm]
         \cdots \!\!&\!\! \bL_{m,-1} \!\!&\!\! \bL_{m,0} \!\!&\!\! \bL_{m,+1} \!\!&\!\!  \cdots
         \\[0.15cm]
         \cdots \!\!&\!\!  \bL_{m+1,-2} \!\!&\!\! \bL_{m+1,-1} \!\!&\!\! \bL_{m+1,0} \!\!&\!\! \cdots
         \\[0.1cm]
         \rotatebox{90}{$\ddots$} \!\!&\!\! \vdots \!\!&\!\! \vdots \!\!&\!\! \vdots \!\!&\!\! \ddots
    	\ea
	\right]
	\left[
	\ba{c}
	\vdots
	\\[0.1cm]
	\bar{U}_{-1}
	\\[0.1cm]
	\bar{U}_{0}
	\\[0.1cm]
	\bar{U}_{+1}
	\\[0.1cm]
	\vdots
	\ea
	\right]
	=
	\left[
	\ba{c}
	\vdots
	\\[0.1cm]
	0
	\\[0.1cm]
	Re_\tau\, \bar{P}_{x}
	\\[0.1cm]
	0
	\\[0.1cm]
	\vdots
	\ea
	\right].
\end{align}

\begin{figure}
\begin{center}
\begin{tabular}{cc}
\begin{tabular}{c}
\vspace{.4cm}
\rotatebox{90}{\normalsize $y$}
\end{tabular}
&\hspace{-.3cm}
\begin{tabular}{c}
\includegraphics[width=.56\textwidth]{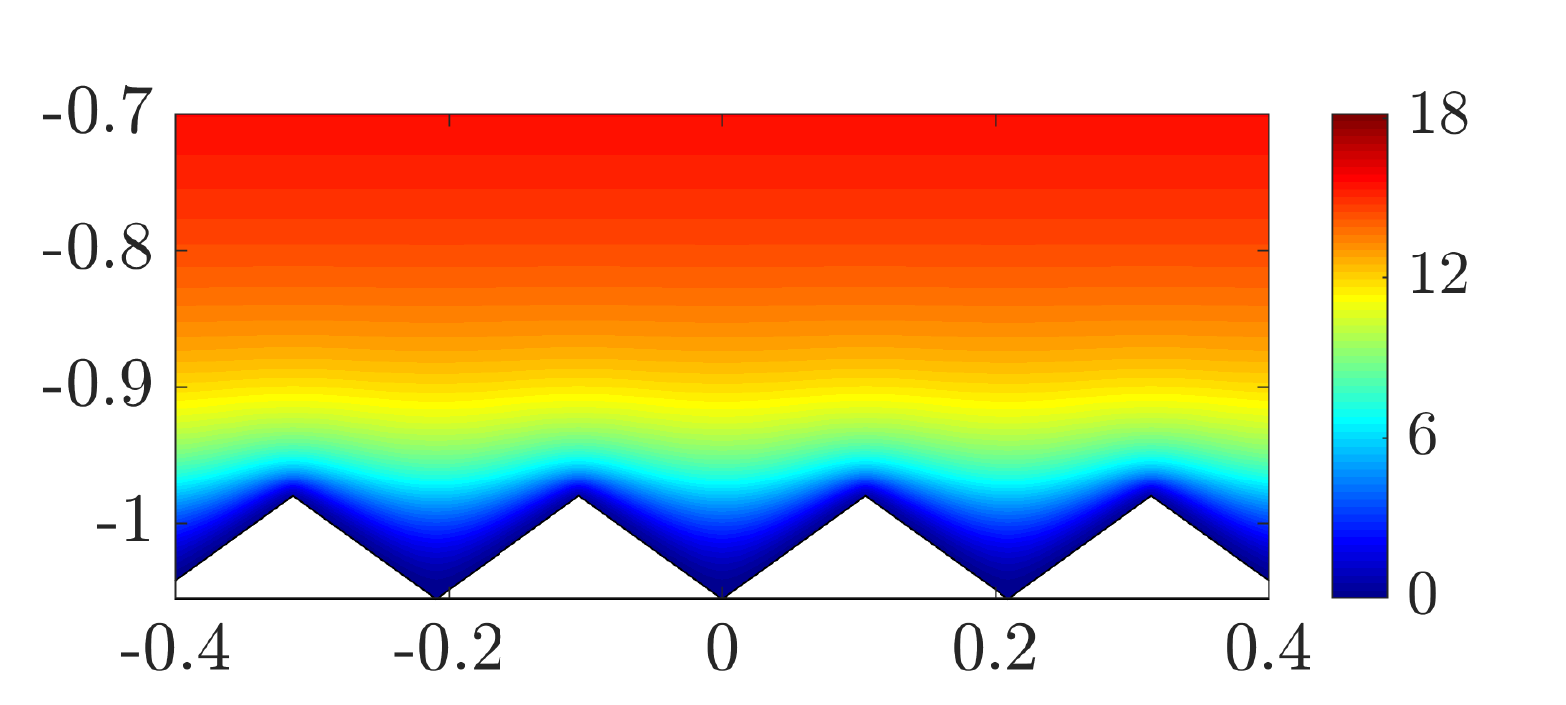}
\\[-.1cm]
\hspace{-.6cm}{\normalsize $z$}
\end{tabular}
\end{tabular}
\end{center}
\caption{The streamwise mean velocity {$\bar{U}(y,z)$} for turbulent channel flow with $Re_\tau=186$ over the triangular riblets {given in~figure~\ref{fig.invKomz30}}.}
\label{fig.baseflow}
\end{figure}

A pseudospectral scheme with Chebyshev polynomials~\citep{weired00} is used to discretize the differential operators in the wall-normal direction. To avoid numerical oscillations in the solution to equations~\eqref{eq.baseflow-U-matrix}, we divide the wall-normal extent of the computational domain into two parts using block operators~\citep{aurtre17} and use $N_{i}$ collocation points for $y\in[-1,1]$ and $N_{o}$ collocation points for $y\in[-1-r_p h,-1]$. We impose no-slip boundary conditions~\eqref{eq.BCs} on the upper wall. Ideally, the adopted volume penalization method should automatically enforce immersed boundary conditions on the non-smooth lower wall without the need for additional boundary conditions. However, in practice, since the resistance rate $R$ in~\eqref{eq.invKform} is a finite number, the immersed boundary conditions cannot be exactly enforced. To ensure that the operators in~\eqref{eq.baseflow-U-matrix} are well-defined, we employ additional no-slip conditions at the lower boundary ($y=-1-r_p h$). The boundary conditions at the intersection of the aforementioned wall-normal regimes ($y=-1$) enforce smoothness on physical quantities, i.e.,
\begin{align*}
	\ba{rcl}
        \bar{U}(y\,=\,-1^+,z)
        &\!\!=\!\!&
        \bar{U}(y\,=\,-1^-,z)
        \\[.15cm]
	\dfrac{\partial \bar{U}}{\partial y}(y\,=\,-1^+,z)
	&\!\!=\!\!&
	\dfrac{\partial \bar{U}}{\partial y}(y\,=\,-1^-,z).
        \ea
\end{align*}
Figure~\ref{fig.baseflow} shows the solution $\bar{U}$ to equation~\eqref{eq.NSvisriblet} for a turbulent channel flow with $Re_\tau=186$ subject to a streamwise pressure gradient $\bar{P}_{x}=-1$ over triangular riblets. Here, $N_i=179$, $N_o=20$, $25$ harmonics have been used to approximate the solution, i.e., $m = -12,\dots,12$, and the triangular riblets are characterized by $K^{-1}(y,z)$ with $h=0.0804$, $\omega_z=30$, $R = 1.5\times 10^5$, $s_f = 141$, and $r_p = 0.7434$. Figure~\ref{fig.baseflow} demonstrates that the solution respects the shape of riblets and is approximately zero in the solid region.

	\vspace*{-1ex}
\subsection{Prediction of drag reduction}
\label{sec.drag}

In what follows, subscripts $s$ and $c$ are used to signify channel flow with smooth walls and the correction that arises from surface corrugation. We use a variation in the driving pressure gradient to enforce a constant bulk flux requirement. This introduces a correction to the $0$th harmonic of  {the} mean velocity as
\begin{align}
\label{eq.Uc}
	\bar{U}_c(y)
	\;=\;
	 \bar{U}_0(y) \;-\; \left(1\,-\, \bar{P}_{x,c}\right)\bar{U}_s(y).
\end{align}
Here, $\bar{U}_s(y)$ denotes the mean velocity profile in channel flow with smooth walls and the additional bulk introduced by the $0$th harmonic of the solution to equation~\eqref{eq.baseflow-U} is used to compute the correction to pressure gradient $\bar{P}_{x,c}$, i.e., 
\begin{align}
\label{eq.Pxc}
	\bar{P}_{x,c}
	\;=\;
	1 \;-\; \dfrac{1}{U_B} \int_{-1-h\, r_p}^1\bar{U}_0(y)\, \mrd y.
\end{align}
The form of $\bar{P}_{x,c}$ ensures that the mean velocity correction $\bar{U}_{c}(y)$ in equation~\eqref{eq.Uc} has zero bulk. The corrections to the pressure gradient and mean velocity are used to compute the variation in skin-friction drag.

The rate of drag reduction caused by riblets is given by
\begin{align*}
        \Delta D
       \;\DefinedAs\;
        \left(
        D \,-\, D_s
        \right)
        /D_s,
\end{align*}
where $D_s$ denotes the slope of the mean velocity at the lower wall in a flow without riblets. In a flow with riblets, the skin-friction drag at the lower wall can be computed using the pressure gradient, which maintains a constant bulk, and the well-defined slope of the mean velocity at the upper wall,
\begin{align*}
	D
	\;=\; 
	\bar{P}_{x} \;-\;
	\dfrac{\omega_z}{2\pi} \int_0^{{2\pi}/{\omega_z}}
	\dfrac{\partial {\bar{U}}}{\partial y}(y=1,z) \, \mrd z.
\end{align*}
Since $\bar{P}_{x}=2D_s$, $ \Delta D$ is determined by the difference between the pressure gradient adjustment and the drag reduction at the upper wall, i.e.,
\begin{align}
\label{eq.DR}
        \Delta D
       \;=\;
        \frac{1}{D_s}\left[
        \bar{P}_{x,c} \,-\, \left(\dfrac{\omega_z}{2\pi} \int_0^{{2\pi}/{\omega_z}} \dfrac{\partial \bar{U}}{\partial y}(y=1,z)\, \mrd z \,-\, D_s\right)
        \right].
\end{align}

\begin{figure}
\begin{center}
\begin{tabular}{cc}
\begin{tabular}{c}
\vspace{.4cm}
\rotatebox{90}{\normalsize  $\Delta D(\%)$}
\end{tabular}
&\hspace{-.5cm}
\begin{tabular}{c}
\includegraphics[width=.5\textwidth]{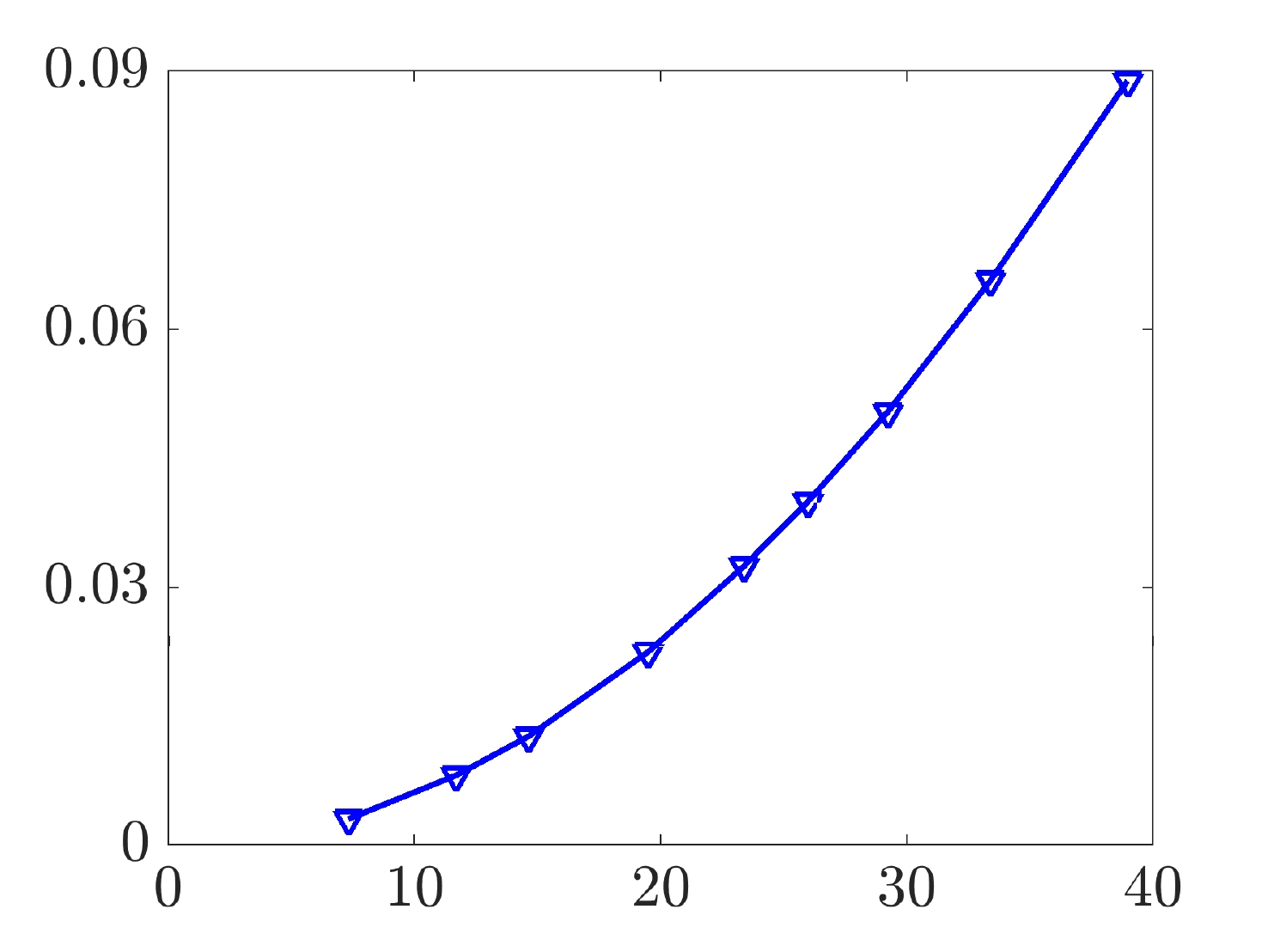}
\\[-.1cm]
{\normalsize $s^+$}
\end{tabular}
\end{tabular}
\end{center}
\caption{Prediction of drag reduction in a turbulent channel flow with $Re_\tau=186$ resulting from the steady-state solution of equations~\eqref{eq.NSvisriblet} with {$\nu_{Ts} (y)$} given by~\eqref{eq.nuT0}. Triangular riblets, shown in figure~\ref{fig.riblet-sketch}, with different peak-to-peak spacing but a constant height to spacing ratio $h/s = 0.38$ are considered and the spacing is reported in inner viscous units, i.e., $s^+ = Re_\tau s$.}
\label{fig.DRnuT0}
\end{figure}

Even though the mean velocity profile shown in figure~\ref{fig.baseflow} respects the shape of riblets and goes to zero within the solid region, the resulting drag reduction does not follow trends reported in literature. As demonstrated in figure~\ref{fig.DRnuT0}, the mean velocity profile resulting from the use of $\nu_{Ts}$ implies a reduction in drag regardless of the size of riblets. Furthermore, no optimal spacing that maximizes drag reduction is identified. To improve predictions of the mean velocity and the resulting skin-friction~drag, in \S~\ref{sec.fluctuations} we extend the framework proposed in~\cite{moajovJFM12} to account for the effect of velocity fluctuations in a flow over riblets on the turbulent viscosity $\nu_T$.

	\vspace*{-3ex}
\section{Stochastically-forced dynamics of velocity fluctuations}
\label{sec.fluctuations}

\begin{figure}
\begin{center}
         %_______________________________________________________________________________
%
%   tikz figure for inclusion in exercises:
%
%   clp_2dof_output_pert_config:  closed loop system with perturbed models.  This uses
%                                              and input perturbation and the 2-DOF structure used for
%                                              the satellite example.
%
%   Marcello Colombino, 4 May 2015
%
%_______________________________________________________________________________
%
% TikZ styles for drawing
%
\input{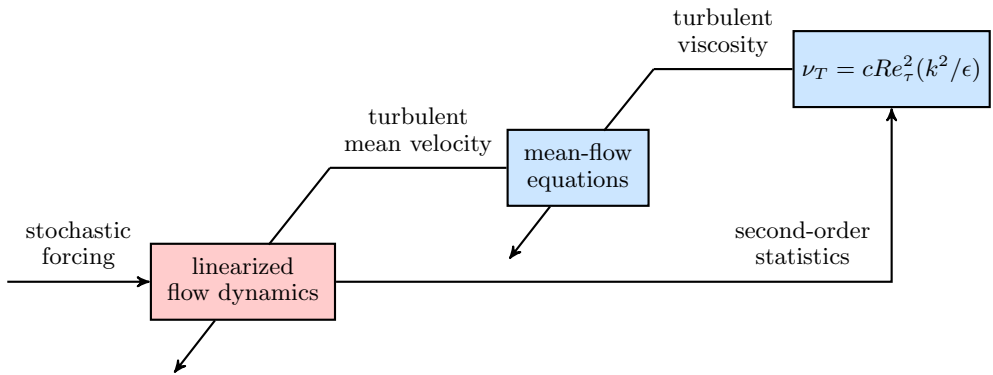}
%
%   set a filename for externalization
% \tikzsetnextfilename{clp_2dof_input_pert_config}
%
\noindent
\begin{tikzpicture}[scale=1, auto, >=stealth']

    \footnotesize

   % specify with respect to the plant
     \node[block, minimum height = 1cm, top color=red!20, bottom color=red!20] (sys0) {\begin{tabular}{c}linearized\\ flow dynamics\end{tabular}};     
     
     \node[block, minimum height = 1cm, top color=RoyalBlue!20, bottom color=RoyalBlue!20] (sys1) at ($(sys0.east) + (3.2cm, 1.5cm)$) {\begin{tabular}{c}mean-flow\\ equations\end{tabular}};
     
     \node[block, minimum height = 1cm, top color=RoyalBlue!20, bottom color=RoyalBlue!20] (sys2) at ($(sys1.east) + (3.2cm, 1.3cm)$) {$\nu_T=cRe_\tau^2(k^2/\epsilon)$};

     \node[] (input-node) at ($(sys0.west) + (-2.0cm, 0cm)$) {};

     %\node[] (output-node) at ($(sys1.west) + (-4cm, 0cm)$) {$\mathrm{turbulent \atop drag}$};
     
     \node[] (in-node1) at ($(sys0.north) + (0.25cm, -0.14cm)$) {};
     
     \node[] (end-node0) at ($(sys0.south) + (-0.28cm, 0.12cm)$) {};
     
     \node[] (end-node1) at ($(sys0.south) + (-1.0cm, -0.8cm)$) {};

     %\node[] (mid1) at ($(sys2.east) + (2.3cm,0cm)$) {};
     
     \node[] (mid2) at ($(sys1.west) + (-2.45cm,0cm)$) {};
     
     \node[] (mid20) at ($(sys2.west) + (-1.95cm,0cm)$) {};
     
     \node[] (mid21) at ($(sys1.north) + (0.25cm, -0.14cm)$) {};
     
     \node[] (mid22) at ($(sys1.south) + (-0.28cm, 0.12cm)$) {};
     
     \node[] (mid23) at ($(sys1.south) + (-1.0cm, -0.8cm)$) {};
    % now link the nodes
    
    \draw [connector, thick] (input-node) -- node [midway, above] {\begin{tabular}{c}stochastic\\ forcing\end{tabular}} (sys0.west);
    
    \draw [connector, thick] (sys0.east) -| node [pos=0.42, above] {\begin{tabular}{c}second-order\\ statistics\end{tabular}} (sys2.south);
	
    %\draw [line, double] (sys2.east) -- node [midway, above] {$\mathrm{turbulent \atop mean\ viscosity}$} (mid1.east);

    %\draw [connector, double] (mid1.east) |- (sys1.east);
    
    \draw [line, thick] (sys2.west) -- node [midway, above] {\begin{tabular}{c}turbulent\\ viscosity\end{tabular}} (mid20.east);

    \draw [line, thick] (mid20.east) -- (mid21);

    \draw [connector, thick] (mid22) -- (mid23);

    \draw [line, thick] (sys1.west) -- node [midway, above] {\begin{tabular}{c}turbulent\\ mean velocity\end{tabular}} (mid2.east);

    \draw [line, thick] (mid2.east) -- (in-node1);

    \draw [connector, thick] (end-node0) -- (end-node1);

\end{tikzpicture}
%_______________________________________________________________________________
\end{center}
\caption{Block diagram of our simulation-free approach for determining the influence of riblets on skin-friction drag in turbulent flows. The slanted lines represent coefficients into the mean flow and linearized equations.}
\label{fig.blockdiagram}
\end{figure}

In this section, we compute a correction to the turbulent viscosity and, subsequently, the mean velocity of a turbulent channel flow over riblets using second-order statistics of velocity fluctuations. To this end, we examine the dynamics of fluctuations around the mean velocity profile computed in \S~\ref{sec.drag}. {As illustrated in figure~\ref{fig.blockdiagram},} our model-based framework for studying the effect of riblets involves the following steps: 
\vsp
\begin{enumerate}
  \item{{[\S~\ref{sec.baseflow}]} The turbulent mean velocity $\bar{\bu}$ is obtained from equations~ {\eqref{eq.NSvisriblet}}, where closure is achieved using the turbulent viscosity $\nu_{Ts}$ for the channel flow with smooth~walls.}
   \vspace{.1cm}
  \item{[\S~\ref{sec.correctnuT}] The stochastically forced linearized NS equations around the mean flow $\bar{\bu}$ resulting from step~(i) are used to compute the second-order statistics of the fluctuating velocity field and provide a correction to $\nu_{Ts}$.}
	 \vspace{.1cm}
  \item{{[\S~\ref{sec.baseflow} and \S~\ref{sec.drag}]} The modification to turbulent viscosity is used to correct the mean velocity and compute skin-friction drag.}
\end{enumerate}
	\vsp

{In \S~\ref{sec.dragreduction}, we show that the correction to the mean velocity  {significantly improves our prediction of} the optimal size of triangular riblets for drag reduction. 
The separation of steps (i) and (iii), in which the mean velocity is updated, from step (ii), in which the statistics of velocity fluctuations are computed, is justified by the slower time evolution of the mean velocity compared to fluctuations~\citep{moajovJFM12}.
While the turbulent viscosity and the mean velocity can be updated in an iterative manner, a theoretical justification for the convergence of such an iterative procedure requires additional examination and is outside of the scope of the current study. Even though our discussion focuses on spanwise-periodic triangular riblets, the methodology and theoretical framework that we develop can be used to study turbulent flows over a much broader class of periodic surface corrugations.}

\subsection{Model equation for $\nu_T$}
\label{sec.modelnuT}

As described in \S~\ref{sec.drag}, $\nu_{Ts}$ does not provide the proper eddy viscosity model for the channel flow with riblets. Establishing a relation between $\nu_T$ and the second-order statistics of velocity fluctuations represents the main challenge for identifying the appropriate eddy viscosity model. With appropriate choices of velocity and length scales, turbulent viscosity can be expressed as~\citep{pop00}
\begin{align}
\label{eq.nuT}
	\nu_T(y)
	\;=\;
	c \, Re_\tau^2 \, \dfrac{k^2(y)}{\epsilon(y)}
\end{align}
where $c = 0.09$, $k$ is the turbulent kinetic energy, and $\epsilon$ is the rate of dissipation. The $k$-$\eps$ model~\citep{jonlau72,lausha74} provides two differential transport equations for $k$ and $\eps$, but is computationally demanding and does not offer insight into analysis, design, and optimization. On the other hand, wall-normal profiles for $k$ and $\eps$ can be obtained by averaging the second-order statistics of velocity fluctuations over the streamwise coordinate and one period of the spanwise surface corrugation:
{
\begin{align}
\label{eq.kandepsilon}
 \ba{rcl}
 k(y)
 &\!\!=\!\!&
 \dfrac{1}{2}\,(\overline{\langle uu\rangle} \;+\;\overline{\langle vv\rangle} \;+\;\overline{\langle ww\rangle})
 \\[0.35cm]
 \epsilon(y)
 &\!\!=\!\!&
 2\,(\overline{\langle u_xu_x\rangle}\;+\;\overline{\langle v_yv_y\rangle}\;+\;\overline{\langle w_zw_z\rangle}\;+\;\overline{\langle u_yv_x\rangle}\;+\;\overline{\langle u_zw_x\rangle}\;+\;\overline{\langle v_zw_y\rangle})
 \\[0.15cm]
 &&
 +\;\overline{\langle u_yu_y\rangle}\;+\;\overline{\langle w_yw_y\rangle}\;+\;\overline{\langle v_xv_x\rangle}\;+\;\overline{\langle w_xw_x\rangle}\;+\;\overline{\langle u_zu_z\rangle}\;+\;\overline{\langle v_zv_z\rangle}.
 \ea
\end{align}
Here, overline denotes averaging in $x$ and one period in $z$.} We next demonstrate how second-order statistics, e.g., $uu$, can be computed using the stochastically forced linearized NS equations.

	\vspace*{-2ex}
\subsection{Stochastically forced linearized Navier-Stokes equations}
\label{sec.LNSE}

	The dynamics of infinitesimal velocity $\bv=[\,u\,~v\,~w\,]^T$ and pressure $p$ fluctuations around $\bar{\bu} = [\,{\bar{U}}(y,z)\,~0\,~0\,]^T$ and $\bar{P}$ are governed by the linearized NS and continuity equations:
\begin{align}
\label{eq.turblinriblet}
	\ba{rclcll}
    		\partial_t \bv
    		&\!\!=\!\!&
    		-
    		\left( \nabla \cdot \bar{\bu} \right) \bv
     		&\!\!\!\! - \!\!\!\!&
    		\left( \nabla \cdot \bv \right) \bar{\bu}
    		\; - \;
    		\nabla p \;-\; {K}^{-1}\bv
            	\\[.15cm]
		&&
    		&\!\!\!\! + \!\!\!\!&\dfrac{1}{Re_\tau} \, \nabla\cdot\left((1\,+\,\nu_T)( \nabla\bv \;+\; (\nabla\bv)^T )\right)
		\;+\;
		\mathbf{f},
    		\\[.35cm]	
    		0
		&\!\!=\!\!&
		\nabla \cdot \bv,&&
    \ea
\end{align}
where $\mathbf{f}$ is a zero-mean white-in-time additive stochastic forcing. The normal modes in $x$ are given by $\mre^{\mri k_x x}$, where $k_x$ is the streamwise wavenumber, and the normal modes in $z$ are given by the Bloch waves~\citep{odekel64,benliopap78}, which are determined by the product of $\mre^{\mri\theta z}$ with $\theta\in[0,\omega_z/2)$ and a $2\pi/\omega_z$ periodic function in $z$. For example, the forcing field in~\eqref{eq.turblinriblet} can be represented as
\begin{align}
\label{eq.bloch1}
        \left.
        \ba{rcl}
        \mathbf{f}(x,y,z,t)
        &\!\!=\!\!&
        \mre^{\mri k_x x}\, \mre^{\mri\theta z}
        \,
        \hat{\mathbf{f}}(k_x, y,z,t)
        \\[0.15cm]
        \hat{\mathbf{f}}(k_x,y,z,t)
        &\!\!=\!\!&
        \hat{\mathbf{f}}(k_x, y, z + 2\pi/\omega_z, t)
        \ea
        \right\}
        \quad
        k_x \,\in\,\bbR,\quad
        \theta \,\in\, [0,\,\omega_z/2),
\end{align}
where real parts are used to represent physical quantities. The Fourier series expansion of the $2\pi/\omega_z$-periodic function $\hat{\mathbf{f}}(k_x,y,z,t)$ can be used to obtain,
\begin{align}
\label{eq.bloch2}
        \ds{
        \mathbf{f}(x,y,z,t)
        \;=\;
        \sum_{n \, \in \, \bbZ}
        \hat{\mathbf{f}}_n (k_x,y,\theta,t)
        \,
        \mre^{\mri(k_x x \,+\, \theta_n z)},
        }
        \ba{ll}
        &\theta_n \;=\; \theta \,+\, n\omega_z,\\[0.1cm]
        &k_x\,\in\,\bbR,\;
        \theta\,\in\, [0,\,\omega_z/2)
        \ea
\end{align}
where $\{\hat{\mathbf{f}}_n (k_x,y,\theta,t)\}_{n \, \in \, \bbZ}$ are the Fourier coefficients of the function $\hat{\mathbf{f}}(k_x,y,z,t)$ in~\eqref{eq.bloch1}.

Substituting~\eqref{eq.bloch2} into the linearized equations~\eqref{eq.turblinriblet} and eliminating pressure through a standard conversion~\citep{schhen01} yields the evolution form
\begin{align}
	\label{eq.evolutionform}
	\ba{rcl}
		\partial_t
		\bvarphi_\theta(k_x,y,t)
		&\!\!=\!\!&
		[\bA_\theta(k_x)\, \bvarphi_\theta (k_x, \, \cdot \, ,t)](y)
		\;+\;
		\bd_\theta (k_x,y,t),
		\\[.15cm]
		\bv_\theta(k_x,y,t)
		&\!\!=\!\!&
		[\bC_\theta(k_x)\, \bvarphi_\theta(k_x, \, \cdot \,  ,t)](y),
	\ea
\end{align}
with the state $\bvarphi_\theta$ consisting of the wall-normal velocity $v$ and vorticity $\eta = \partial_z u - \partial_x w$. The state-space representation~\eqref{eq.evolutionform} is parameterized by the streamwise wavenumber $k_x$ and the spanwise wavenumber offset $\theta$: for each $k_x$ and $\theta$, $\bvarphi_\theta$, $\bv_\theta$, and $\bd_\theta \DefinedAs \bB_\theta \mathbf{f}_{\theta}$ are bi-infinite column vectors, e.g., $\bvarphi_{\theta}(k_x, y, t) = \text{col}\{ \hat{\bvarphi}_n (k_x, y, \theta, t) \}_{n \, \in \, \bbZ}$, and $\bA_{\theta}(k_x)$, $\bB_{\theta}(k_x)$, and $\bC_{\theta}(k_x)$ are bi-infinite matrices whose elements are operators in $y$, e.g.,
\begin{align}
\label{eq.Aform}
	\bA_\theta
	\;\DefinedAs\;
	\left[
	\ba{ccccc}
         \ddots \!\!&\!\! \vdots \!\!&\!\! \vdots \!\!&\!\! \vdots \!\!&\!\! \rotatebox{90}{$\ddots$}
         \\[0.1cm]
         \cdots \!\!&\!\! \bA_{n-1,0} \!\!&\!\! \bA_{n-1,+1} \!\!&\!\! \bA_{n-1,+2} \!\!&\!\! \cdots
         \\[0.15cm]
         \cdots \!\!&\!\! \bA_{n,-1} \!\!&\!\! \bA_{n,0} \!\!&\!\! \bA_{n,+1} \!\!&\!\!  \cdots
         \\[0.15cm]
         \cdots \!\!&\!\!  \bA_{n+1,-2} \!\!&\!\! \bA_{n+1,-1} \!\!&\!\! \bA_{n+1,0} \!\!&\!\! \cdots
         \\[0.1cm]
         \rotatebox{90}{$\ddots$} \!\!&\!\! \vdots \!\!&\!\! \vdots \!\!&\!\! \vdots \!\!&\!\! \ddots
    	\ea
	\right]
\end{align}
where the off-diagonal term $\bA_{n,m}$ denotes the influence of the $(n+m)$th harmonic $\hat{\bvarphi}_{n+m}$ on the dynamics of the $n$th harmonic $\hat{\bvarphi}_{n}$. {Apart from accounting for an extended wall-normal region, the block operators on the main diagonal of $\bA_\theta$ are identical to the operators for the channel flow without riblets;} see Appendix~\ref{sec.A-C} for details. At the {upper wall} of the channel, homogenous Dirichlet boundary conditions are imposed on $\eta$, and homogeneous Dirichlet and Neumann boundary conditions are imposed on $v$. Similar to the mean flow equations~\eqref{eq.NSvisriblet}, the boundary conditions at the corrugated surface are automatically satisfied via volume penalization. Finally, smoothness of all physical quantities at the intersection of the inner and outer wall-normal regimes ($y=-1$) is imposed by enforcing the following conditions:
\begin{align*}
	\ba{rclrcl}
        {v}(y\,=\,-1^+,z)
        &\!\!=\!\!&
        {v}(y\,=\,-1^-,z),
        &
	\dfrac{\partial v}{\partial y}(y=-1^+,z)
	&\!\!=\!\!&
	\dfrac{\partial v}{\partial y}(y=-1^-,z),
        \\[.3cm]
	\dfrac{\partial ^2v}{\partial y^2}(y=-1^+,z)
	&\!\!=\!\!&
	\dfrac{\partial ^2v}{\partial y^2}(y=-1^-,z),
        &
        \dfrac{\partial ^3v}{\partial y^3}(y=-1^+,z)
	&\!\!=\!\!&
	\dfrac{\partial ^3v}{\partial y^3}(y=-1^-,z),
        \\[0.3cm]
        \eta(y=-1^+,z)
        &\!\!=\!\!&
        \eta(y=-1^-,z),
        &
	\dfrac{\partial \eta}{\partial y}(y=-1^+,z)
	&\!\!=\!\!&
	\dfrac{\partial \eta}{\partial y}(y=-1^-,z).
        \ea
\end{align*}

A pseudospectral scheme used for discretizing the mean flow equations~\eqref{eq.NSvisriblet} is utilized to discretize the wall-normal operators in~\eqref{eq.evolutionform}. In addition, a change of variables is employed to obtain a state-space representation in which the kinetic energy is determined by the Euclidean norm of the state vector in a finite-dimensional approximation of the evolution model~\citep[Appendix A]{zarjovgeoJFM17},
\begin{align}
	\label{eq.state-space}
	\ba{rcl}
		\dot{\bpsi_\theta}(k_x,t)
		& \!\!=\!\! &
		A_\theta(k_x)\, \bpsi_\theta (k_x,t)
		\;+\;
		{\bd}_{\theta} (k_x,t),
		\\[.15cm]
		\bv_\theta (k_x,t)
		& \!\!=\!\! &
		C_\theta (k_x)\, \bpsi_\theta(k_x,t).
	\ea
\end{align}
For $N_i$ and $N_o$ collocation points in the inner and outer wall-normal regimes, respectively, and a Fourier series expansion~\eqref{eq.invK} with $M$ harmonics, $\bpsi_\theta(k_x,t)$ and $\bv_\theta(k_x,t)$ are vectors with $2\times M \times (N_i+N_o)$ and $3\times M \times (N_i+N_o)$ complex-valued entries, respectively. The state-space matrices $A_\theta(k_x)$ and $C_\theta(k_x)$ are discretized versions of the operators in~\eqref{eq.evolutionform} that incorporate the aforementioned change of coordinates.

	\vspace*{-2ex}
\subsection{Second-order statistics of velocity fluctuations and forcing}
\label{sec.correctX}

Let the linearized dynamics~\eqref{eq.state-space} be driven by zero-mean {stochastic forcing $\mathbf{d}_\theta (k_x,t)$ that is white in time,} with covariance matrix $M_\theta(k_x)=M^*_\theta(k_x) \succeq 0$, i.e.,
\begin{align}
\label{eq.forcing-spectrum}
	\langle \bd_\theta(k_x,t_1)\, \bd^*_\theta(k_x,t_2) \rangle
	\;=\;
	M_\theta(k_x)\,\delta(t_1\,-\,t_2),
\end{align}
where $\delta$ is the Dirac delta function. {Following the bi-infinite structure of $\bd_\theta(k_x,t)$, $M_\theta(k_x)$ takes the bi-infinite form,
\begin{align}
\label{eq.Mform}
	M_\theta(k_x)
	\;\DefinedAs\;
	\left[
	\ba{ccccc}
         \ddots \!\!&\!\! \vdots \!\!&\!\! \vdots \!\!&\!\! \vdots \!\!&\!\! \rotatebox{90}{$\ddots$}
         \\[0.1cm]
         \cdots \!\!&\!\! M(k_x, \theta_{n-1}) \!\!&\!\! 0 \!\!&\!\! 0 \!\!&\!\! \cdots
         \\[0.15cm]
         \cdots \!\!&\!\! 0 \!\!&\!\! M(k_x, \theta_{n}) \!\!&\!\! 0 \!\!&\!\!  \cdots
         \\[0.15cm]
         \cdots \!\!&\!\!  0 \!\!&\!\! 0 \!\!&\!\! M(k_x, \theta_{n+1}) \!\!&\!\! \cdots
         \\[0.1cm]
         \rotatebox{90}{$\ddots$} \!\!&\!\! \vdots \!\!&\!\! \vdots \!\!&\!\! \vdots \!\!&\!\! \ddots
    	\ea
	\right]
\end{align}
with the block operator $M(k_x, \theta_{n})$ representing the covariance of the $n$th harmonic of the forcing $\bd_\theta(k_x,t)$. The off-diagonal blocks of $M_\theta(k_x)$ are zero because the stochastic forcing is uncorrelated over various spanwise hamronics.} 

The steady-state covariance of the state in equations~\eqref{eq.state-space} can be determined from
the solution $X_\theta(k_x)$ to the Lyapunov equation~\citep{farjovbam08,moajovJFM10}
\begin{align}
\label{eq.lyap}
    A_\theta(k_x)\,{X}_\theta(k_x)
    \;+\;
    {X}_\theta(k_x)\,A_\theta^*(k_x)
    \;=\;
    -M_\theta(k_x),
\end{align}
where the ($i,j$)th block of ${X}_\theta(k_x)$ determines the correlation matrix associated with the $i$th and $j$th harmonics of the state $\bpsi_\theta$.

As mentioned in \S~\ref{sec.LNSE}, the block operators on the main diagonal of $A_\theta$ contain the dynamical generators of the turbulent channel flow with smooth walls at various wavenumber pairs $(k_x,\theta_n)$. Based on this, $A_\theta$ can be decomposed as
\begin{align}
\label{eq.decomposed-dynamics}
	A_\theta(k_x) 
	\;=\; 
	A_{\theta, s}(k_x) \;+\; A_{\theta, c}(k_x).
\end{align}
where $A_{\theta, s} = \diag\{ \, \ldots,\, A_s(k_x,\theta_{n-1}),\, A_s(k_x,\theta_n),\, A_s(k_x,\theta_{n+1}), \, \ldots \}$ accounts for the dynamical generator of the turbulent channel flow with smooth walls and $A_{\theta, c}$ captures the contribution of the spatially periodic surface corrugation. The block-diagonal structure of $M_\theta(k_x)$ implies that the solution $X_\theta(k_x)$ to {~\eqref{eq.lyap}} can also be decomposed~as 
\begin{align}
\label{eq.decomposed-covariance}
	X_\theta(k_x) 
	\;=\; 
	X_{\theta, s}(k_x) \;+\; X_{\theta, c}(k_x).
\end{align}
Here, $X_{\theta, s} = \diag\{ \, \ldots,\, X_s(k_x,\theta_{n-1}), \, X_s(k_x,\theta_n), \, X_s(k_x,\theta_{n+1}), \, \ldots \}$ is a block-diagonal covariance operator whose entries are determined by the steady-state covariance matrix of turbulent channel flow over smooth walls parameterized by $(k_x,\theta_n)$, and $X_{\theta, c}$ denotes the modification resulting from the presence of riblets. This follows from substitution of~\eqref{eq.decomposed-dynamics} and~\eqref{eq.decomposed-covariance} into the Lyapunov equation~\eqref{eq.lyap},
\begin{align*}
	\ba{l}
        (A_{\theta,s}(k_x) \,+\, A_{\theta, c}(k_x))(X_{\theta,s}(k_x) \,+\, X_{\theta, c}(k_x)) 
        \\[.15cm]
        \hspace{3cm}
        +\; 
        (X_{\theta,s}(k_x) \,+\, X_{\theta, c}(k_x))(A_{\theta,s}(k_x) \,+\, A_{\theta, c}(k_x))^* 
        \;=\; 
        -M_\theta(k_x),
        \ea
\end{align*}
from which the Lyapunov equation corresponding to the turbulent channel flow with smooth walls can be extracted as
\begin{align*}
	A_{\theta, s}(k_x)\,X_{\theta,s} (k_x)
	\;+\; 
	X_{\theta, s}(k_x)\,A_{\theta,s}(k_x) 
	\;=\; 
	-M_\theta(k_x).
\end{align*}
Following~\cite{moajovJFM12}, we select the block-diagonal covariance matrix $M_\theta(k_x)$ to guarantee equivalence between the two-dimensional energy spectrum of the turbulent channel flow with smooth walls and the flow governed by the stochastically-forced NS equations linearized around $\bar{\bu}=[\,{\bar{U}_s}(y)~ 0~ 0\,]^T$. This is achieved by scaling the block covariances in~\eqref{eq.Mform} as
\begin{align*}
        M(k_x,\theta_{n})
        \;=\;
        \dfrac{\bar{E}_s(k_x, \theta_{n})}{\bar{E}_{s,0}(k_x, \theta_{n})} \; M_s(k_x, \theta_{n}).
\end{align*}
Here, $\bar{E}_s(k_x,\theta_{n}) = \int_{-1}^{1} E_s(y, k_x, \theta_{n}) \, \mrd y$ is the two-dimensional energy spectrum of turbulent channel flow with smooth walls, which is obtained from the DNS-based energy spectrum {$E_s(y, k_x, \theta_{n})$}~\citep{deljim03,deljimzanmos04}, and $\bar{E}_{s,0}(k_x, \theta_{n})$ is the energy spectrum resulting from the linearized NS equations~\eqref{eq.state-space} subject to white-in-time stochastic forcing with the covariance matrix

\begin{align*}
M_s(k_x, \theta_{n})
	=
        \tbt{\!\!\!\sqrt{E_s(y,k_x,\theta_{n})} \, I\!\!\!}{\!0\!\!}{\!0\!\!\!}{\!\sqrt{E_s(y,k_x,\theta_{n})} \, I\!\!\!}\!\!
        \tbt{\!\!\!\sqrt{E_s(y,k_x,\theta_{n})} \, I\!\!\!}{\!0\!}{\!0\!\!\!}{\!\sqrt{E_s(y,k_x,\theta_{n})} \, I\!\!\!}^*.
\end{align*}
Finally, the energy spectrum of velocity fluctuations is determined from the solution to the Lyapunov equation~\eqref{eq.lyap} as
\begin{align}
\label{eq.Ebar}
        {\bar{E}(k_x,\theta)
        \;=\;
        \sum_{n \, \in \, \bbZ} \trace \left(X_d(k_x,\theta_n) \right).}
\end{align}
where $X_d(k_x,\theta_n)$ {represent the block covariance matrices on the main diagonal of $X_\theta(k_x)$ confined to the wall-normal range of $y\in[-1, 1]$. The} correction to the energy spectrum that arises from the presence of riblets is determined by 
{
\begin{align}
\label{eq.Ec}
\bar{E}_c(k_x,\theta) 
	\;=\;
	 \bar{E}(k_x,\theta) \,-\, \bar{E}_s(k_x,\theta)
\end{align}
where
\begin{align}
\label{eq.Es}
        \bar{E}_s(k_x,\theta)       
        \;=\;
        \sum_{n \, \in \, \bbZ} \bar{E}_s(k_x,\theta_{n})
\end{align}
denotes the reference energy spectrum from DNS of channel flow in the absence of riblets.}

	\vspace*{-2ex}
\subsection{Correction to turbulent viscosity}
\label{sec.correctnuT}

The turbulent viscosity $\nu_T(y)$ is determined by the second-order statistics of velocity fluctuations, i.e., the kinetic energy $k(y)$ and its rate of dissipation $\eps(y)$; see equation~\eqref{eq.nuT}. The statistics can be computed using the covariance matrix  $X_d(k_x,\theta_n)$ and $k(y)$, $\eps(y)$ can be decomposed as
\begin{align}
\label{eq.k-eps-corr}
    k(y) \;=\; k_s(y) \,+\, k_c(y),
    \quad
    \epsilon(y) \;=\; \epsilon_s(y) \,+\, \epsilon_c(y),
\end{align}
where{, again,} the subscript $s$ signifies channel flow with smooth walls, and the subscript $c$ quantifies the influence of fluctuations in the flow over riblets. The DNS results for turbulent channel flow yield $k_s$. {On the other hand,} $\eps_s$ is computed using $\eps_s (y) = c Re_\tau^2\, k_s^2 (y)/\nu_{Ts} (y)$ {and} the corrections $k_c$ and $\epsilon_c$ can be determined from the second-order statistics in {$X_{\theta,c}(k_x)$}; see Appendix~\ref{sec.kecorrection} for details. Substitution of $k(y)$ and $\eps(y)$ from~\eqref{eq.k-eps-corr} into equation~\eqref{eq.nuT} and application of the Neumann series expansion yields
\begin{align}
\label{eq.nuT-total}
	\nu_T(y)
	\;=\;
	\nu_{Ts}(y) \,+\, \nu_{Tc}(y),
\end{align}
where the correction $\nu_{Tc}(y)$ to turbulent viscosity $\nu_{Ts}(y)$ is given by

\begin{align}
\label{eq.nuTc}
	\nu_{Tc}(y)
	\;=\;
	\nu_{Ts}(y) \left(\dfrac{2 \, k_c(y)}{k_s(y)} \; - \; \dfrac{\epsilon_c(y)}{\epsilon_s(y)}\right).
\end{align}
Since we primarily focus on small size riblets, this expression is obtained by neglecting higher-order terms that involve multiplication of $k_c(y)$ and $\eps_c(y)$.

The influence of fluctuations on the turbulent mean velocity and, consequently, skin-friction drag can be evaluated by substituting $\nu_T(y)$ from~\eqref{eq.nuT-total} and solving equations~\eqref{eq.NSvisriblet}; see {\S~\ref{sec.baseflow}} for details regarding the correction to the mean flow profile {and \S~\ref{sec.drag} for the subsequent computation of  the drag.}

\section{Turbulent drag reduction and energy suppression}
\label{sec.result}

\begin{figure}
	\begin{center}
                \includegraphics[width=.4\textwidth]{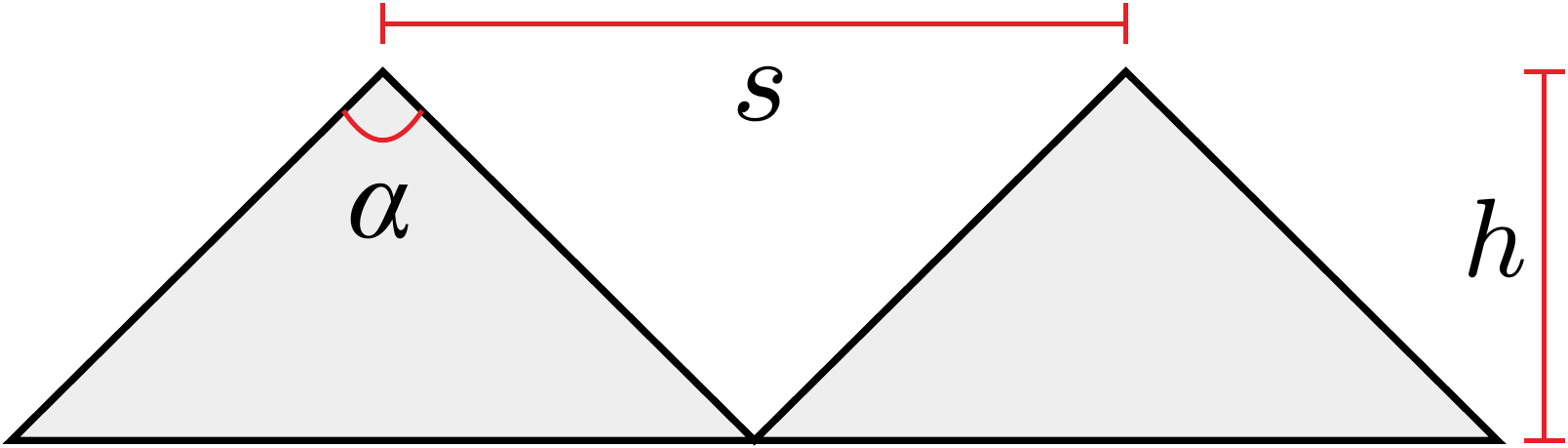}
        \end{center}
        \caption{Triangular riblets with height $h$, spacing $s=2\pi/\omega_z$, and tip angle $\alpha$.}
        \label{fig.riblet-sketch}
\end{figure}

In this section, we use the framework developed in \S~\ref{sec.fluctuations} to examine the effect of triangular riblets shown in figure~\ref{fig.riblet-sketch} on the mean velocity, skin-friction drag, and kinetic energy in turbulent channel flow with $Re_\tau = 186$. We assume that the influence of small-size riblets on the channel height and shear velocity is negligible, thereby implying that the Reynolds number remains unchanged over various case studies. By letting the ratio between the height and spacing of riblets be fixed, the riblets of different sizes are obtained by modifying the frequency $\omega_z$; see Table~\ref{table.comp-case} for a list of cases considered in our study. In the absence of riblets, DNS results~\citep{deljim03,deljimzanmos04} provide second-order statistics which are used to determine the covariance of stochastic forcing $\bd_\theta(k_x,t)$ in equation~\eqref{eq.forcing-spectrum} and to compute the kinetic energy {$k_s(y)$}; see \S~\ref{sec.correctX}.

We use a total of $199$ Chebyshev collocation points to discretize the operators in the wall-normal direction ($N_i = 179$, $N_o=20$). Furthermore, we parameterize the linearized equations~\eqref{eq.state-space} using $48$ logarithmically spaced streamwise wavenumbers with $0.03 < k_x < 40$ and utilize $25$ harmonics of $\omega_z$ ($n=-12,\dots,12$) with $50$ equally spaced offset points $\theta\in[0,\omega_z/2)$ to parameterize $\theta_n = \theta + n \omega_z$. Finally, to capture the triangular shape of riblets via~\eqref{eq.invK}, we use $25$ harmonics in $z$ ($m=-12,\dots,12$).

\begin{table}
\tabcolsep 0pt
\begin{center}
{\rule{0.7\textwidth}{1pt}}
\begin{tabular*}{0.7\textwidth}{@{\extracolsep{\fill}} cccc}
 {$Re_\tau$} & $h/s$ & $\alpha$ & $\omega_z$ \\[-0.2cm]
  \hline \\[-0.4cm]
 	& $0.38$ & $105\degree$ & $30, 35, 40, 45, 50, 60, 80, 100, 160$ \\[0.1cm]
 	& $0.5$ & $90\degree$ & $30, 35, 40, 45, 50, 60, 80, 100, 160$ \\[0.1cm]
 {186} & $0.65$ & $75\degree$ & $35, 40, 45, 50, 60, 80, 100, 160$  \\[0.1cm]
 	& $0.87$ & $60\degree$ & $50, 60, 70, 80, 100, 120, 160$ \\[0.1cm]
  & $1.2$ & $45\degree$ & $60, 80, 100, 120, 160, 210$\\[-0.2cm]
  \hline \\[-0.4cm]
  {547} & {$0.5$} & {$90\degree$} & {$90, 115, 145, 175, 210, 250, 300, 360$}
\end{tabular*}
{\rule{0.7\textwidth}{1pt}}
\caption{Triangular riblets with different height to spacing ratios ($h/s$), tip angles $\alpha$, and spanwise frequencies $\omega_z$ that we examine in our study.}
\label{table.comp-case}
\end{center}
\end{table}

	\vspace*{-2ex}
\subsection{Drag reduction}
\label{sec.dragreduction}

\begin{figure}
        \begin{center}
        \begin{tabular}{cccc}
        \hspace{-.8cm}
        \subfigure[]{\label{fig.tridragsp}}
        &&
        \hspace{-.6cm}
        \subfigure[]{\label{fig.tridraghp}}
        &
        \\[-.5cm]
        \hspace{-.6cm}
	\begin{tabular}{c}
        \vspace{.5cm}
        \small{\rotatebox{90}{$\Delta D (\%)$}}
       \end{tabular}
       &\hspace{-.3cm}
	\begin{tabular}{c}
       \includegraphics[width=0.4\textwidth]{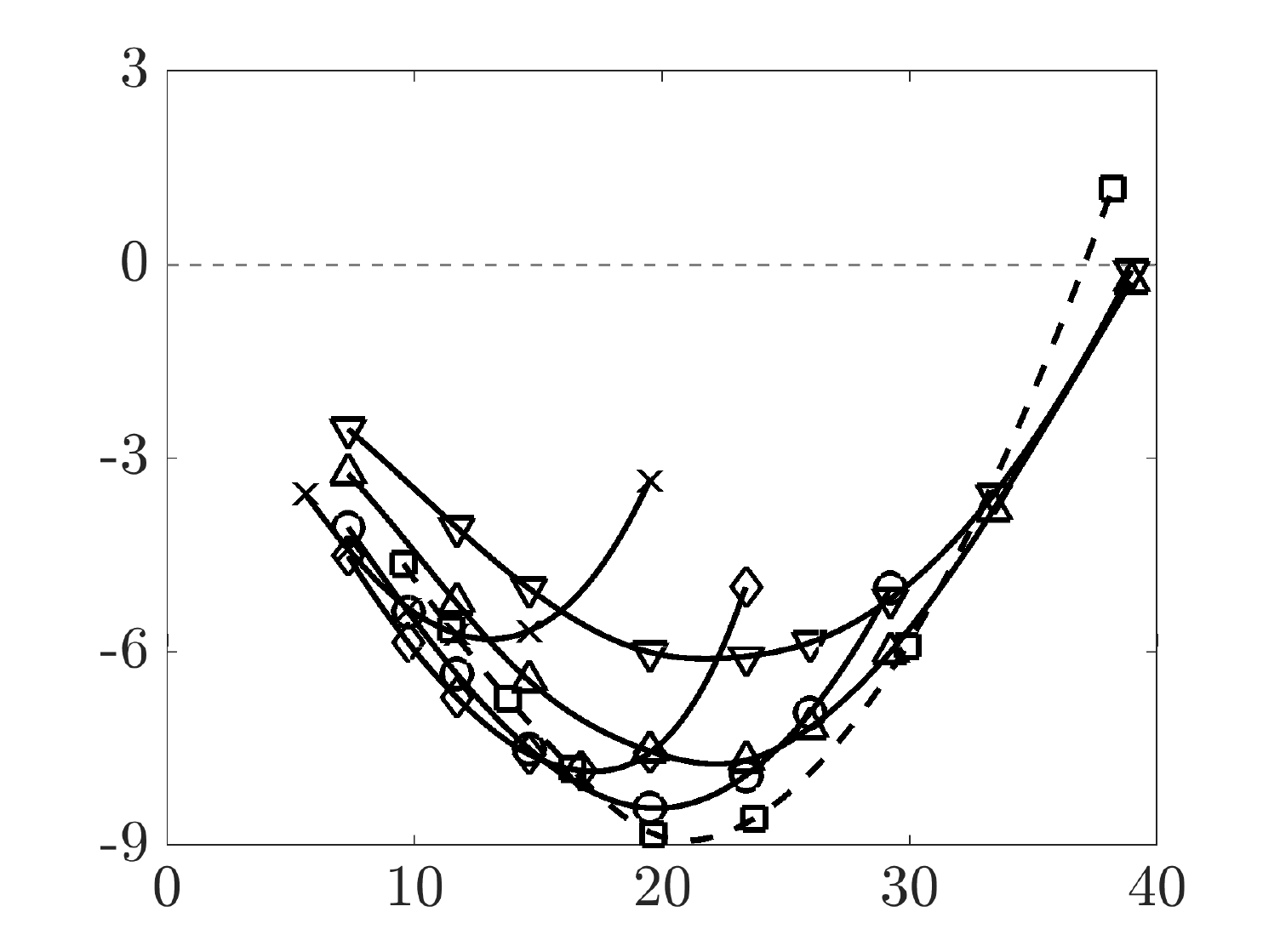}
       \\ {\small $s^+$}
       \end{tabular}
       \hspace{-.3cm}
       &
    \begin{tabular}{c}
        \vspace{.5cm}
        \small{\rotatebox{90}{$\Delta D(\%)$}}
       \end{tabular}
       &\hspace{-.3cm}
    \begin{tabular}{c}
       \includegraphics[width=0.4\textwidth]{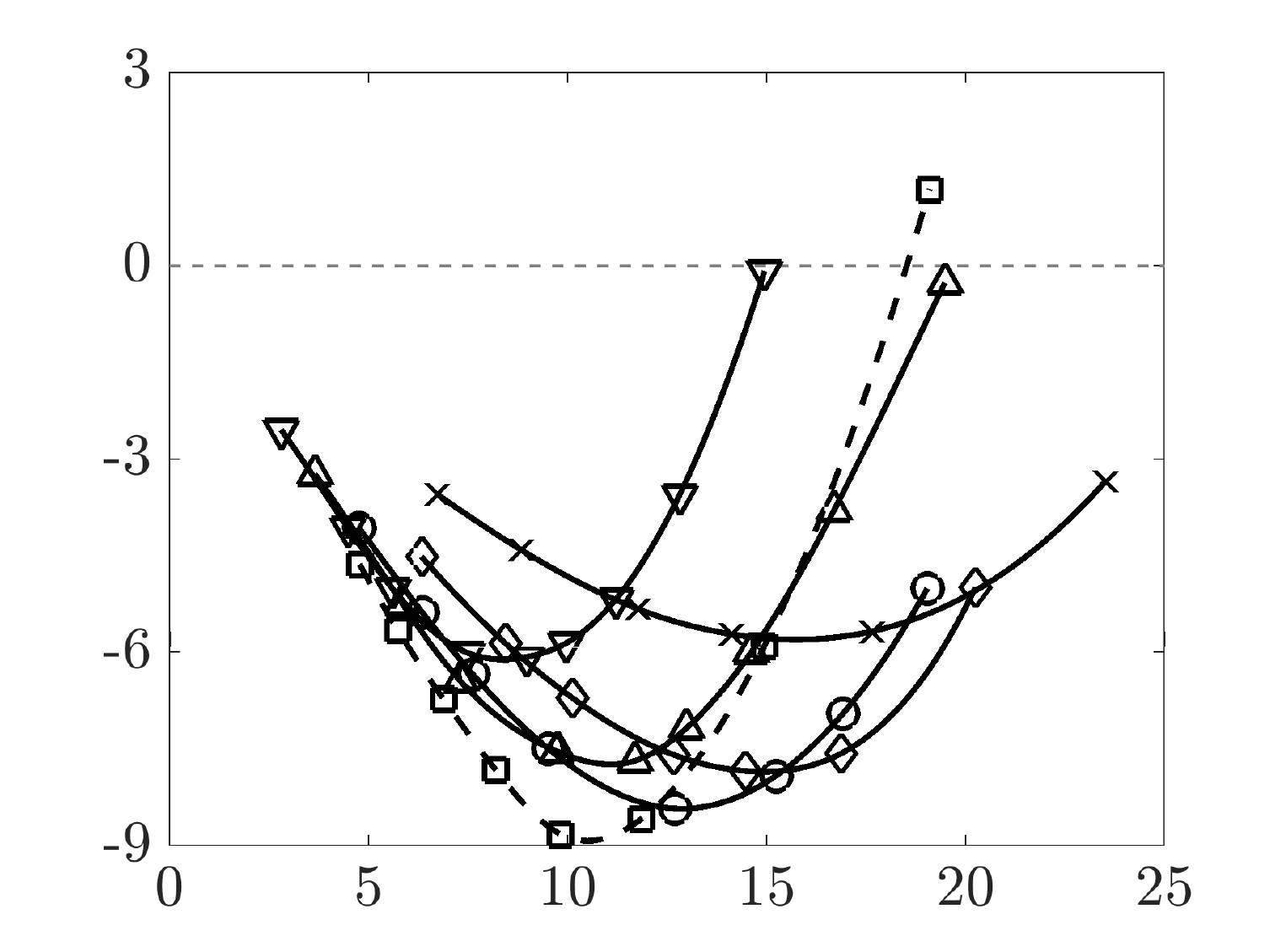}
       \\ {\small $h^+$}
       \end{tabular}
       \end{tabular}
       \end{center}
        \caption{Turbulent drag reduction for triangular riblets with different tip angles $\alpha$ as a function of (a) spacing $s^+$; and (b) height $h^+$ in a channel flow with $Re_\tau=186$ and $\alpha=105\degree$ ($\bigtriangledown$); $90\degree$ ($\bigtriangleup$); $75\degree$ ($\bigcirc$); $60\degree$ ($\lozenge$); and $45\degree$ ($\times$),  {as well as $Re_\tau=547$ and $\alpha=90\degree$ ($\square$).}}
        \label{fig.trish}
\end{figure}

We first examine the effect of riblet size on turbulent drag. In our parametric study, we follow~\cite{garjim11b} and refer to the regime of vanishing riblet spacing, in which the drag reduction is proportional to the size of riblets, as the viscous regime. For a turbulent channel flow with $Re_\tau=186$ subject to triangular riblets on the lower wall, figure~\ref{fig.trish} shows the influence of the height and peak-to-peak spacing of riblets on $\Delta D$ in equation~\eqref{eq.DR}. In this figure, the height and spacing are reported in inner viscous units, i.e. $h^+ = Re_\tau h$ and $s^+ = Re_\tau s$, and various curves represent different tip angles $\alpha$ as a measure of riblet geometry. As shown in figure~\ref{fig.riblet-sketch}, a particular tip angle $\alpha$ corresponds to a specific height to spacing ratio.

Clearly, small-size riblets can indeed reduce skin-friction drag.  {Figure~\ref{fig.trish} shows the percentage of drag reduction obtained in turbulent channel flows with $Re_\tau=186$ and $Re_\tau=547$ over the different triangular riblets listed in Table~\ref{table.comp-case}. Herein, we focus on the channel flow with $Re_\tau=186$ to analyze the dependence of drag reduction on the spacing and height of riblets.} Figure~\ref{fig.tridragsp} demonstrates that, for $s^+<20$, the drag reduction first increases as $h^+$ increases, saturates, and then decreases. This trend, however, slows down for smaller values of $s^+$. For $\alpha=90\degree$ and $\alpha=60\degree$, the drag reduction trends and optimal $s^+$ values resulting from our method reliably capture the trends reported in experimental studies~\citep{becbruhaghoehop97}. On the other hand, as shown in figure~\ref{fig.tridraghp}, for a fixed height, as the spacing $s^+$ of riblets decreases, drag reduction increases, saturates, and then decreases. Furthermore, figures~\ref{fig.tridragsp} and~\ref{fig.tridraghp} show that as the riblet tip angle $\alpha$ decreases, maximum drag reduction is achieved for less separated and taller riblets, respectively. Finally, as $\alpha$ decreases, the maximum value of drag reduction first increases and then decreases, which is also in agreement with experimental observations~\citep{becbruhaghoehop97}. The trends predicted by our framework indicate an optimal height to spacing ratio of $h/s\approx0.65$ $(\alpha=75\degree)$ for triangular riblets, which over-predicts the previously reported optimal tip angle $\alpha=54^\degree$~\citep{deabhu10}. 

As demonstrated in figure~\ref{fig.trish}, the optimal height and spacing can be quite different for riblets of different shape (i.e., different values of $\alpha$), thereby indicating that the height and spacing may not be suitable metrics for characterizing the breakdown of the linear viscous regime. Instead, the groove cross-section area $l_g^+ \DefinedAs\sqrt{A^+}$ (for triangular riblets, $A^+ = h^+s^+/2$) provides the best collapse of the critical breakdown dimension across different riblet shapes~\citep{garjim11a,garjim11b}. Furthermore, to remove the effect of riblets' shape on their slope in the viscous regime $m_l \DefinedAs \lim_{l_g^+ \, \rightarrow \, 0}{\Delta D}/{l_g^+}$, we normalize the drag reduction curves by $m_l$~\citep{garjim11b}.

\begin{figure}
	\begin{center}
        \begin{tabular}{cc}
        \begin{tabular}{c}
                \vspace{.4cm}
                \rotatebox{90}{\normalsize $-\Delta D/m_l$}
        \end{tabular}
        &\hspace{-.5cm}
        \begin{tabular}{c}
                \includegraphics[width=.5\textwidth]{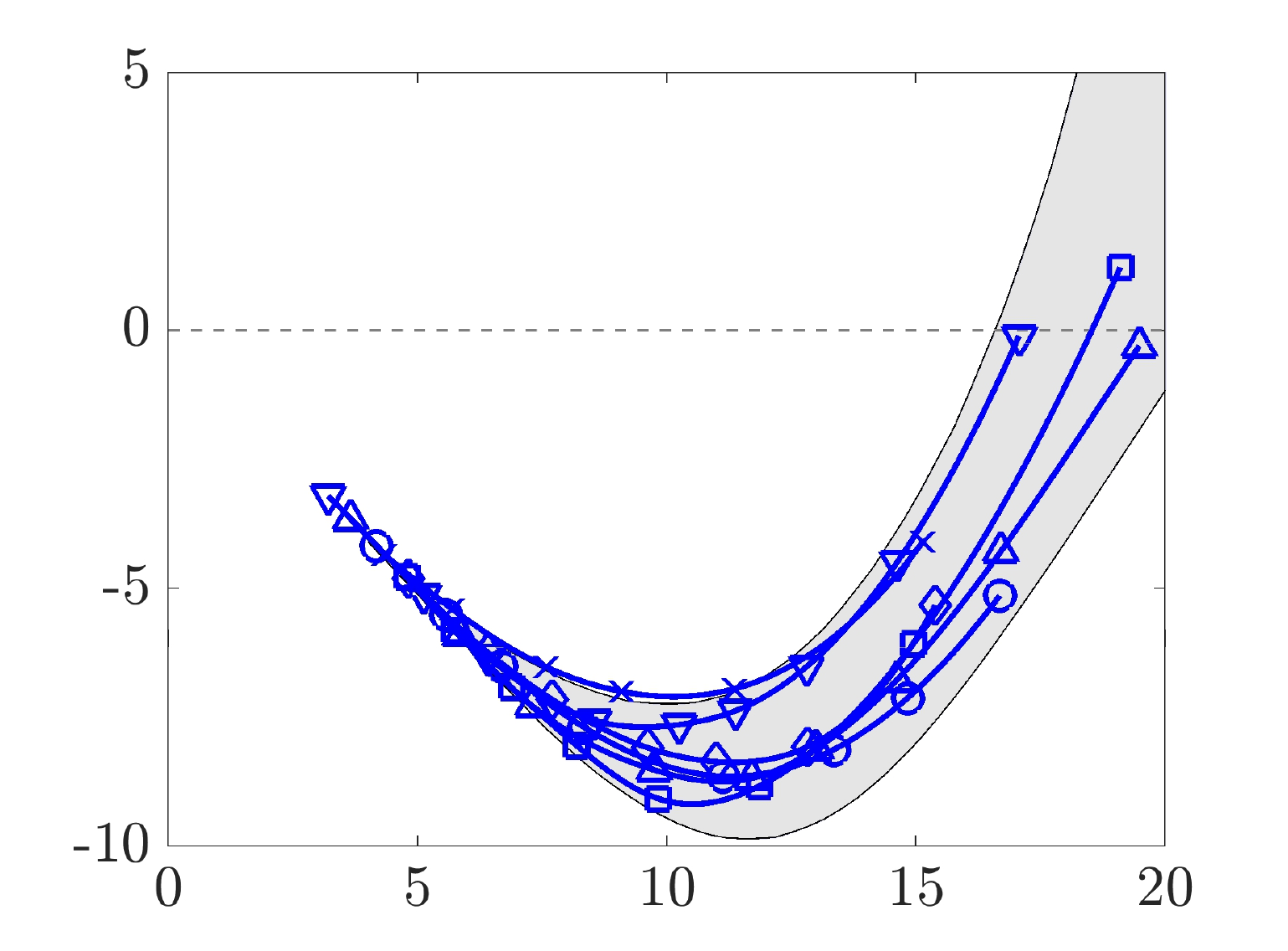}
                \\[-.1cm]
                {\normalsize $l_g^+$}
        \end{tabular}
        \end{tabular}
        \end{center}
        \caption{Drag reduction normalized with its slope in the viscous regime $m_l$. Different shapes of triangular riblets are represented by $\alpha=105\degree$ ($\bigtriangledown$); $90\degree$ ($\bigtriangleup$); $75\degree$ ($\bigcirc$); $60\degree$ ($\lozenge$); $45\degree$ ($\times$) {for $Re_\tau=186$; and $\alpha = 90\degree$ ($\square$) for $Re_\tau=547$}. The shaded area shows the envelope of experimental and numerical results~\citep{becbruhaghoehop97,garjim11b}.}
        \label{fig.DRml}
\end{figure}

For turbulent channel flow over triangular riblets, figure~\ref{fig.DRml} shows the $m_l$-normalized drag reduction as a function of $l_g^+$. The normalization factor $m_l$ is computed by averaging the slope obtained from the first two points on each curve, which are both in the viscous regime. The shaded region represents the envelope of normalized drag reduction values resulting from prior experimental and numerical studies~\citep{garjim11b}.  For riblets with $\alpha= 105\degree,\,90\degree,\,75\degree,\,60\degree$, and $45\degree$, figure~\ref{fig.DRml} shows  {the} collapse of drag reduction curves with the largest drag reduction occurring within a tight range of cross-section areas; {$l_g^+=9.7,\,11.1,\,11.3,\,11.3$, and $10.2$,} for $\alpha= 105\degree,\,90\degree,\,75\degree,\,60\degree$, and $45\degree$, respectively. This prediction agrees well with the values reported by~\cite{garjim11b}, $l_g^+ \in [ 9.7, 11.7]$. Moreover, the drag reduction curves resulting from our framework  {are located} within the shaded region and they reliably predict the overall trend. {For turbulent channel flow over triangular riblets with $Re_\tau=547$, $h/s=0.5$, and $\alpha= 90\degree$, our predictions of the normalized drag reduction remain within this shaded region and are very similar to the results obtained for $Re_\tau=186$; see square symbols in figure~\ref{fig.DRml}.}

\begin{figure}
        \begin{center}
        \begin{tabular}{cccc}
        \hspace{-.8cm}
        \subfigure[]{\label{fig.tridraglg60d}}
        &&
        \hspace{-.6cm}
        \subfigure[]{\label{fig.tridraglg90d}}
        &
        \\[-.5cm]
        \hspace{-.6cm}
	\begin{tabular}{c}
        \vspace{.5cm}
        \small{\rotatebox{90}{$-\Delta D/m_l$}}
       \end{tabular}
       &\hspace{-.3cm}
	\begin{tabular}{c}
       \includegraphics[width=0.4\textwidth]{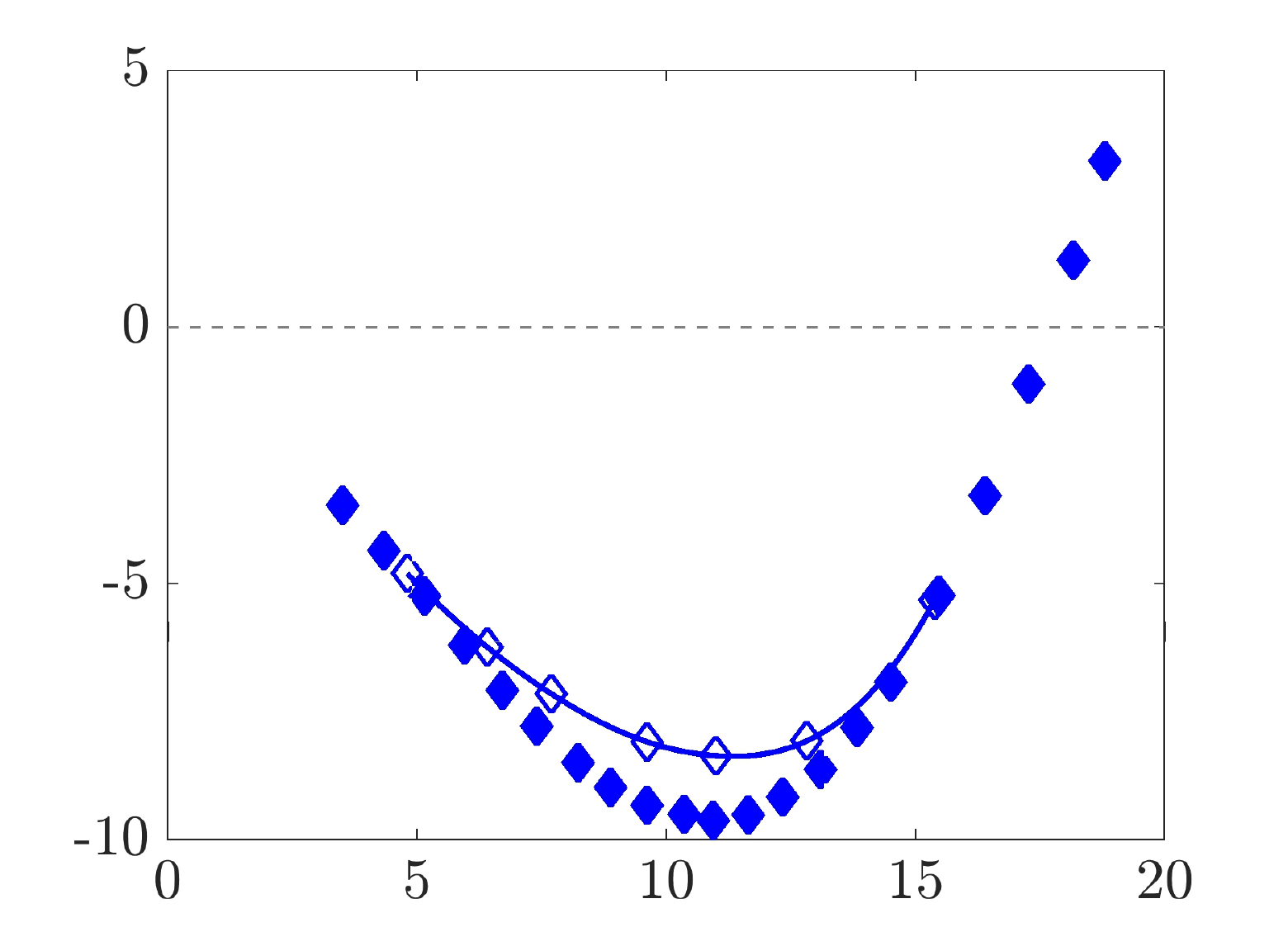}
       \\ {\small $l_g^+$}
       \end{tabular}
       \hspace{-.3cm}
       &
    \begin{tabular}{c}
        \vspace{.5cm}
        \small{\rotatebox{90}{$-\Delta D/m_l$}}
       \end{tabular}
       &\hspace{-.3cm}
    \begin{tabular}{c}
       \includegraphics[width=0.4\textwidth]{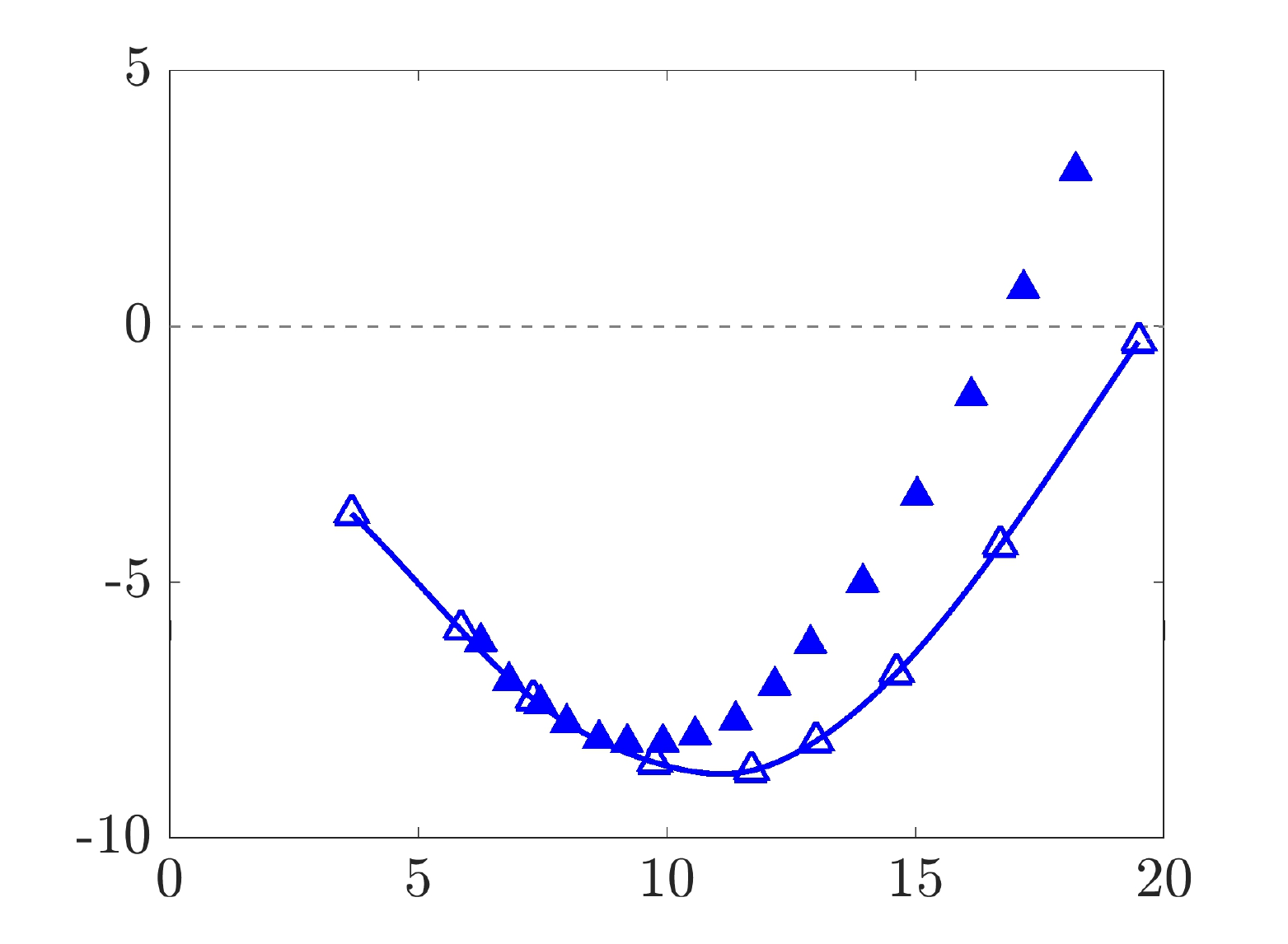}
       \\ {\small $l_g^+$}
       \end{tabular}
       \end{tabular}
       \end{center}
        \caption{Drag reduction normalized with its slope in the viscous regime $m_l$ resulting from our framework (open symbols) and experiments (solid symbols)~\citep{becbruhaghoehop97} with tip angles (a) $\alpha = 60\degree$; and (b) $90\degree$ for $Re_\tau=186$.}
        \label{fig.DRmlexp}
\end{figure}

Figure~\ref{fig.DRmlexp}  {compares} the $m_l$-normalized drag reduction resulting from our framework and the experimental results of~\cite{becbruhaghoehop97}.  {Our method captures the} overall trends and even the optimal size of riblets with tip angle $\alpha=60\degree$  {($2.9\%$ relative error). For riblets with tip angle $\alpha=90\degree$, the optimal size is overpredicted by $14.3\%$. We note that while the maximum amount of drag reduction is reasonably captured in both cases ($13.1\%$ and $7.6\%$ relative errors for $\alpha=60\degree$ and $\alpha=90\degree$, respectively), the quality of our predictions deteriorates for larger riblets with $\alpha=90\degree$.} As we  {discuss} in \S~\ref{sec.nuT_mean}, an overpredicted turbulence suppression in wall-normal regions away from the riblet-mounted lower surface  {can be a source of} this mismatch.

The performance deterioration for large riblets observed in figures~\ref{fig.DRml} and~\ref{fig.DRmlexp} is associated with the breakdown of the viscous regime within the grooves. This breakdown arises from the lodging of near-wall vortices~\citep{leelee01,suzkas94}, the generation of secondary flow vortices~\citep{goltua98}, or the emergence of spanwise coherent rollers~\citep{garjim11b}. When turbulence moves into the grooves, a turbulence model, which assumes that the wall-normal region with $y<-1$ is laminar (i.e., $\nu_T=0$), loses its validity. To support an extended turbulent regime and go beyond the breakdown of the viscous regime, our model of surface corrugation assumes the tip of riblets to be located within the original channel region (i.e., $y>-1$); see \S~\ref{sec.ribletmodel}. The parameter $r_p$ in~\eqref{eq.rz}, which controls the level of protrusion into the turbulent regime, is determined to satisfy a constant bulk assumption. Because of this, our model remains valid even for riblets that are larger than the optimal ($l_g^+\lesssim20$).

	\vspace*{-2ex}
\subsection{Effect of riblets on turbulent viscosity and turbulent mean velocity}
\label{sec.nuT_mean}

\begin{figure}
        \begin{center}
        \begin{tabular}{cccc}
        \hspace{-.6cm}
        \subfigure[]{\label{fig.omz0nuT}}
        &&
        \subfigure[]{\label{fig.omz0U}}
        &
        \\[-.5cm]\hspace{-.3cm}
	\begin{tabular}{c}
        \vspace{.5cm}
        \normalsize{\rotatebox{90}{$\nu_{Ts}$}}
       \end{tabular}
       &\hspace{-.3cm}
	\begin{tabular}{c}
       \includegraphics[width=0.4\textwidth]{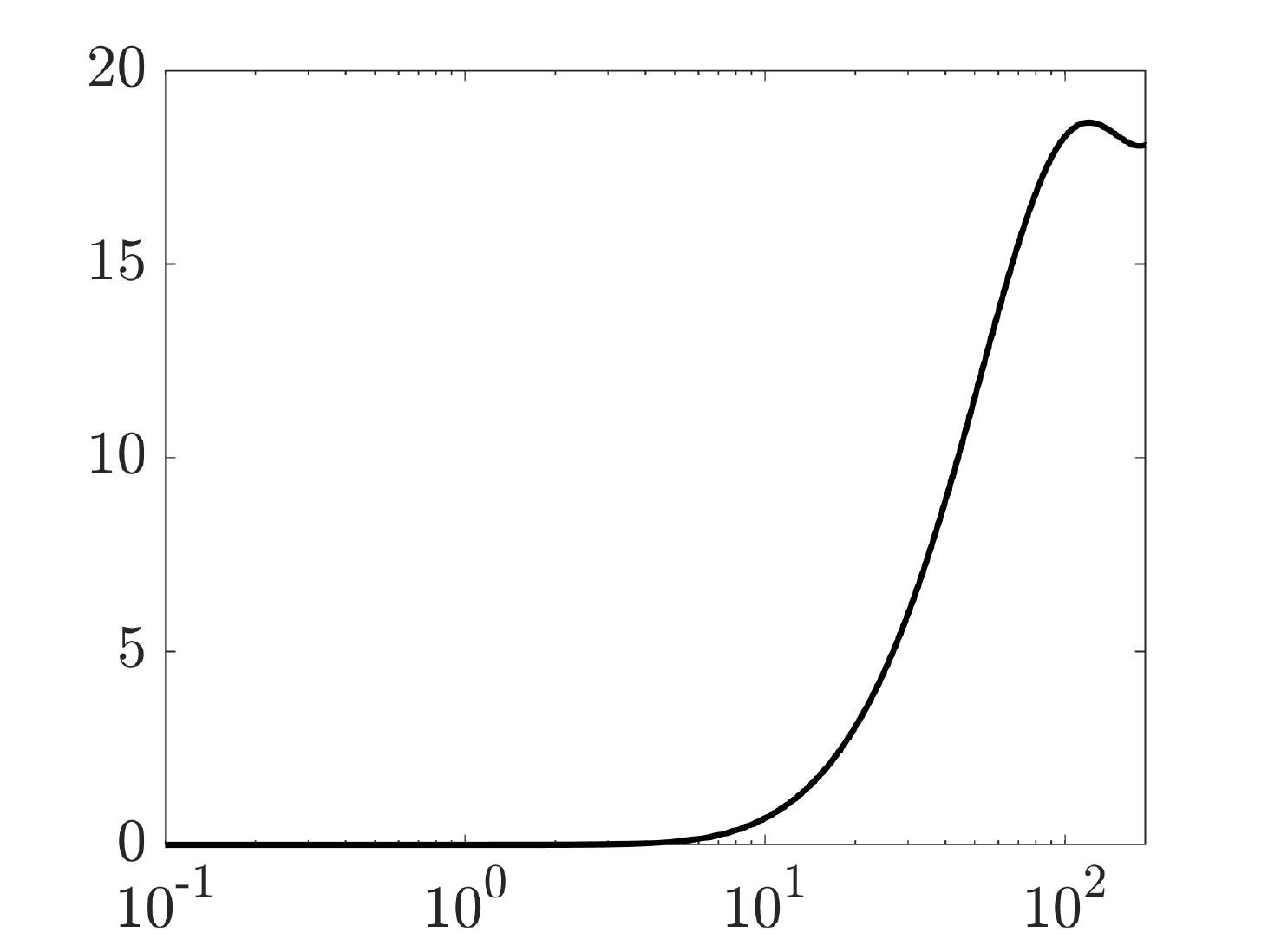}
       \end{tabular}
       &\hspace{.2cm}
       \begin{tabular}{c}
        \vspace{.5cm}
        \normalsize{\rotatebox{90}{{$\bar{U}_s$}}}
       \end{tabular}
       &\hspace{-.3cm}
    \begin{tabular}{c}
       \includegraphics[width=0.4\textwidth]{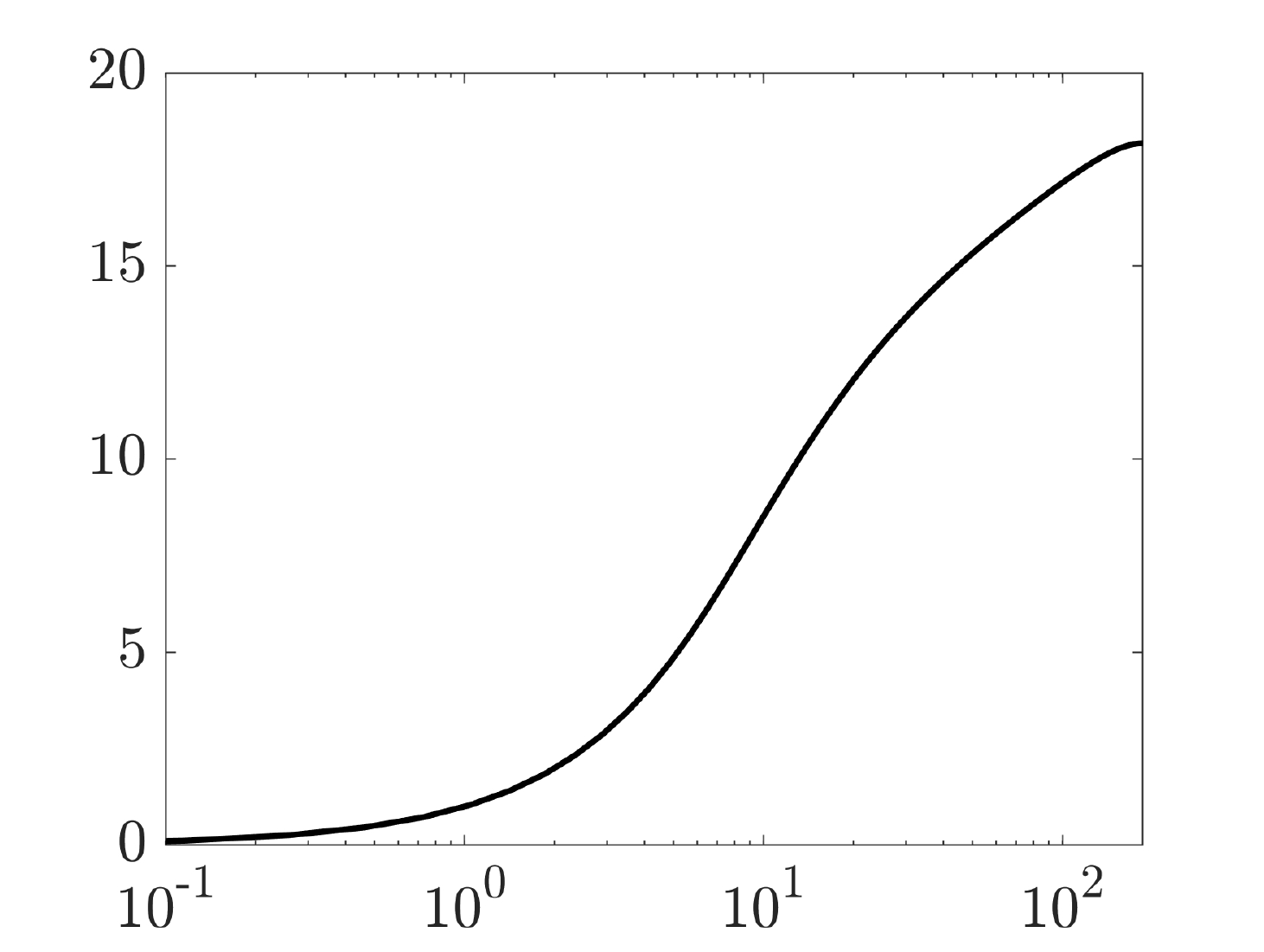}
       \end{tabular}
       \\[-0.1cm]
       \hspace{-.6cm}
        \subfigure[]{\label{fig.omz3050nuTc}}
        &&
        \subfigure[]{\label{fig.omz3050Uc}}
        &
        \\[-.5cm]\hspace{-.3cm}
	\begin{tabular}{c}
        \vspace{.5cm}
        \normalsize{\rotatebox{90}{$\nu_{Tc}$}}
       \end{tabular}
       &\hspace{-.3cm}
	\begin{tabular}{c}
       \includegraphics[width=0.4\textwidth]{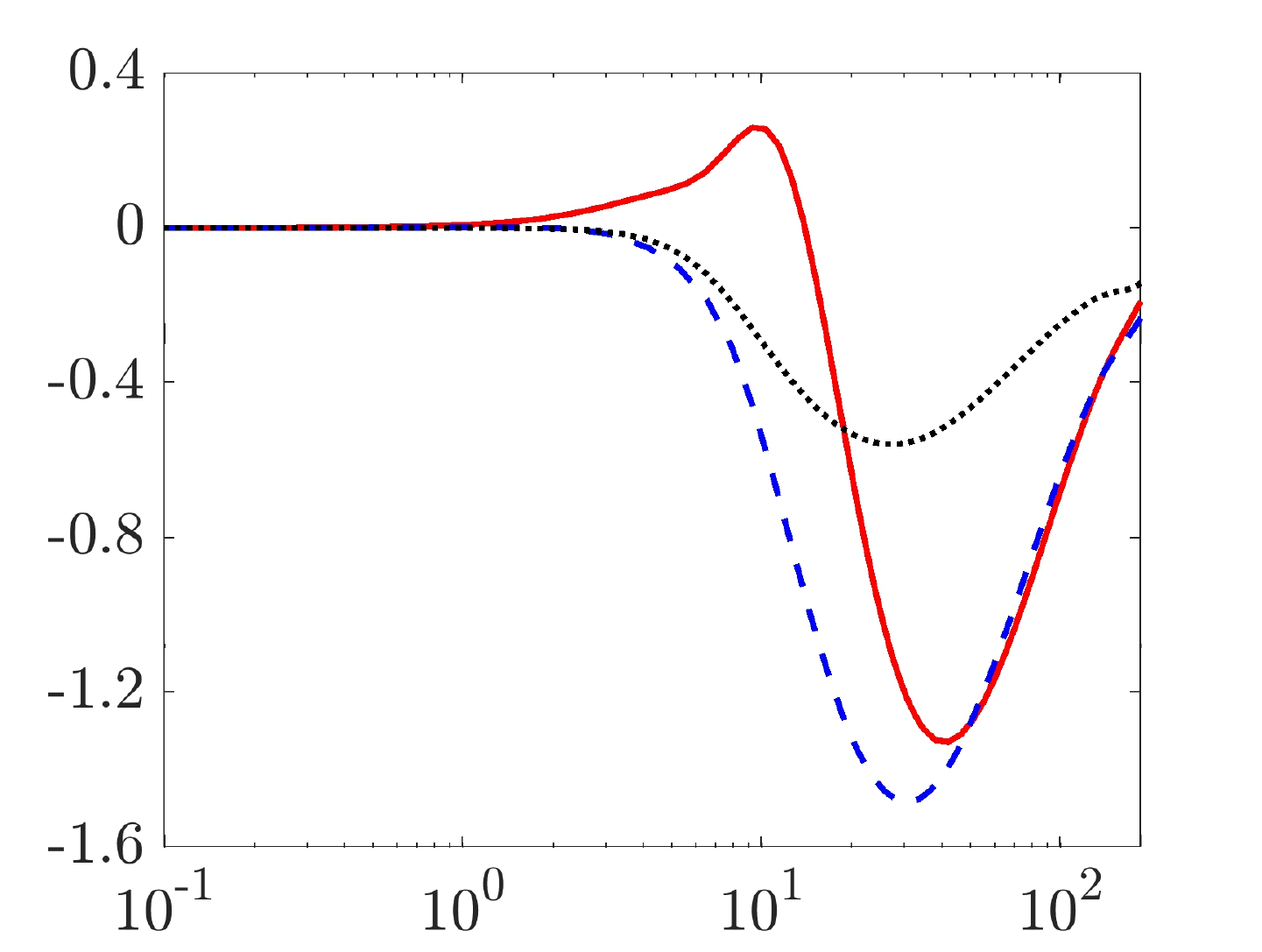}
       \\ $y^+$
       \end{tabular}
       &\hspace{.2cm}
       \begin{tabular}{c}
        \vspace{.5cm}
        \normalsize{\rotatebox{90}{{$\bar{U}_c$}}}
       \end{tabular}
       &\hspace{-.3cm}
    \begin{tabular}{c}
       \includegraphics[width=0.4\textwidth]{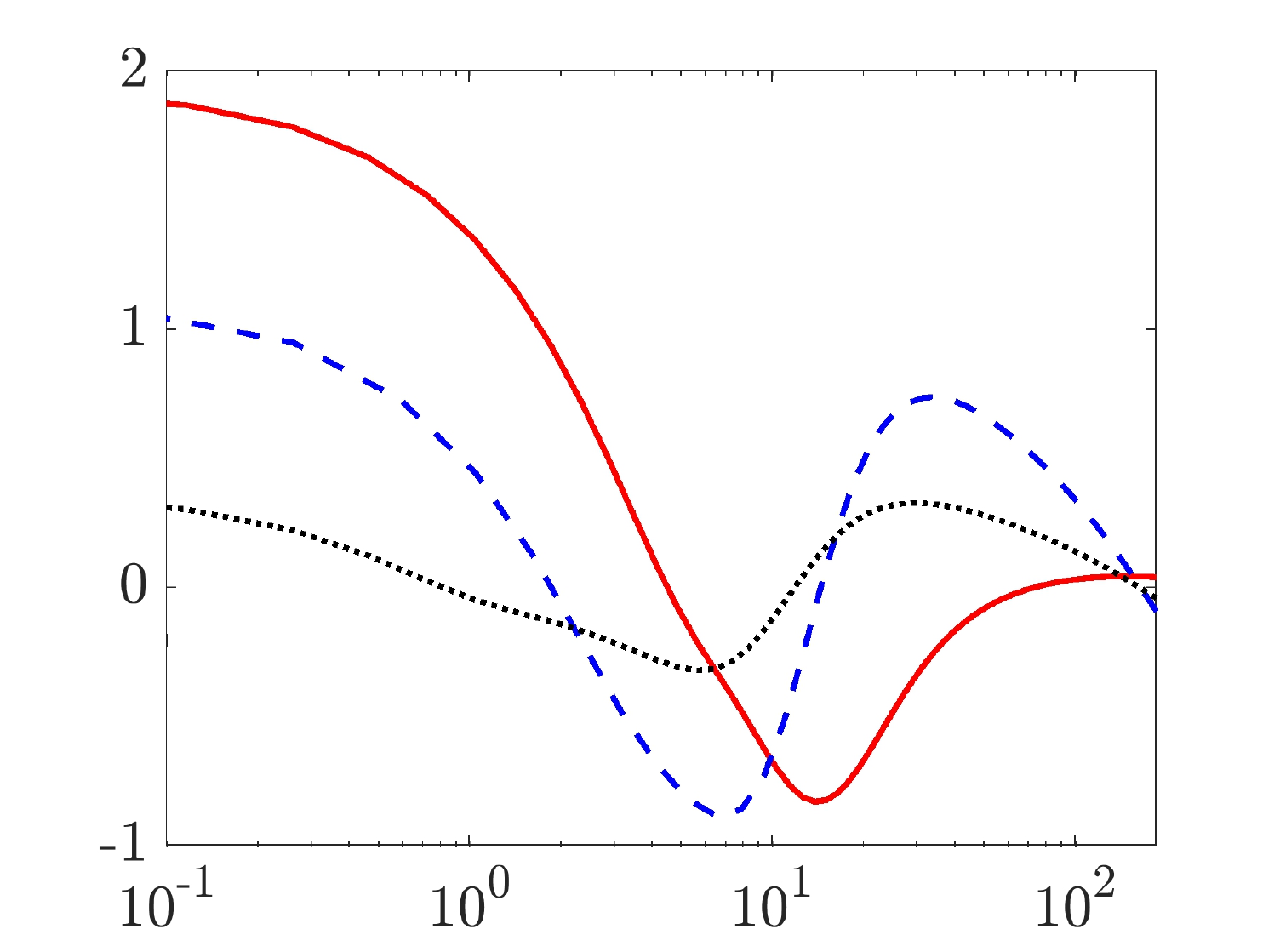}
       \\ $y^+$
       \end{tabular}
       \end{tabular}
       \end{center}
        \caption{(a) The turbulent viscosity, $\nu_{Ts}(y^+)$; and (b) the turbulent mean velocity ${\bar{U}_s}(y^+)$, in uncontrolled channel flow with $Re_\tau=186$. The correction to (c) turbulent viscosity, $\nu_{Tc}(y^+)$; and (d) the mean velocity, {$\bar{U}_c(y^+)$}, in the presence of riblets with $\alpha=90\degree$. Red solid lines correspond to riblets with $\omega_z=30$ (minimum drag reduction), blue dashed lines correspond to riblets with $\omega_z=50$ (maximum drag reduction), {and black dotted lines correspond to riblets with $\omega_z=160$ (smallest riblets).}}
        \label{fig.omz3560correction}
\end{figure}

We next examine the effect of riblets on turbulent viscosity and mean velocity. Figure~\ref{fig.omz3560correction} shows the turbulent eddy viscosity $\nu_{Ts}$ and mean velocity {$\bar{U}_s$} of channel flow over smooth walls with $Re_\tau=186$ along with the corresponding corrections, $\nu_{Tc}$ and {$\bar{U}_c$}, introduced by riblets with $\alpha=90\degree$ on the lower wall. {The wall-normal coordinate is given in inner (viscous) units, i.e., $y^+=Re_\tau(1+y)$.} Among the cases listed in table~\ref{table.comp-case}, {cases with maximum drag reduction $(\omega_z=50)$, minimum drag reduction $(\omega_z=30)$, and smallest riblets $(\omega_z=160)$} are chosen. For $\omega_z=30$, figure~\ref{fig.omz3050nuTc} shows that turbulence is promoted at the beginning of the buffer layer, but is then suppressed in regions farther away from the wall. On the other hand, for $\omega_z=50$, turbulence is always suppressed ($\nu_{Tc} \leq 0$) and the region of suppression shifts closer to the wall ($y^+ \gtrsim 3$). {We observe similar trends for riblets with $\omega_z=160$ to riblets with $\omega_z=50$, but with smaller amplitude.} Figure~\ref{fig.omz3050Uc} shows that, {in all cases,} riblets reduce the mean velocity gradient in the immediate vicinity of the wall ($y^+ \lesssim 6$). These results demonstrate that for the same shape of riblets (i.e., same tip angle $\alpha$), riblets of sizes that are larger than the optimal yield smaller amounts of turbulence suppression and mean shear reduction. 

\begin{figure}
        \begin{center}
        \begin{tabular}{cccc}
        \hspace{-.8cm}
        \subfigure[]{\label{fig.nuTopt}}
        &&
        \hspace{-.6cm}
        \subfigure[]{\label{fig.Ucopt}}
        &
        \\[-.5cm]
        \hspace{-.4cm}
\begin{tabular}{c}
        \vspace{.5cm}
        {\normalsize \rotatebox{90}{$\nu_{Tc}$}}
       \end{tabular}
       &\hspace{-.3cm}
\begin{tabular}{c}
       \includegraphics[width=0.4\textwidth]{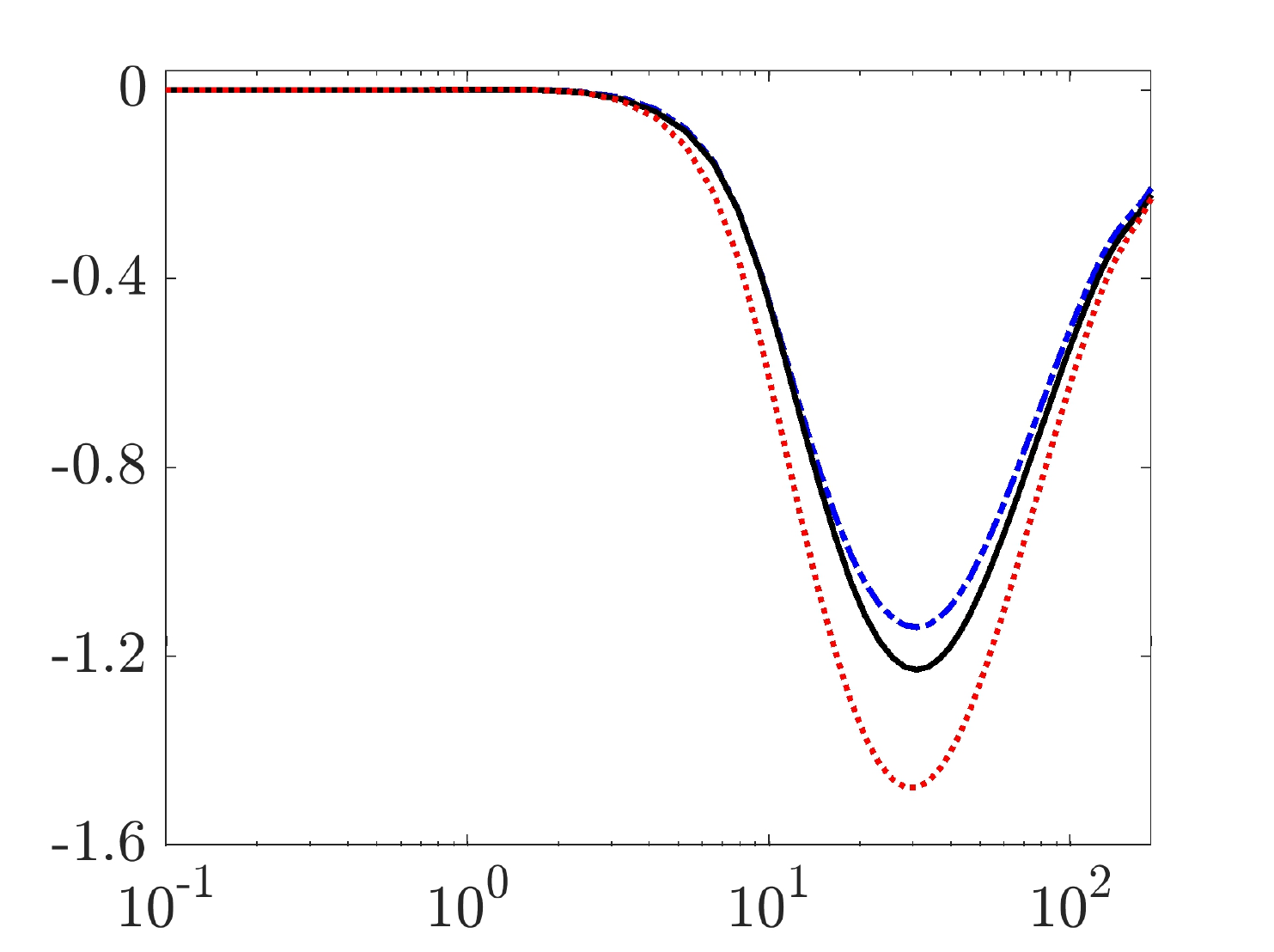}
       \\
       {\normalsize $y^+$}
       \end{tabular}
       \hspace{-.3cm}
       &
    \begin{tabular}{c}
        \vspace{.5cm}
        \normalsize{\rotatebox{90}{{$\bar{U}_c$}}}
       \end{tabular}
       &\hspace{-.3cm}
    \begin{tabular}{c}
       \includegraphics[width=0.4\textwidth]{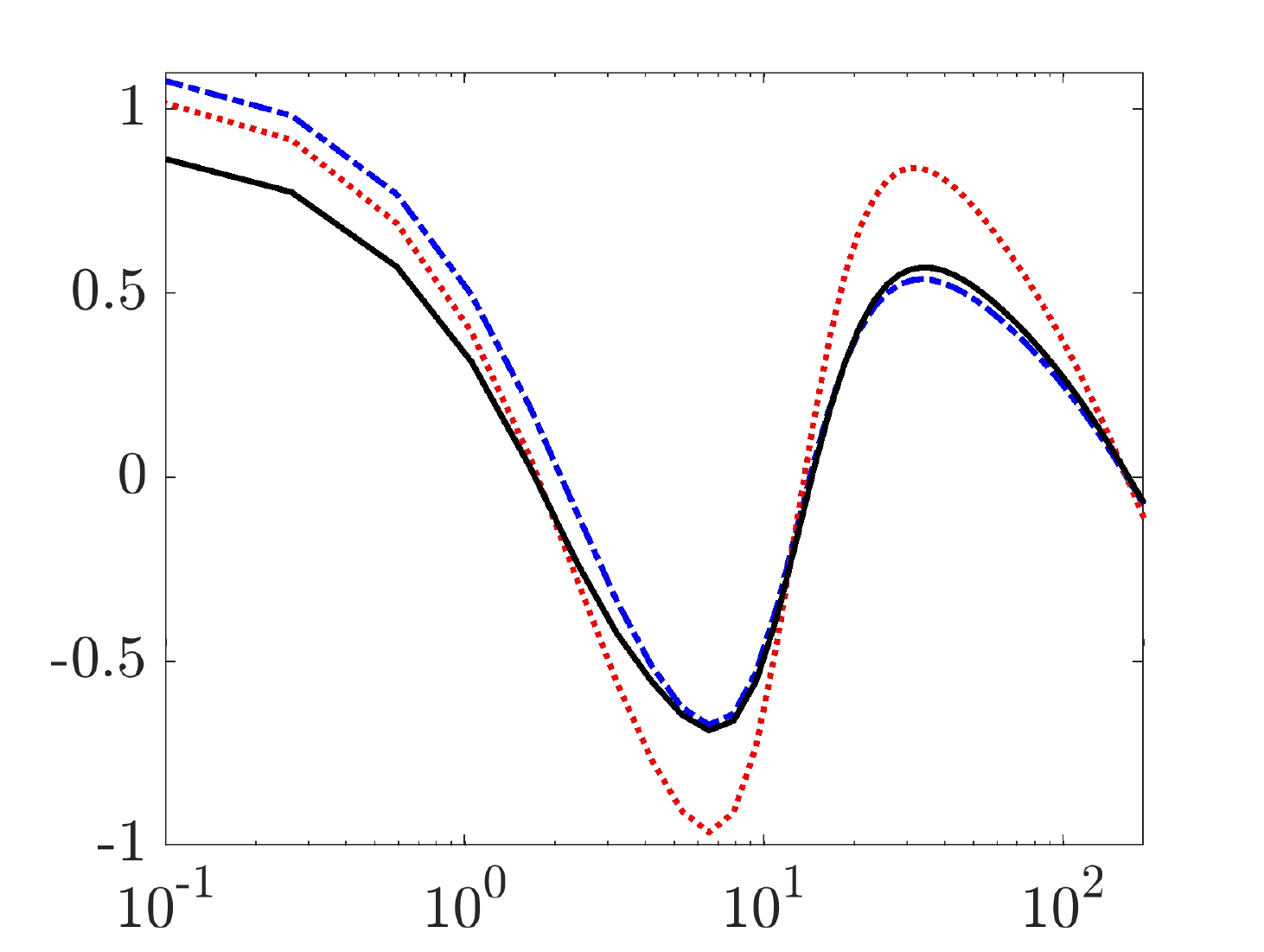}
       \\
       {\normalsize $y^+$}
       \end{tabular}
       \end{tabular}
       \end{center}
         \caption{Correction to (a) turbulent viscosity $\nu_{Tc}(y^+)$; and (b) mean velocity {$\bar{U}_c(y^+)$} in a turbulent channel flow with $Re_\tau=186$ over triangular riblets on the lower wall. The spanwise frequency $\omega_z$ associated with different shapes is selected to maximize drag reduction: {$\alpha=105\degree$, $\omega_z=50$ (solid black); $\alpha=75\degree$; $\omega_z=60$ (dotted red); $\alpha=45\degree$, $\omega_z=100$ (dot-dashed blue)}.}
        \label{fig.nuTUcopt}
\end{figure}

To illustrate the {influence of the shape of riblets on} turbulent viscosity and mean velocity, figure~\ref{fig.nuTUcopt} shows $\nu_{Tc}$ and {$\bar{U}_c$} for turbulent channel flow over riblets with different tip angles $\alpha$ and with spanwise frequencies $\omega_z$ that correspond to the maximum drag reduction. The largest turbulence suppression and mean velocity reduction is achieved for $\alpha=75\degree$, which is in agreement with the drag reduction trends observed in figure~\ref{fig.trish}. Between $\alpha=45\degree$ and $\alpha=105\degree$, the reduction in turbulent viscosity is more pronounced for the latter, which again reflects the drag reduction trends reported in figure~\ref{fig.trish}. 

\begin{figure}
        \begin{center}
        \begin{tabular}{cccc}
        \hspace{-.8cm}
        \subfigure[]{\label{fig.meanflowalp60s20}}
        &&
        \hspace{-.6cm}
        \subfigure[]{\label{fig.meanflowalp60s20cross}}
        &
        \\[-.5cm]
        \hspace{-.4cm}
\begin{tabular}{c}
        \vspace{.5cm}
        \normalsize{\rotatebox{90}{{$\bar{U}/U_l$}}}
       \end{tabular}
       &\hspace{-.5cm}
\begin{tabular}{c}
       \includegraphics[width=0.63\textwidth]{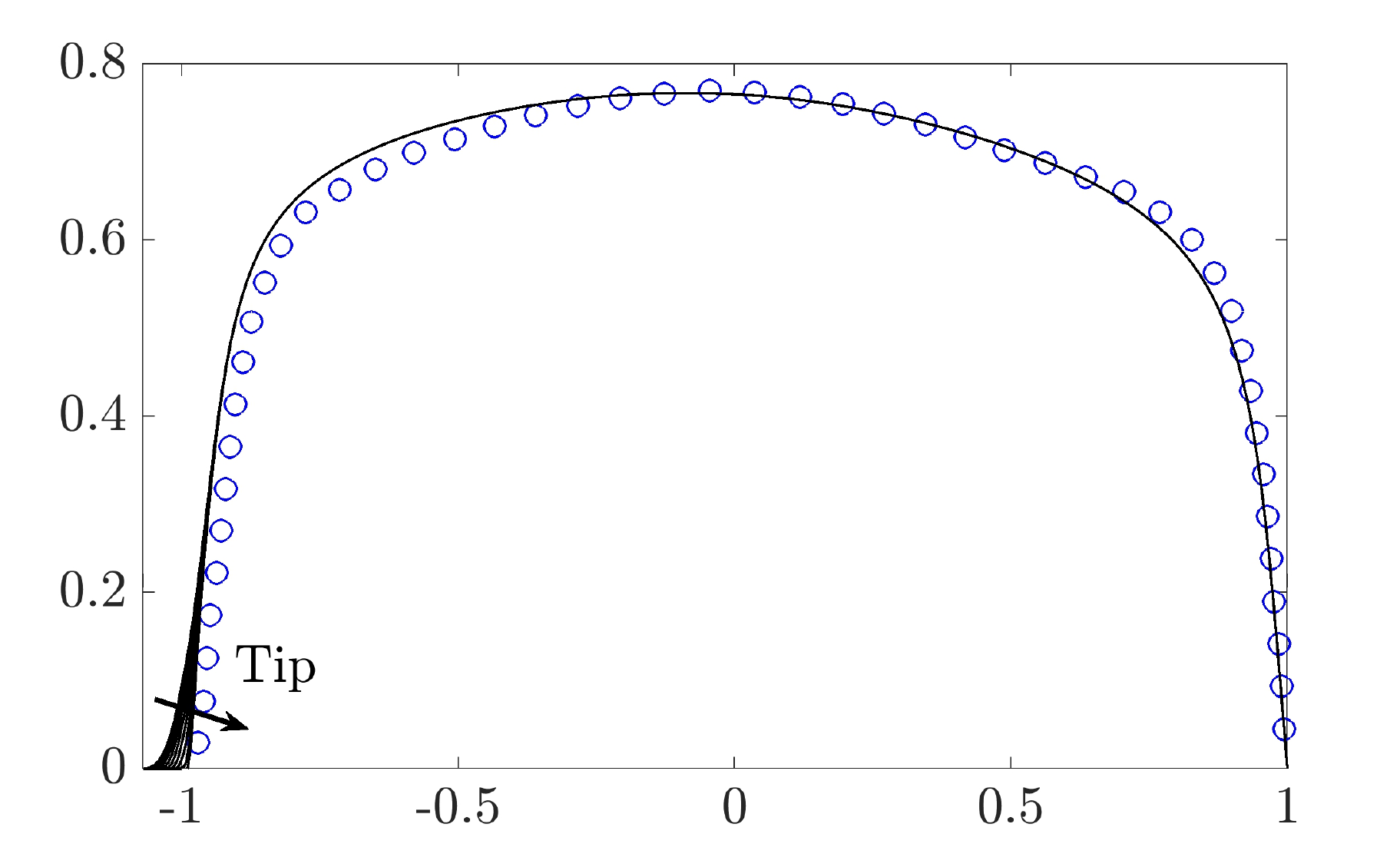}
       \\[-.2cm]
       \hspace{0.3cm} {\normalsize $y$}
       \end{tabular}
       &
    \begin{tabular}{c}
        \vspace{.6cm}
        \normalsize{\rotatebox{90}{$y$}}
       \end{tabular}
       &
    \begin{tabular}{c}
       \includegraphics[width=0.17\textwidth]{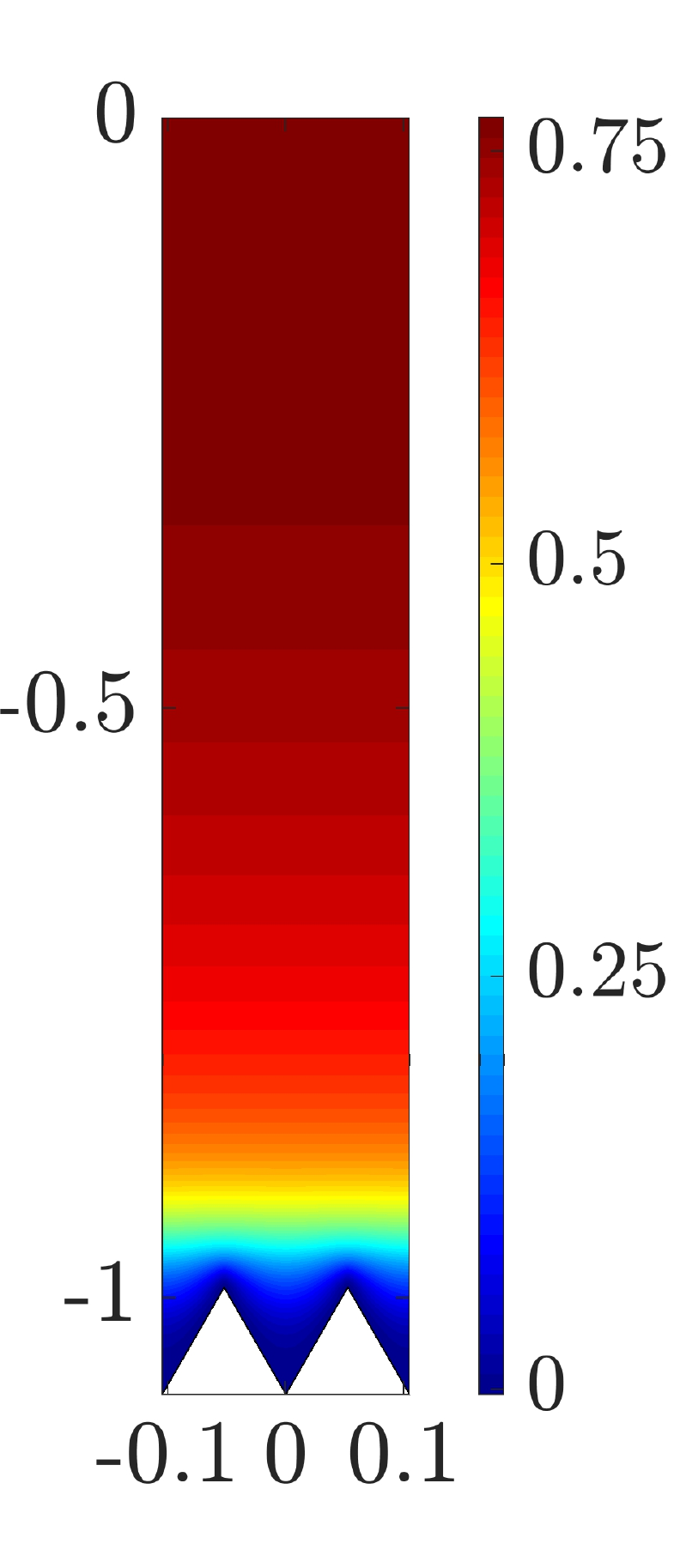}
       \\[-.3cm]
       \hspace{-0.35cm}{\normalsize $z$}
       \end{tabular}
       \end{tabular}
       \end{center}
         \caption{Mean velocity profiles {$\bar{U}(y,z)$ normalized with the laminar centreline velocity $U_l$} for $\alpha=60^\degree$ and $\omega_z=60$ ({$s^+\approx20$}): (a) One-dimensional view for different spanwise locations {($z\in[0,\pi/60]$) from our model (black curves) and the profile corresponding to $z\approx\pi/120$ resulting from DNS of~\cite{chomoikim93} (blue circles).} The direction of the arrow points to velocity profiles corresponding to spanwise locations farther away from the tip of riblets. (b) Color-plot of the streamwise mean velocity in the cross-plane.}
        \label{fig.meanflowalp60}
\end{figure}

For riblets with $\alpha=60\degree$ and $\omega_z=60$ ({$s^+\approx20$}), figure~\ref{fig.meanflowalp60} shows the variation of the mean velocity in the spanwise plane. No variation is found above $y>-0.9$, which is in agreement with the result of numerical simulations~\citep{chomoikim93}. {However, our predictions of the mean velocity profiles deviate from the result of DNS in the vicinity of the wall. This is because we have set the location of riblets to be slightly lower than the DNS computations in order to satisfy the constant bulk criteria. On the other hand, our results show a consistent promotion in the lower-half and suppression in the upper-half of the channel resulting in a slight lack of symmetry with respect to the centerline. This is mainly because of an over-predicted correction to turbulent eddy viscosity $\nu_{Tc}$ in the buffer layer and the inertial sublayer; cf.~figure~\ref{fig.omz3050nuTc}. Such over-predicted levels of turbulence suppression and drag reduction (cf.~figures~\ref{fig.trish} and~\ref{fig.tridraglg90d}) are caused by high amplitude stochastic forcing to the linearized equations~\eqref{eq.state-space}, which is shaped to match the two-dimensional DNS energy spectrum (\S~\ref{sec.correctX}). Analyzing the efficacy of more sophisticated forcing schemes~\citep{zarjovgeoCDC16,zarchejovgeoTAC17,zarjovgeoJFM17} that may refine mean velocity predictions is a topic for future research.}

	\vspace*{-2ex}
\subsection{Effect of riblets on turbulent kinetic energy}
\label{sec.energyspectra}

\begin{figure}
        \begin{center}
        \begin{tabular}{cccc}
        \hspace{-.8cm}
        \subfigure[]{\label{fig.DNSspectrum160}}
        &&
        \subfigure[]{\label{fig.90domz160E2}}
        &
        \\[-.5cm]
        \hspace{-.6cm}
	\begin{tabular}{c}
        \vspace{.4cm}
        \normalsize{\rotatebox{90}{$k_x$}}
       \end{tabular}
       &\hspace{-.3cm}
	\begin{tabular}{c}
       \includegraphics[width=0.4\textwidth]{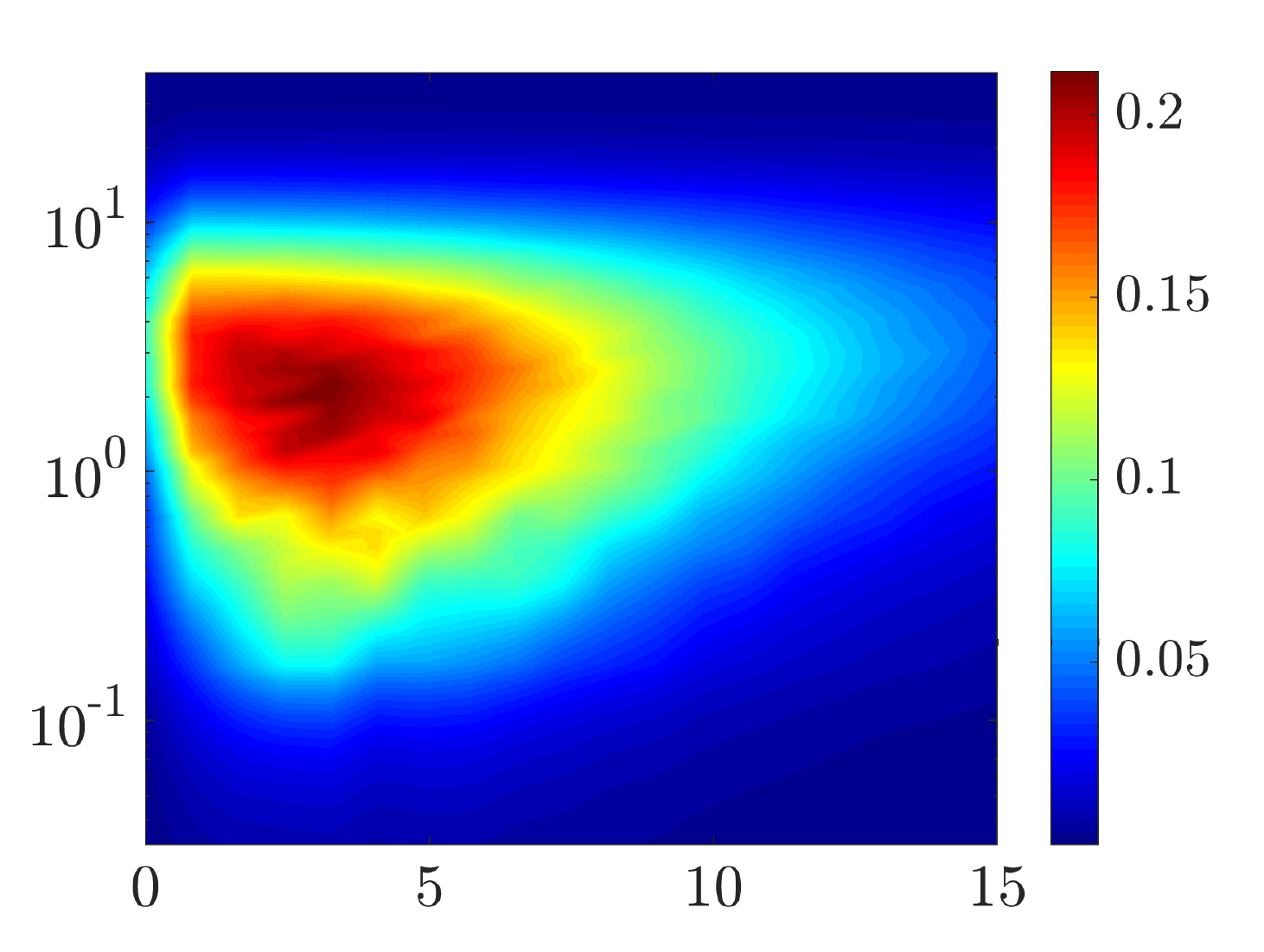}
       \end{tabular}
       &
       &\hspace{.1cm}
    \begin{tabular}{c}
       \includegraphics[width=0.4\textwidth]{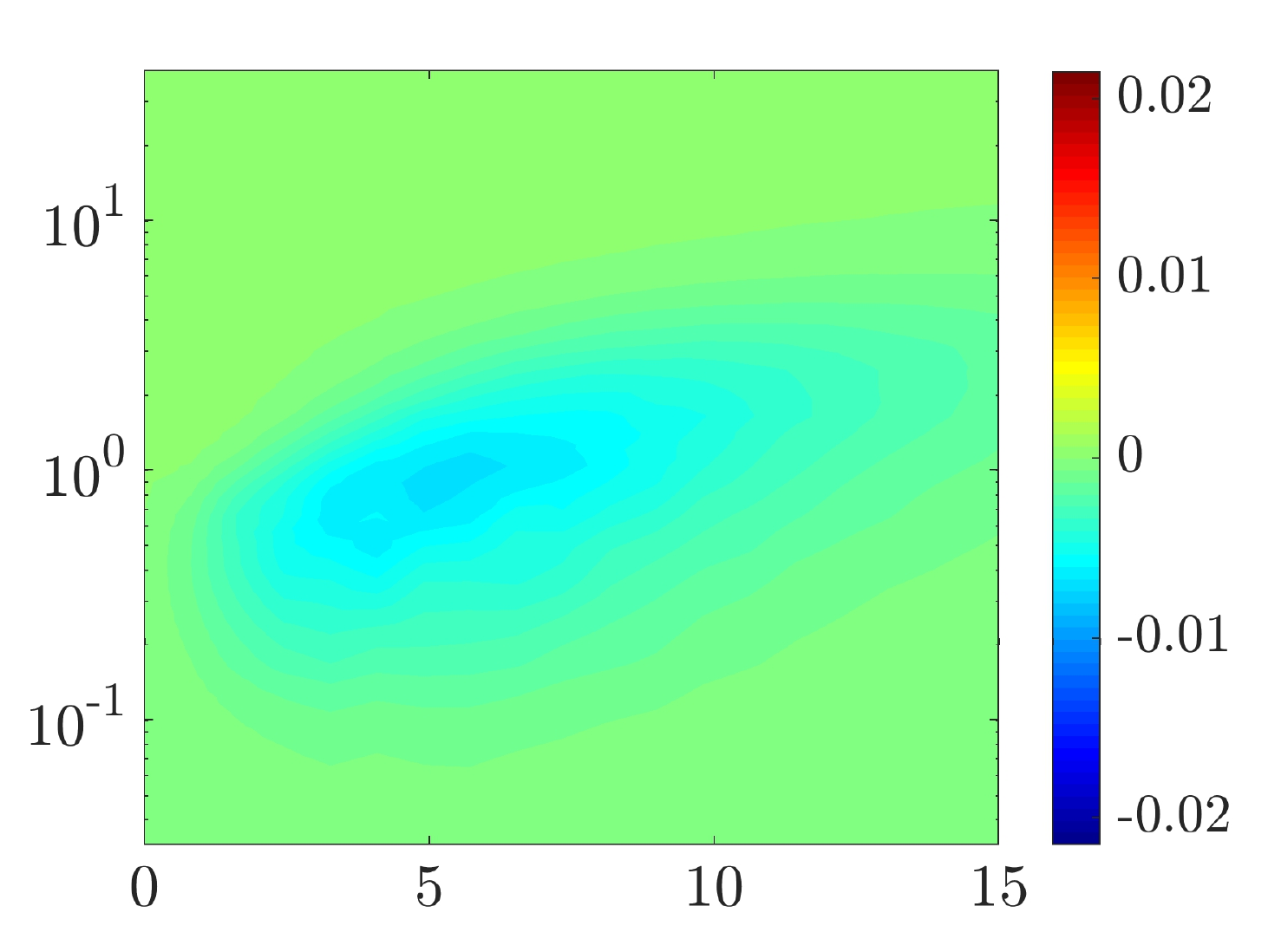}
       \end{tabular}
       \\
       \hspace{-.8cm}
        \subfigure[]{\label{fig.DNSspectrum50}}
        &&
        \subfigure[]{\label{fig.90domz50E2}}
        &
        \\[-.5cm]
        \hspace{-.6cm}
	\begin{tabular}{c}
        \vspace{.4cm}
        \normalsize{\rotatebox{90}{$k_x$}}
       \end{tabular}
       &\hspace{-.3cm}
	\begin{tabular}{c}
       \includegraphics[width=0.4\textwidth]{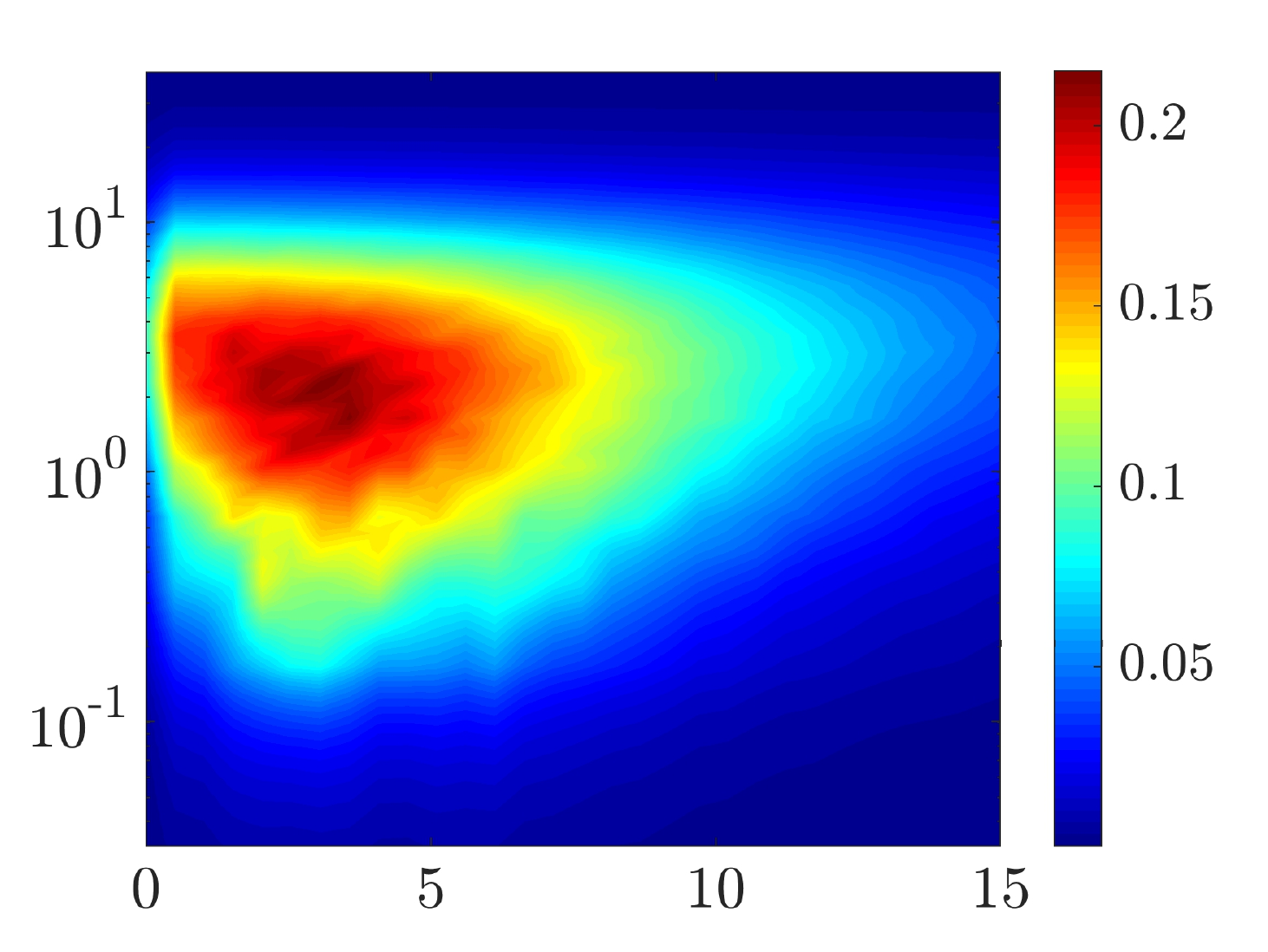}
       \end{tabular}
       &
       &\hspace{.1cm}
    \begin{tabular}{c}
       \includegraphics[width=0.4\textwidth]{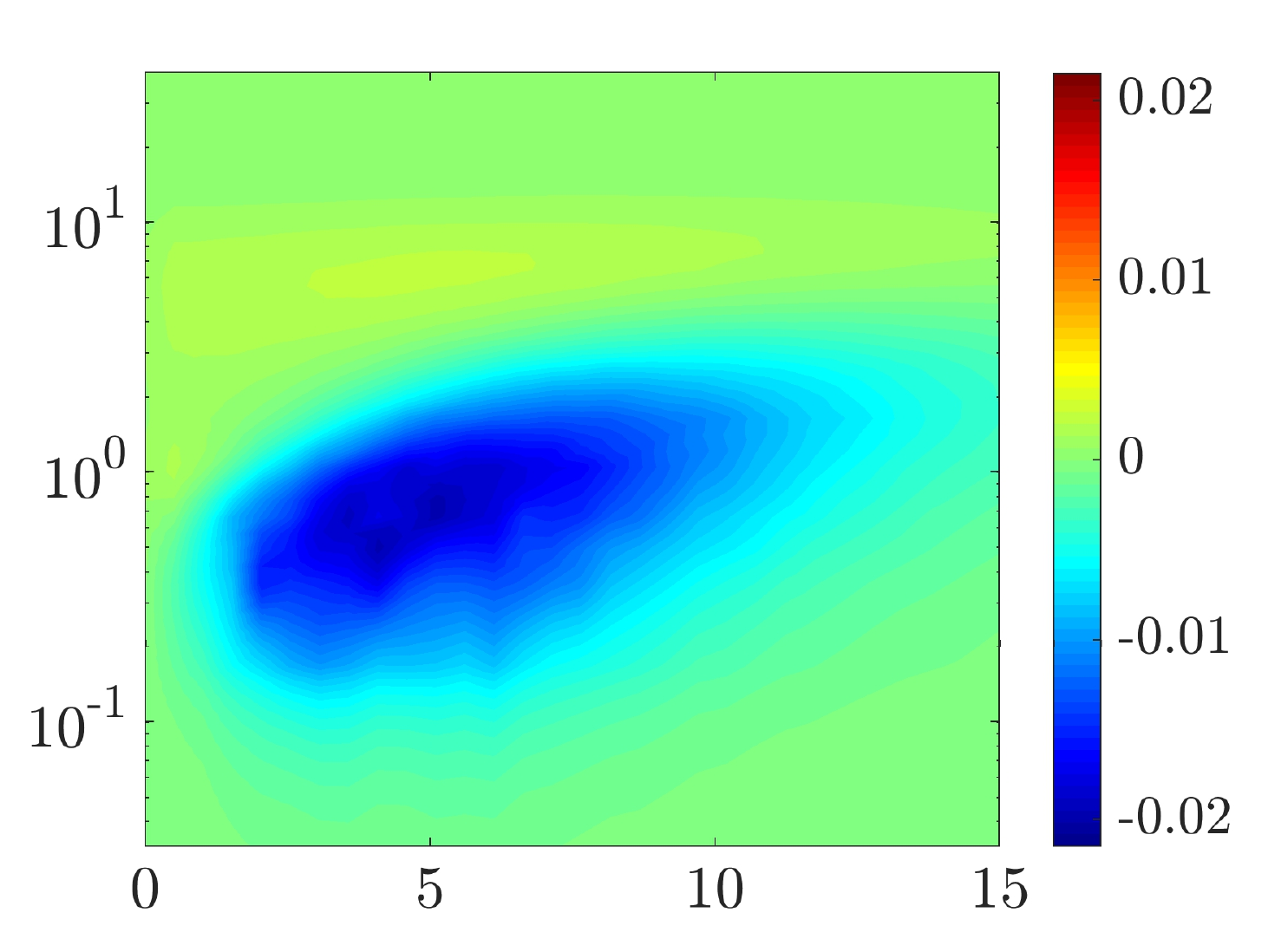}
       \end{tabular}
       \\
       \hspace{-.8cm}
        \subfigure[]{\label{fig.DNSspectrum30}}
        &&
        \subfigure[]{\label{fig.90domz30E2}}
        &
        \\[-.5cm]
        \hspace{-.6cm}
	\begin{tabular}{c}
        \vspace{.4cm}
        \normalsize{\rotatebox{90}{$k_x$}}
       \end{tabular}
       &\hspace{-.3cm}
	\begin{tabular}{c}
       \includegraphics[width=0.4\textwidth]{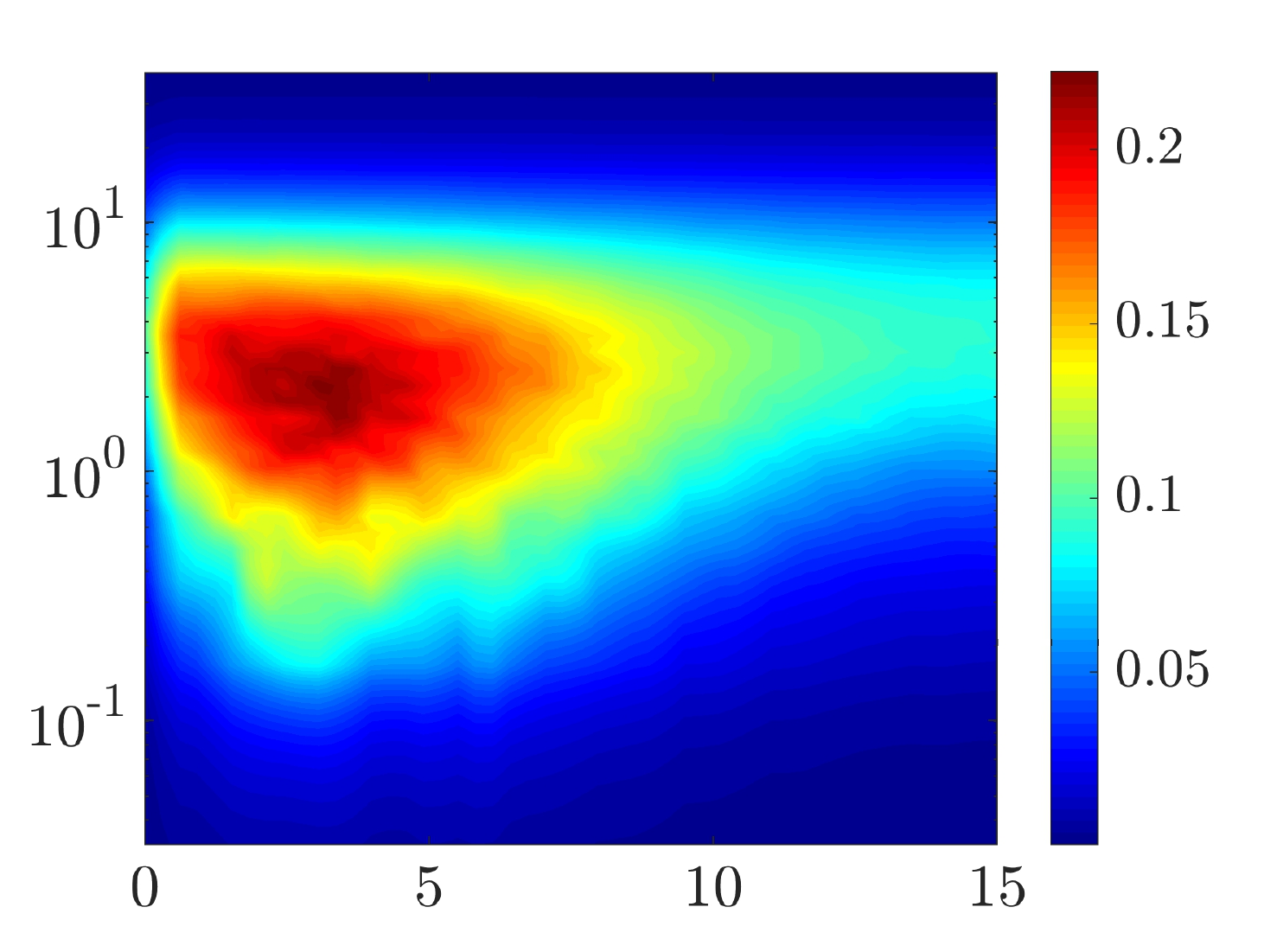}
       \\[-.1cm]
       \hspace{-.3cm}
       {\small $\theta$}
       \end{tabular}       
       &
       &\hspace{.1cm}
    \begin{tabular}{c}
       \includegraphics[width=0.4\textwidth]{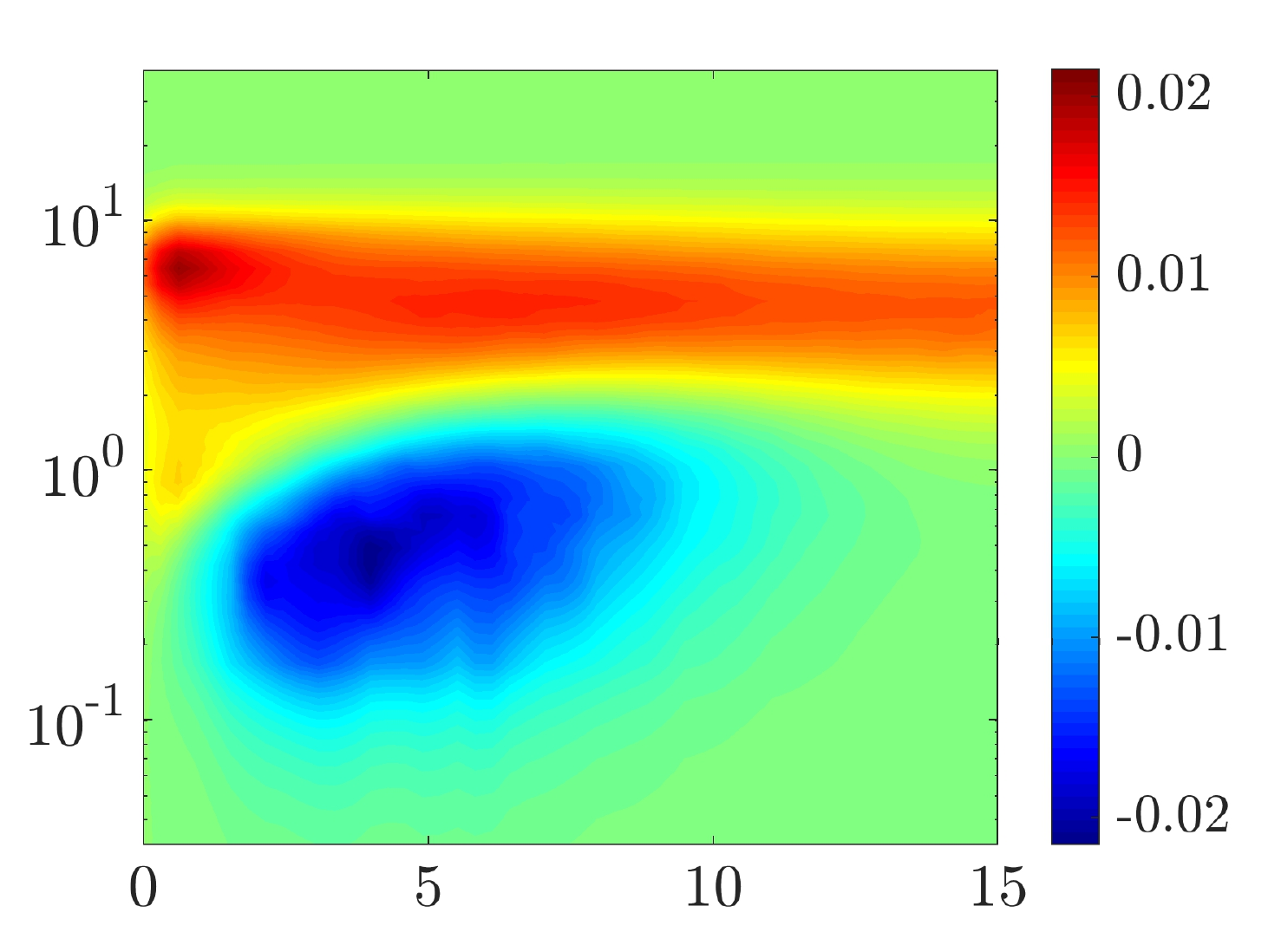}
       \\[-.1cm]
       \hspace{-.3cm}
       {\small $\theta$}
       \end{tabular}       
       \end{tabular}
       \end{center}
        \caption{{The premultiplied reference energy spectrum $k_x {\bar{E}_s}(k_x,\theta)$ (equation~\eqref{eq.Es}) from DNS of a turbulent channel flow with $Re_\tau=186$~\citep{deljim03} (left column), and the correction to the premultiplied spectrum $k_x {\bar{E}_c}(k_x,\theta)$ (equation~\eqref{eq.Ec}) due to the presence of various sizes of triangular riblets with $\alpha=90\degree$ (right column): (a, b) $\omega_z=160$ ($l_g^+=3.6$); (c, d) $\omega_z=50$ ($l_g^+=11.7$, optimal); and (e, f) $\omega_z=30$ ($l_g^+=19.5$).}} 
\label{fig.spectrums}
\end{figure}

We next examine the effect of triangular riblets on the fluctuations' kinetic energy. Figure~\ref{fig.spectrums} compares the {reference energy spectrum of turbulent channel flow with smooth walls (equation~\eqref{eq.Es}) to the changes in the energy spectrum caused by equally shaped riblets of different sizes ($\omega_z = 160$, $50$, and $30$) with $\alpha=90\degree$ (equation~\eqref{eq.Ec}).} The energy spectra are premultiplied by the logarithmically scaled streamwise wavenumber $k_x$ so that the areas under the plots determine the total kinetic energy. Since the spanwise direction involves the parameterization $\theta_n = \theta + n \omega_z$, summation over $n$ is performed to integrate the energetic contribution of various harmonics in $\omega_z$ and identify the dependence of the energy spectrum on $\theta${; see \S~\ref{sec.correctX}}.

\begin{figure}
\begin{tabular}{cccccc}
\hspace{-.4cm}
\subfigure[]{\label{fig.105domz45E2kx}}
&&
\hspace{-.89cm}
\subfigure[]{\label{fig.75domz60E2kx}}
&&
\hspace{-.89cm}
\subfigure[]{\label{fig.45domz80E2kx}}
&
\\[-.2cm]
\hspace{-.3cm}
\begin{tabular}{c}
\vspace{.4cm}
\rotatebox{90}{\small $k_x$}
\end{tabular}
&
\hspace{-.4cm}
\begin{tabular}{c}
\includegraphics[width=0.31\textwidth]{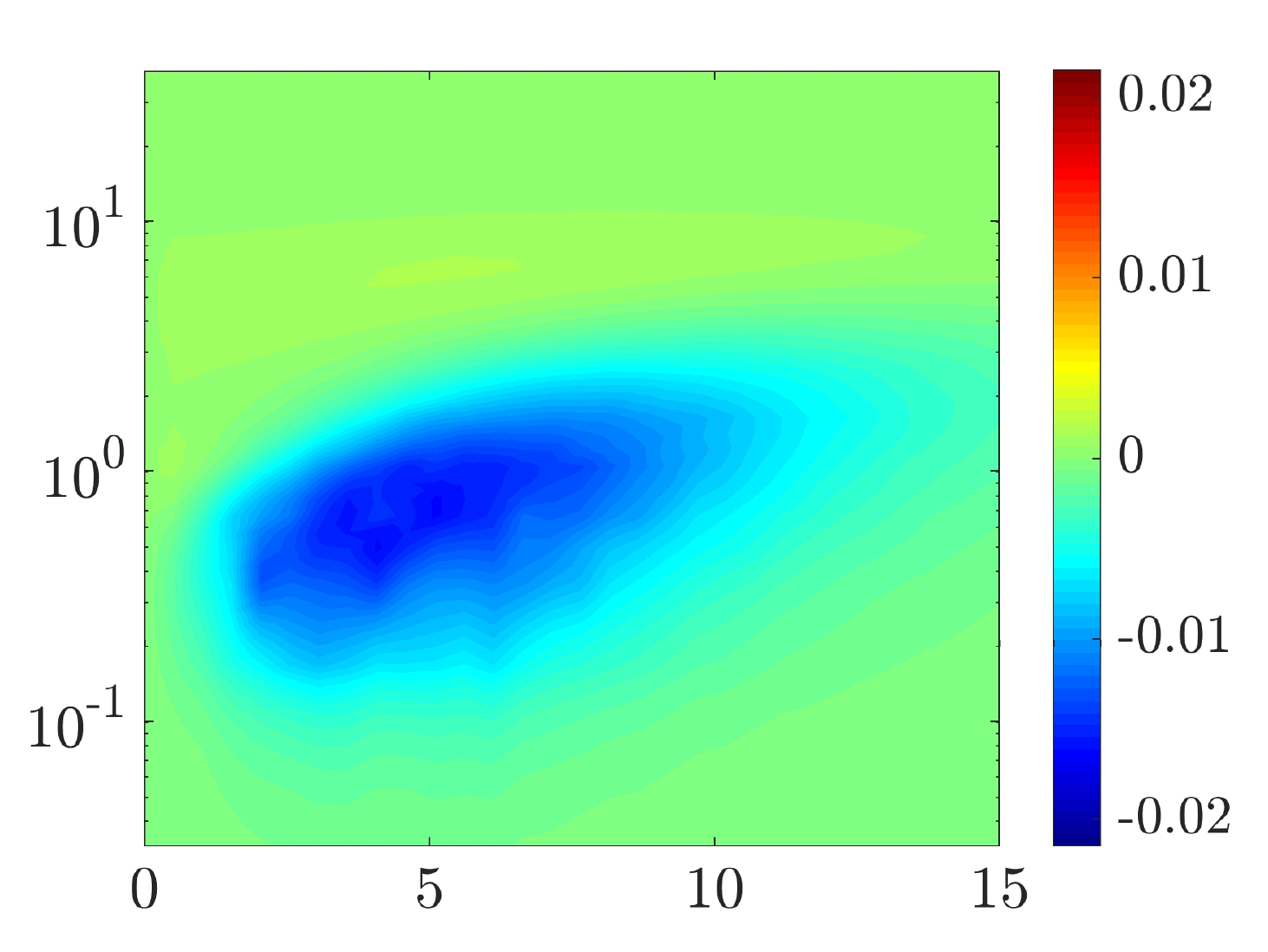}
\end{tabular}
&&
\hspace{-.46cm}
\begin{tabular}{c}
\includegraphics[width=0.31\textwidth]{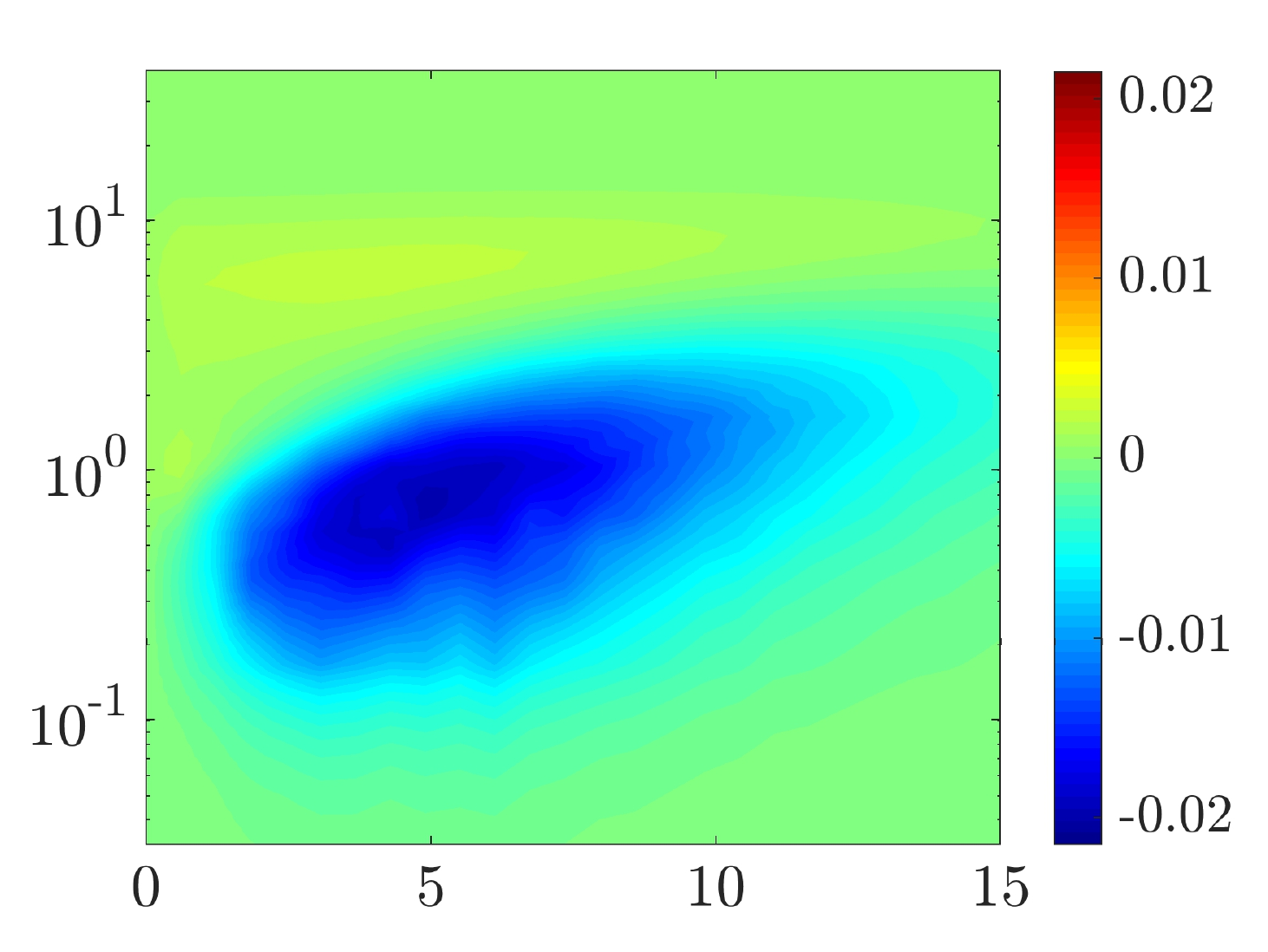}
\end{tabular}
&&\hspace{-.46cm}
\begin{tabular}{c}
\includegraphics[width=0.31\textwidth]{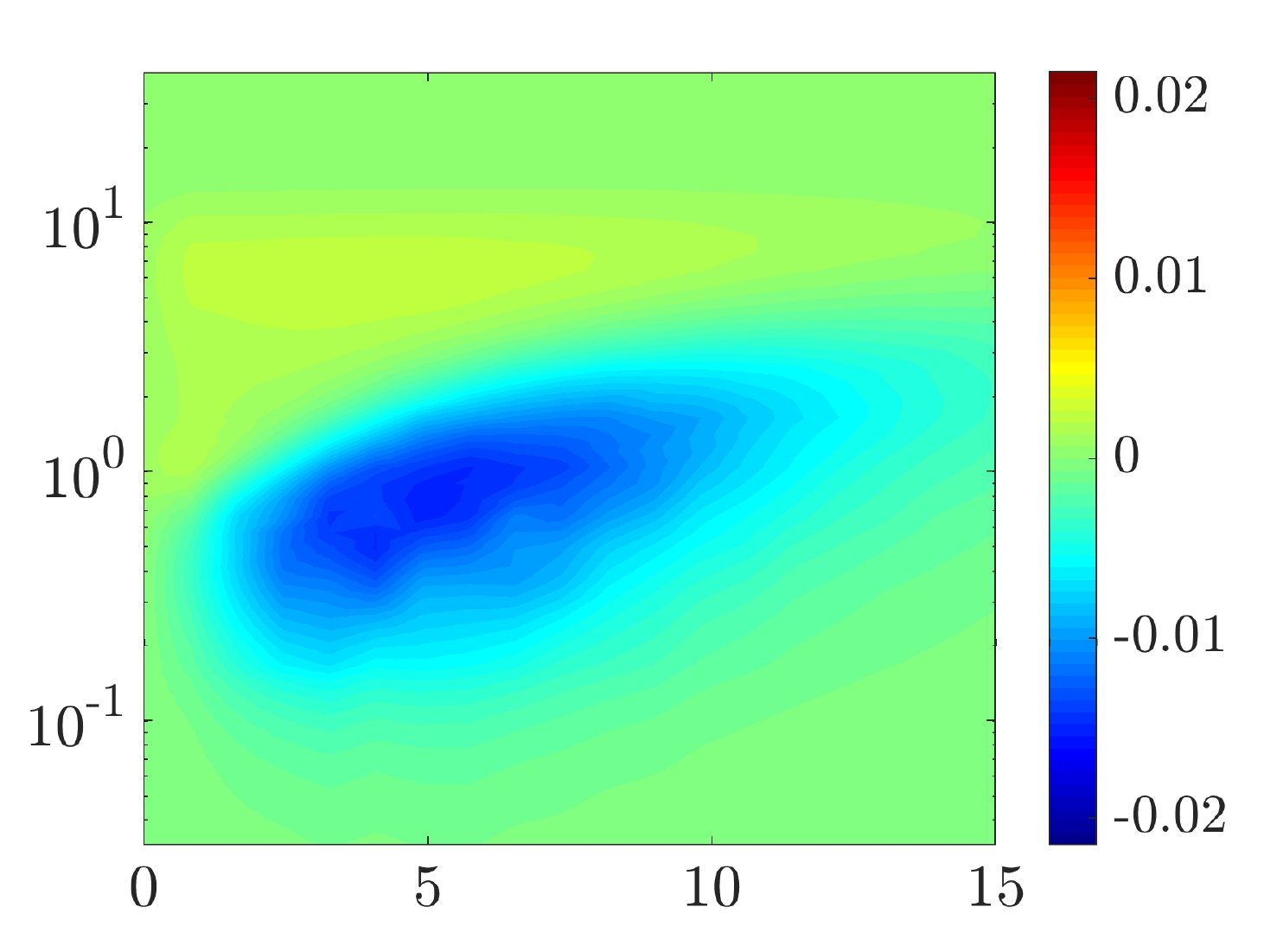}
\end{tabular}
\\[-.1cm]
&\hspace{-.9cm}
{\small $\theta$}
&&
\hspace{-.8cm}
{\small $\theta$}
&&
\hspace{-.8cm}
{\small $\theta$}
\end{tabular}
\caption{Correction to the premultiplied energy spectrum  {$k_x \bar{E}_c(k_x,\theta)$} in a turbulent channel flow with $Re_\tau=186$ triangular riblets of various sizes. The spanwise frequency $\omega_z$ associated with different shapes of riblets corresponds to maximum drag reduction: (a) $\alpha=105^\degree$, {$\omega_z=50$}; (b) $\alpha=75^\degree$, $\omega_z=60$; and (c) $\alpha=45^\degree$, {$\omega_z=100$}.}
\label{fig.optspectrums}
\end{figure}

For channel flow over smooth walls with $Re_\tau=186$, {the color plots in the left column of figure~\ref{fig.spectrums} show} that the most energetic modes take place at $(k_x,\theta)=(2.5,3.5)$. As blue regions in figures~\ref{fig.90domz160E2},~\ref{fig.90domz50E2}, and~\ref{fig.90domz30E2} illustrate, riblets reduce  {the} energy content of flow structures with smaller streamwise wavenumbers. Moreover, yellow and red regions in figures~\ref{fig.90domz50E2} and~\ref{fig.90domz160E2} demonstrate that larger riblets increase  {the} energy content of flow structures with larger streamwise wavenumbers. For these three cases, the largest energy amplification takes place around $(k_x,\theta)=(5.5,2.4)$, $(6.4,4.6)$, and $(6.4,0.6)$, respectively. On the other hand, the maximum energy reduction occurs around $(k_x,\theta)=(4.9,0.9)$, $(4.0,0.5)$, and $(4.0,0.5)$, respectively. Although the peak points are different, figures~\ref{fig.90domz160E2},~\ref{fig.90domz50E2}, and~\ref{fig.90domz30E2} demonstrate similar amplification/suppression trends: riblets suppress/increase energy content of long/short streamwise length scales. These results provide evidence that the analysis of spatially-periodic systems, e.g., the one considered in this paper, cannot be limited to a single horizontal wavenumber pair associated with the peak of the energy spectrum or the dominant near-wall cycle. We note that similar conclusions were reached in the analysis of turbulent channel flow subject to transverse wall oscillations~\citep{moajovJFM12}. Finally, the dependence of correction ${\bar{E}_c}(k_x,\theta)$ on the shape of riblets is shown in figure~\ref{fig.optspectrums}. For all cases shown in this figure, similar modes are affected by the presence of riblets and the suppression of kinetic energy is more pronounced for riblets with $\alpha=75\degree$ and $\omega_z=60$, which also yield the largest drag reduction (cf.\ figure~\ref{fig.trish}). This suggests  {a} synchrony between the dependence of drag reduction and energy suppression on the geometry of triangular riblets.

\begin{figure}
        \begin{center}
        \begin{tabular}{cccc}
        \hspace{-.8cm}
        \subfigure[]{\label{fig.triengylg}}
        &&
        \hspace{-.6cm}
        \subfigure[]{\label{fig.tridraglg}}
        &
        \\[-.5cm]
        \hspace{-.4cm}
        \begin{tabular}{c}
        \vspace{.5cm}
        \normalsize{\rotatebox{90}{$\Delta E(\%)$}}
       \end{tabular}
       &\hspace{-.3cm}
    \begin{tabular}{c}
       \includegraphics[width=0.4\textwidth]{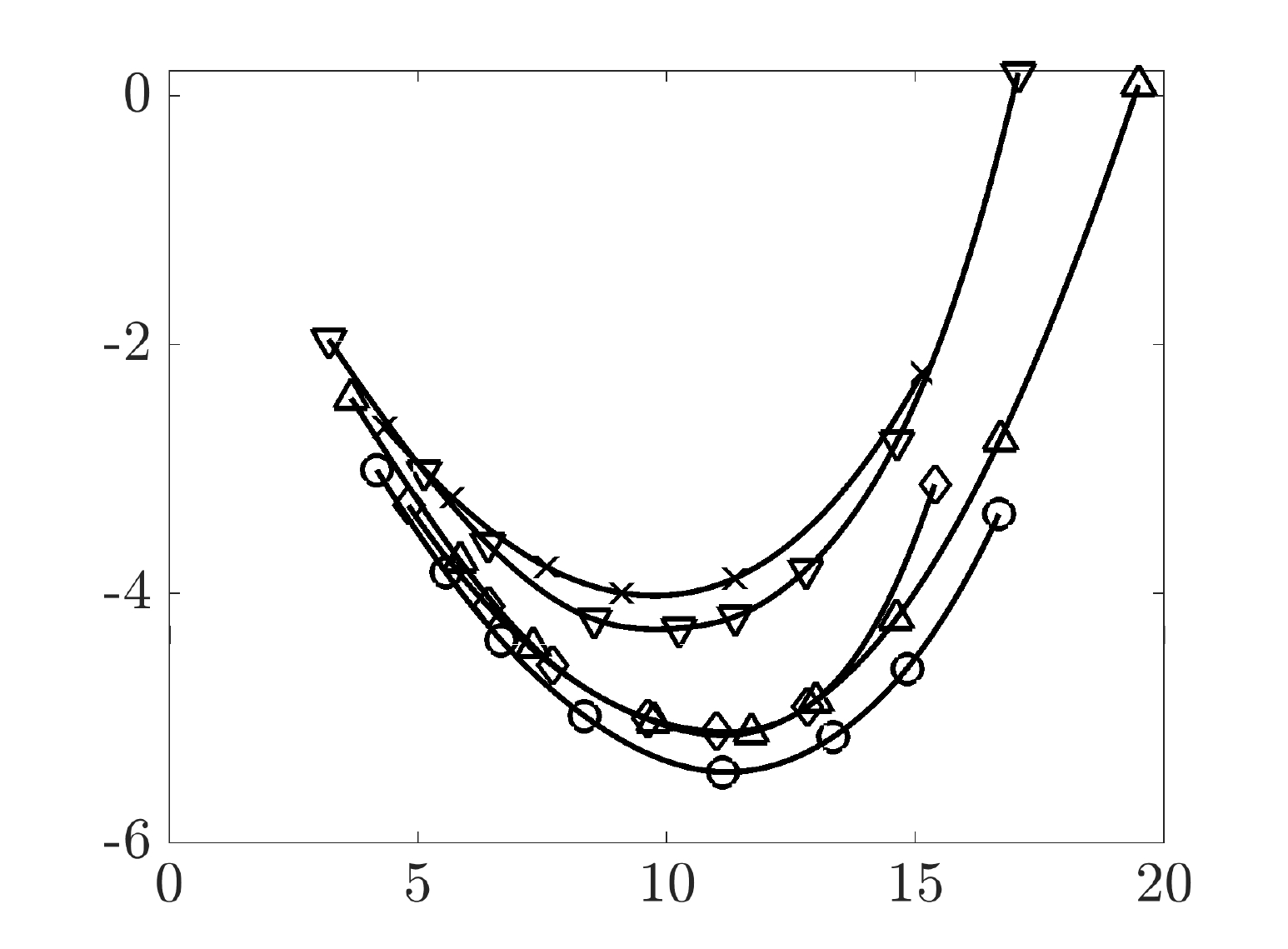}
       \\ {\normalsize $l_g^+$}
       \end{tabular}
        &
	\begin{tabular}{c}
        \vspace{.5cm}
        \normalsize{\rotatebox{90}{$\Delta D (\%)$}}
       \end{tabular}
       &\hspace{-.3cm}
	\begin{tabular}{c}
       \includegraphics[width=0.4\textwidth]{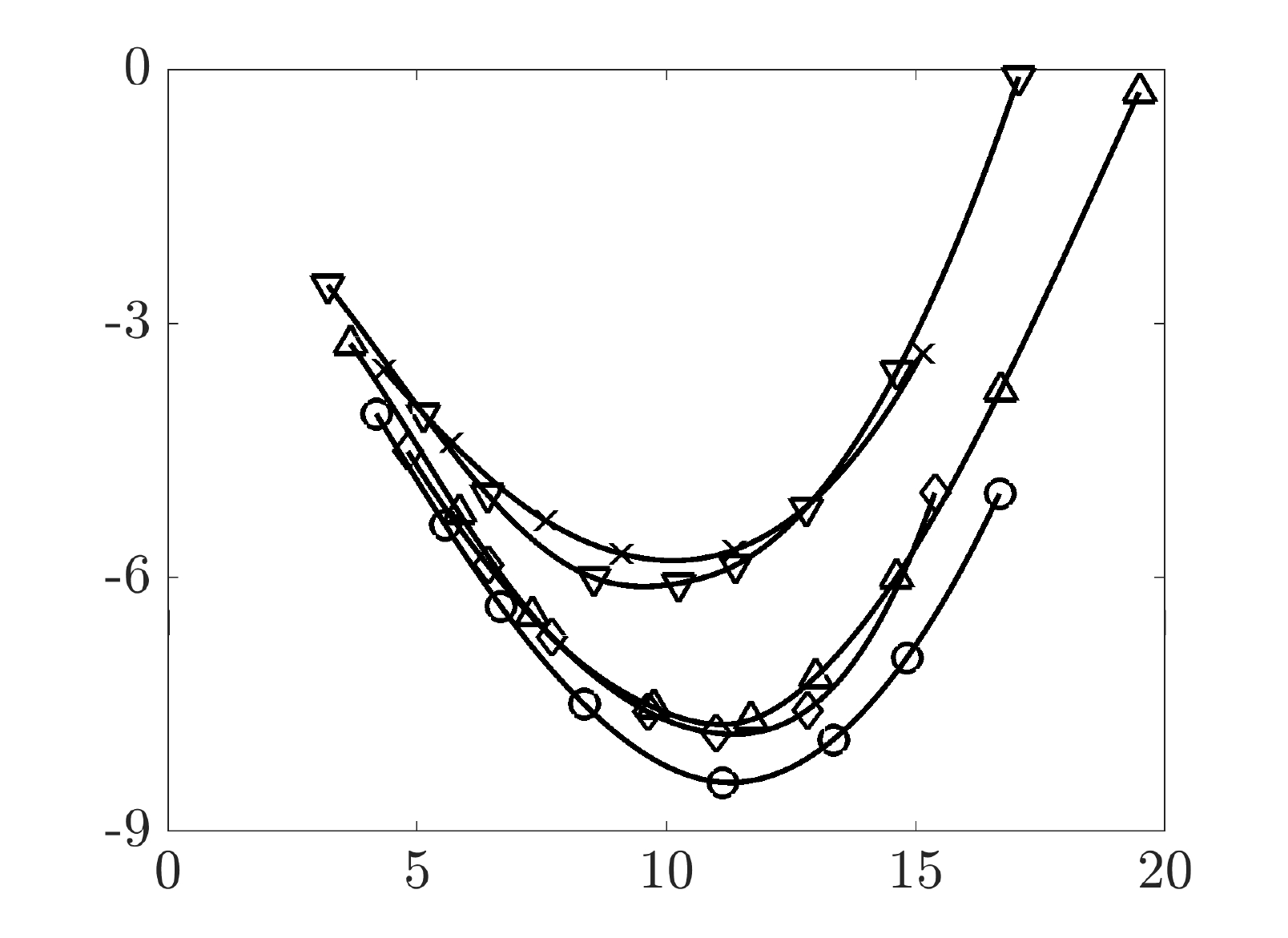}
       \\ {\normalsize $l_g^+$}
       \end{tabular}
       \end{tabular}
       \end{center}
        \caption{(a) Kinetic energy suppression; and (b) turbulent drag reduction in a channel flow with $Re_\tau=186$ over triangular riblets of various sizes. Symbols denote different shapes of triangular riblets with $\alpha=105\degree$ ($\bigtriangledown$); $90\degree$ ($\bigtriangleup$); $75\degree$ ($\bigcirc$); $60\degree$ ($\lozenge$); and $45\degree$ ($\times$).}
        \label{fig.trilg}
\end{figure}

\begin{figure}
	\begin{center}
        \begin{tabular}{cccc}
         \hspace{-.8cm}
        \subfigure[]{\label{fig.ERDR}}
        &&
        \hspace{-.6cm}
        \subfigure[]{\label{fig.Rvslg}}
        &
        \\[-.5cm]
        \hspace{-.4cm}
        \begin{tabular}{c}
                \vspace{.4cm}
                \rotatebox{90}{\normalsize $\Delta D (\%)$}
        \end{tabular}
        &\hspace{-.4cm}
        \begin{tabular}{c}
                \includegraphics[width=.4\textwidth]{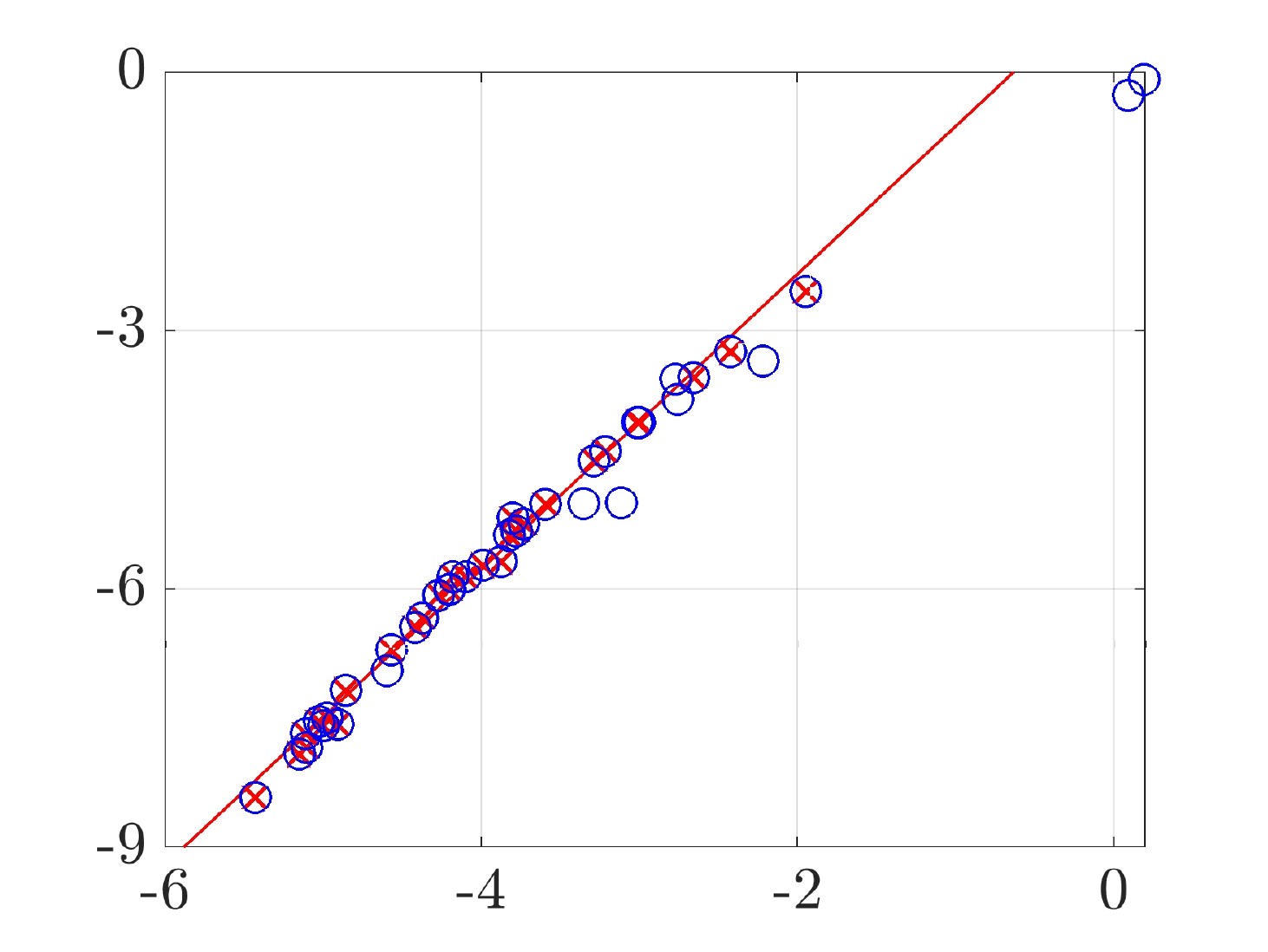}
                \\
                {\normalsize $\Delta E (\%)$}
        \end{tabular}
        &
        \hspace{-.2cm}
        \begin{tabular}{c}
                \vspace{.4cm}
                \rotatebox{90}{\normalsize $R^2$}
        \end{tabular}
        &\hspace{-.3cm}
        \begin{tabular}{c}
                \includegraphics[width=.4\textwidth]{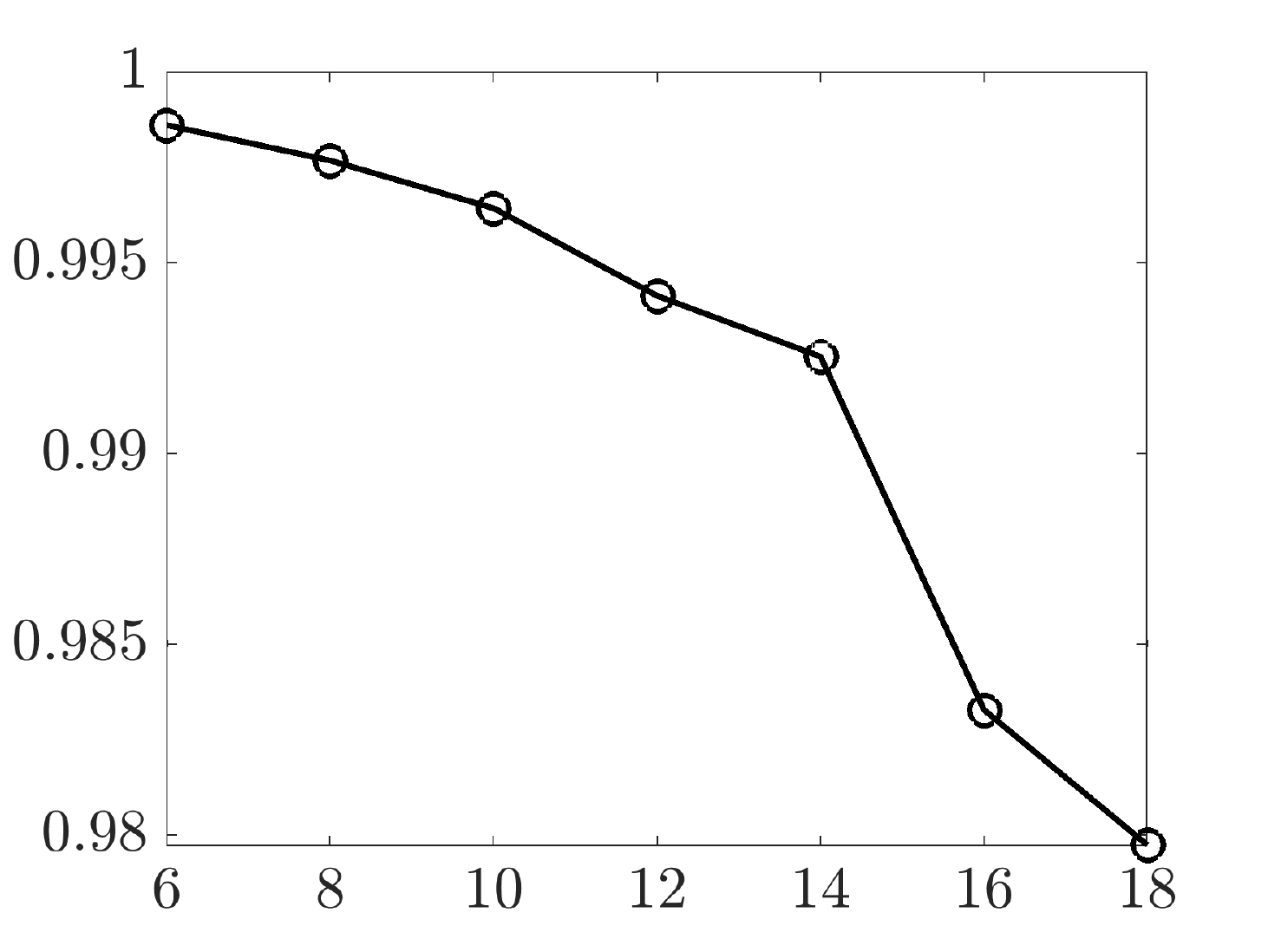}
                \\
                {\normalsize $l_{g_T}^+$}
        \end{tabular}
        \end{tabular}
        \end{center}
        \caption{(a) A linear relation between the turbulent drag reduction $\Delta D$ and the kinetic energy suppression $\Delta E$ for a channel flow with $Re_\tau=186$ over triangular riblets of different size $l_g^+$. Circles mark cases listed in Table~\ref{table.comp-case} and crosses mark data for riblets {with $l_g^+ \leq 14$}. The red line provides linear interpolation for crosses. (b) The coefficient of determination $R^2$ for linear regression models resulting from data with $l_g^+\leq l_{g_T}^+$.}
        \label{fig.linearregression}
\end{figure}

Figure~\ref{fig.triengylg} shows the percentage of kinetic energy variation $\Delta E \DefinedAs \hat{E}_c/\hat{E}_s$ for triangular riblets as a function of the riblet groove area $l_g^+$. Here, $\hat{E}_c$ and $\hat{E}_s$ denote the  {correction to kinetic energy due to} the presence of riblets  {and the kinetic energy of velocity fluctuations in the absence of riblets}, respectively. These two quantities can be computed by integrating the  {corresponding energy spectra, i.e., $\bar{E}_c(k_x,\theta)$ and $\bar{E}_s(k_x,\theta)$,} over all horizontal wavenumbers $k_x$ and $\theta$. 
On the other hand, figure~\ref{fig.tridraglg} shows the percentage of drag reduction for the same values of $l_g^+$. Our computations demonstrate similar trends in the dependence of  {$\Delta {E}$} and $\Delta D$ on $l_g^+$ (cf.\ figures~\ref{fig.triengylg} and~\ref{fig.tridraglg}), especially for the riblets in  {the} viscous regime. Based on the various cases considered in figure~\ref{fig.trilg}, the linear regression model $\Delta D = 1.7152\, {\Delta {E}} + 1.0907$ can be extracted with a coefficient of determination $R^2=0.9925$ for riblets with $l_g^+ \leq 14$. Strong correlation between changes in turbulent drag and kinetic energy suggests that energy can be used as a surrogate for predicting the effect of riblets on skin-friction drag; see figure~\ref{fig.ERDR}. We note that a similar linear relation (but with a different slope) can be observed at $Re_\tau=547$. These results are not reported here for brevity.

As shown in figure~\ref{fig.omz3050nuTc}, riblets can suppress or enhance turbulence near the wall. Small riblets can disturb the near-wall cycle in the turbulent flow by generating and preserving laminar regions within the grooves. On the other hand, for larger riblets, streamwise rollers penetrate into the grooves which enhances turbulence close to the wall. As a result, nonlinear effects take over and the linear relation between drag/energy reduction and any metric of the riblet size (e.g., $l_g$)  {becomes compromised. As illustrated in figure~\ref{fig.Rvslg}, the quality of a linear regression model drops (i.e., $R^2$ becomes smaller) when data for larger-size riblets is taken into account.}

	\vspace*{-4ex}
\section{Turbulent flow structures}
\label{sec.flowstructures}

In this section, we use the stochastically forced linearized model~\eqref{eq.turblinriblet} to examine the effect of riblets of different sizes and shapes on {typical} turbulent flow structures and relevant drag reduction mechanisms. First, we study the distortion of the dominant near-wall cycle that arises from the presence of riblets on the lower wall. We also examine the K-H instability, which is related to the breakdown of the viscous regime, and the performance deterioration for large riblets. Finally, we consider a channel flow with higher-Reynolds number to investigate the wall-normal support of very large scale motions (VLSM) in the presence of riblets. In this section, in addition to the wall-normal coordinate, wavelengths are also given in inner (viscous) units with $\lambda_x^+ = 2\pi Re_\tau/k_x$ and  {$\lambda_z^+ = 2\pi Re_\tau/\theta$}.

Following the proper orthogonal decomposition of~\cite{baklum67,moimos89}, we extract flow structures from our model using the eigenvalue decomposition of the velocity covariance matrix in statistical steady-state,
\begin{align}
\label{eq.output-covariance}
	\Phi_\theta (k_x) \;=\; C_\theta(k_x)\, X_\theta(k_x)\, C_\theta^*(k_x)
\end{align}
where $X_\theta(k_x)$ represents the solution of Lyapunov equation~\eqref{eq.lyap}. The eigenvectors associated with the principal pair of eigenvalues form  {flow structures that are located in the vicinity of the upper and lower channel walls. The first pair of eigenvalues are usually one order of magnitude larger than the second pair. This indicates that the flow structures that correspond to the principal eigenvectors are energetically dominant and representative of the essential dynamics.} The velocity components {of flow structures} are constructed by integrating over all spanwise harmonics and by accounting for the symmetry in the streamwise direction as
\begin{align}
	\label{eq.spatialstruct}
	\ba{rcl}
		u(x,y,z)
		&\!\!=\!\!&
		\phantom{-} \ds{\sum_{n \, \in \, \bbZ}}
		\cos(\theta_n z) \, \mathfrak{Re} \left( \tilde{u}(k_x,\theta_n)\, \mre^{\mri k_x x} \right),
		\\[.5cm]
		v(x,y,z)
		&\!\!=\!\!&
		\phantom{-} \ds{\sum_{n \, \in \, \bbZ}}
		\cos(\theta_n z) \, \mathfrak{Re} \left( \tilde{v}(k_x,\theta_n) \, \mre^{\mri k_x x} \right),
		\\[.5cm]
		w(x,y,z)
		&\!\!=\!\!&
		-\ds{\sum_{n \, \in \, \bbZ}}
		\sin(\theta_n z) \, \mathfrak{Im} \left( \tilde{w}(k_x,\theta_n) \, \mre^{\mri k_x x} \right).
	\ea
\end{align}
Here, $\mathfrak{Re}$ and $\mathfrak{Im}$ denote real and imaginary parts, and $\tilde{u}$, $\tilde{v}$, and $\tilde{w}$ correspond to the streamwise, wall-normal, and spanwise components of {an} eigenvector of the matrix $\Phi_\theta(k_x)$, given in equation~\eqref{eq.output-covariance}. 

In a turbulent channel flow with smooth walls, the dominant eigenmodes of the velocity covariance matrix appear in pairs and represent symmetric flow structures that reside in the vicinity of the upper and lower walls. Surface corrugation on the lower wall breaks this symmetry and can cause a suppression of near-wall structures in the lower half of the channel. In other words, the flow structures that dominate the flow close to the riblets can be less energetic than the flow structures close to the upper wall. As a result, physically relevant flow structures near the riblets, e.g., the dominant flow structures associated with the near-wall cycle over riblets of optimal size, are often associated with the second, less energetic, eigenmode of $\Phi_\theta(k_x)$.

	\vspace*{-2ex}
\subsection{{Near-wall cycle in turbulent channel flow with $Re_\tau=186$}}
\label{sec.NWcycle}

In the absence of riblets, the so-called near-wall cycle dominates the physics of the turbulent channel flow by generating streamwise streaks from the advection of the mean shear by streamwise vortices and the formation of streamwise vortices through streak instability and nonlinear interactions~\citep*{rob91,hamkimwal95,jimpin99}. Riblets can break this near-wall cycle and push the streamwise vortices and streaks away from the wall so that a laminar region is retained within the grooves{. This} ultimately reduces skin-friction drag. The typical wavelength of the flow structures in the near-wall cycle are reported as $(\lambda_x^+, \lambda_z^+) \approx (1000,100)$, which corresponds to {$(k_x,\theta) \approx (1.1687,11.687)$} in a turbulent channel flow with $Re_\tau=186$.

\begin{figure}
        \begin{center}
        \begin{tabular}{cccc}
        \hspace{-.6cm}
        \subfigure[]{\label{fig.d90omz160NWcyclexz}}
        &&
        \subfigure[]{\label{fig.d90omz160NWcycleyz}}
        &
        \\[-.5cm]\hspace{-.3cm}
	\begin{tabular}{c}
        \normalsize{\rotatebox{90}{$z^+$}}
       \end{tabular}
       &\hspace{-.3cm}
	\begin{tabular}{c}
       \includegraphics[width=0.4\textwidth]{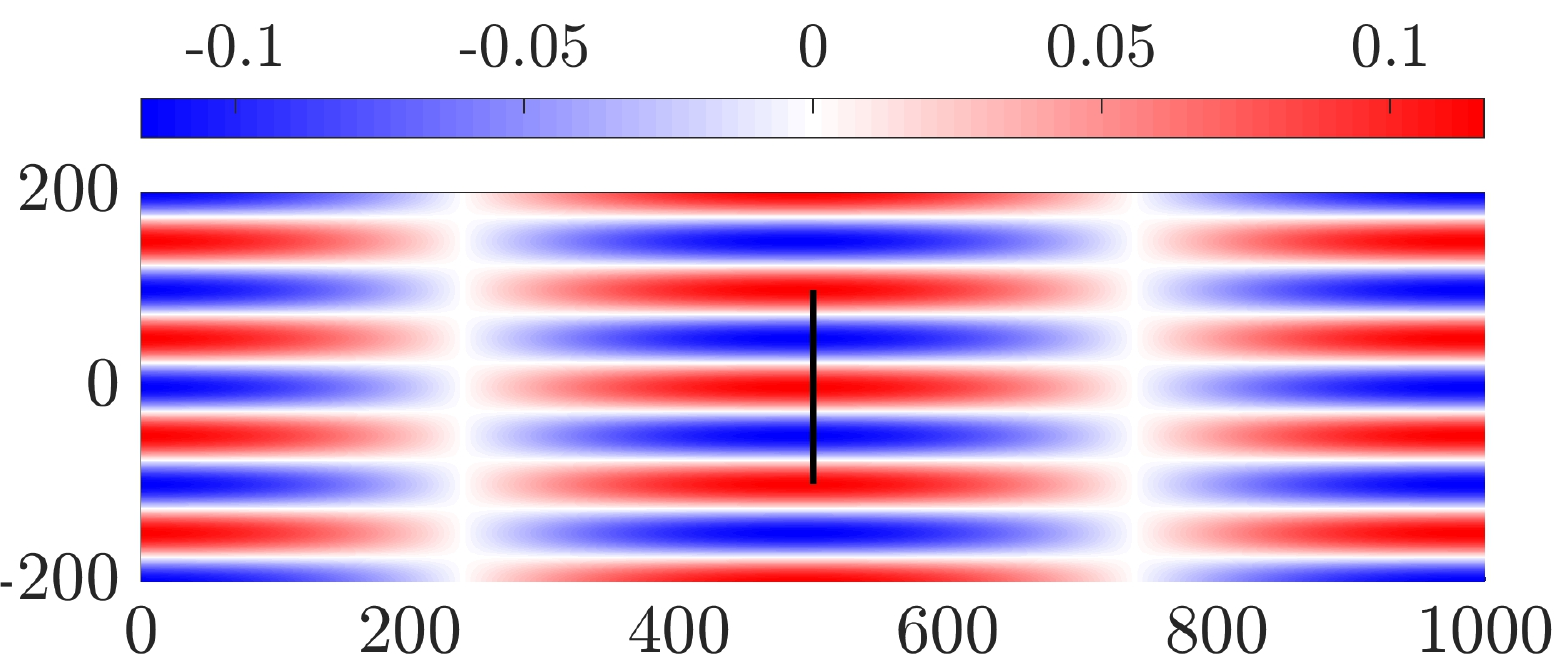}
       \end{tabular}
       &\hspace{.2cm}
       \begin{tabular}{c}
        \normalsize{\rotatebox{90}{$y^+$}}
       \end{tabular}
       &\hspace{-.3cm}
    \begin{tabular}{c}
       \includegraphics[width=0.4\textwidth]{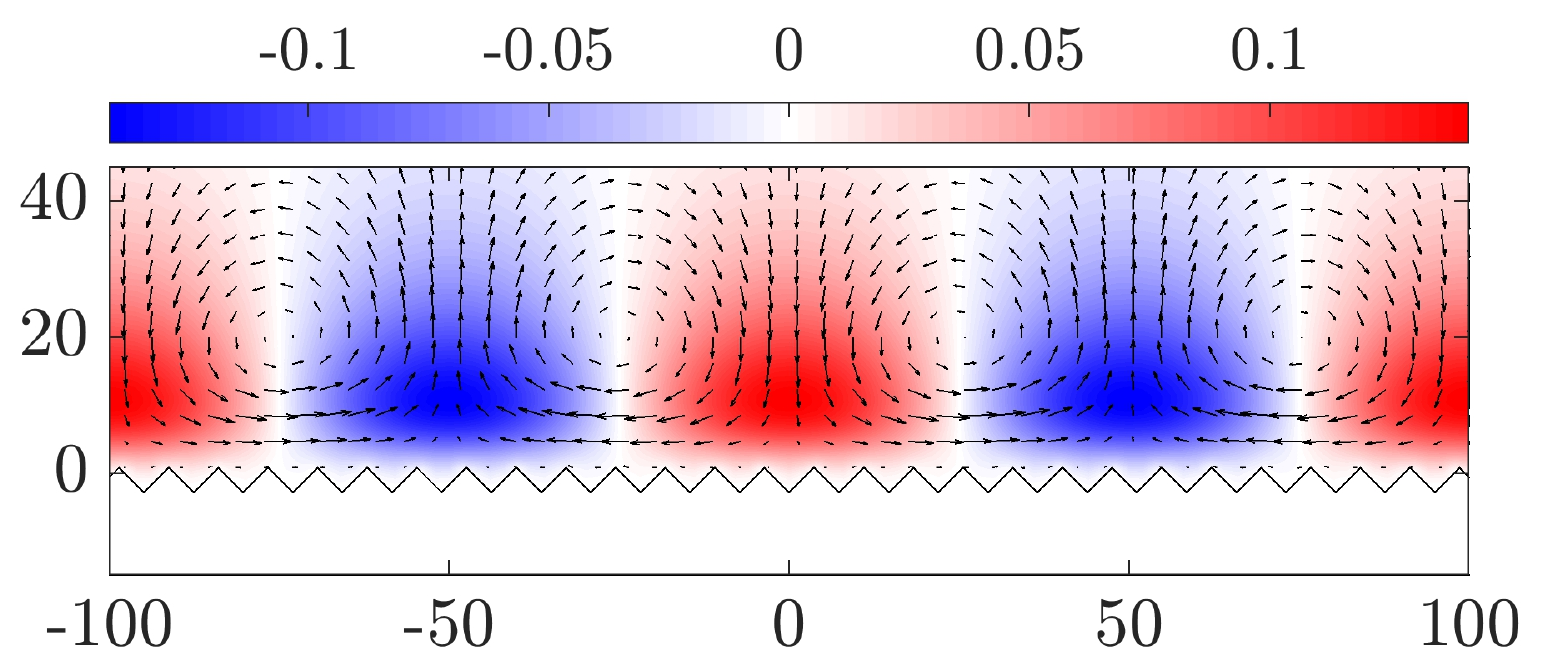}
       \end{tabular}
       \\[-0.1cm]
       \hspace{-.6cm}
        \subfigure[]{\label{fig.d90omz50NWcyclexz}}
        &&
        \subfigure[]{\label{fig.d90omz50NWcycleyz}}
        &
        \\[-.5cm]\hspace{-.3cm}
	\begin{tabular}{c}
        \normalsize{\rotatebox{90}{$z^+$}}
       \end{tabular}
       &\hspace{-.3cm}
	\begin{tabular}{c}
       \includegraphics[width=0.4\textwidth]{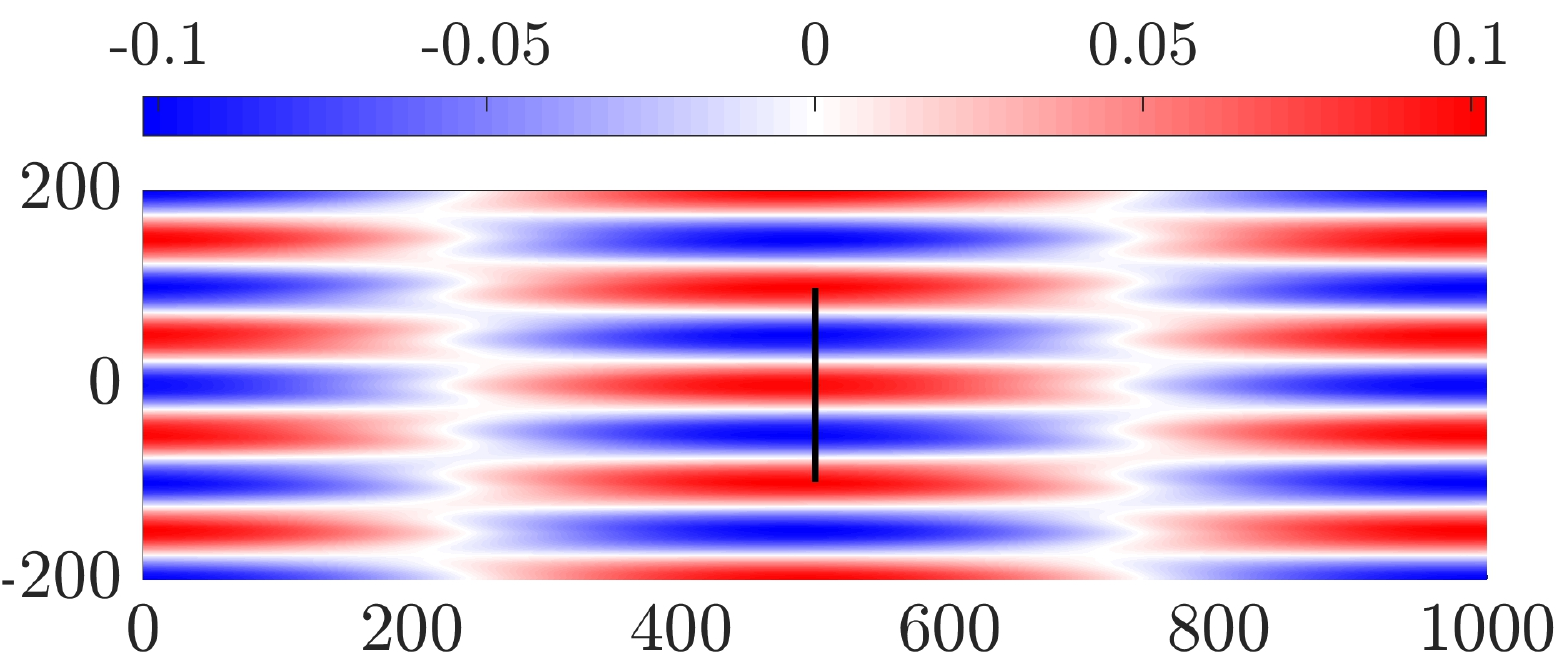}
       \end{tabular}
       &\hspace{.2cm}
       \begin{tabular}{c}
        \normalsize{\rotatebox{90}{$y^+$}}
       \end{tabular}
       &\hspace{-.3cm}
    \begin{tabular}{c}
       \includegraphics[width=0.4\textwidth]{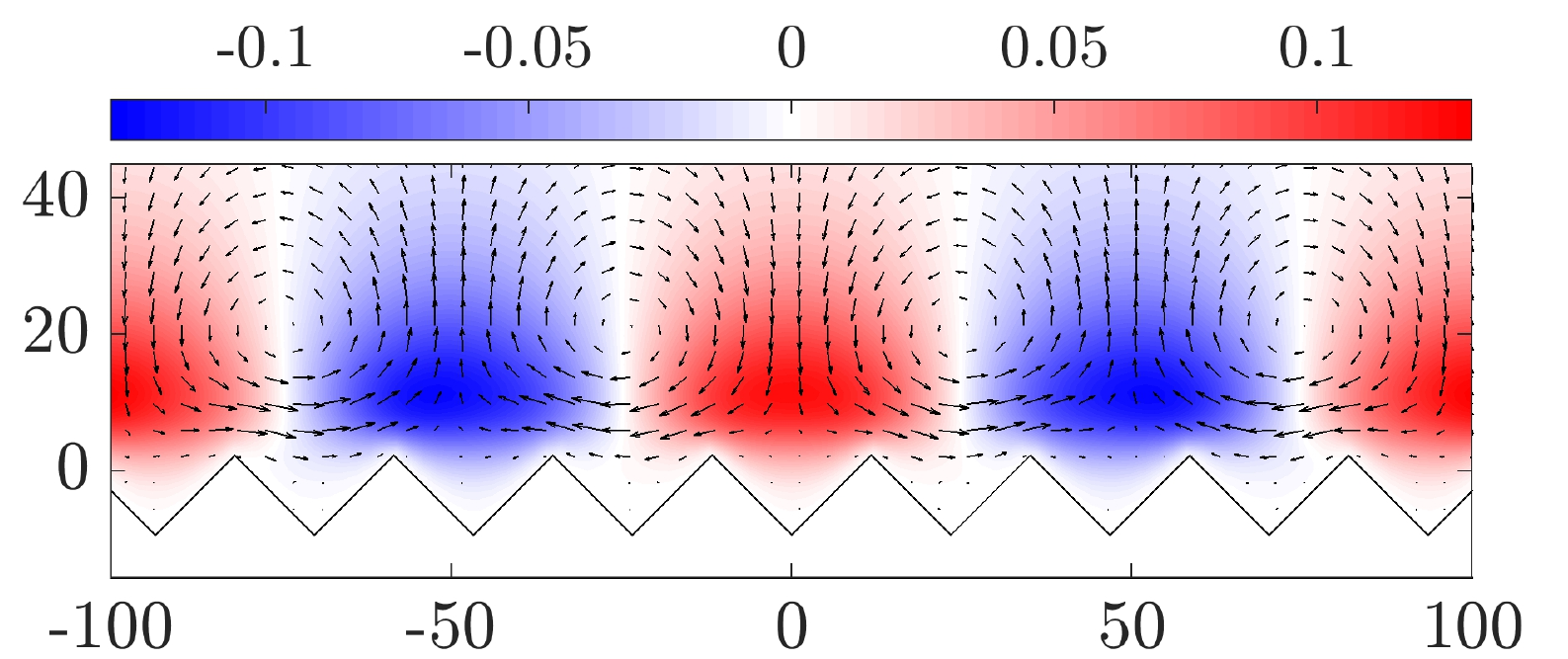}
       \end{tabular}
       \\[-0.1cm]
       \hspace{-.6cm}
        \subfigure[]{\label{fig.d90omz30NWcyclexz}}
        &&
        \subfigure[]{\label{fig.d90omz30NWcycleyz}}
        &
        \\[-.5cm]\hspace{-.3cm}
	\begin{tabular}{c}
        \vspace{.3cm}
        \normalsize{\rotatebox{90}{$z^+$}}
       \end{tabular}
       &\hspace{-.3cm}
	\begin{tabular}{c}
       \includegraphics[width=0.4\textwidth]{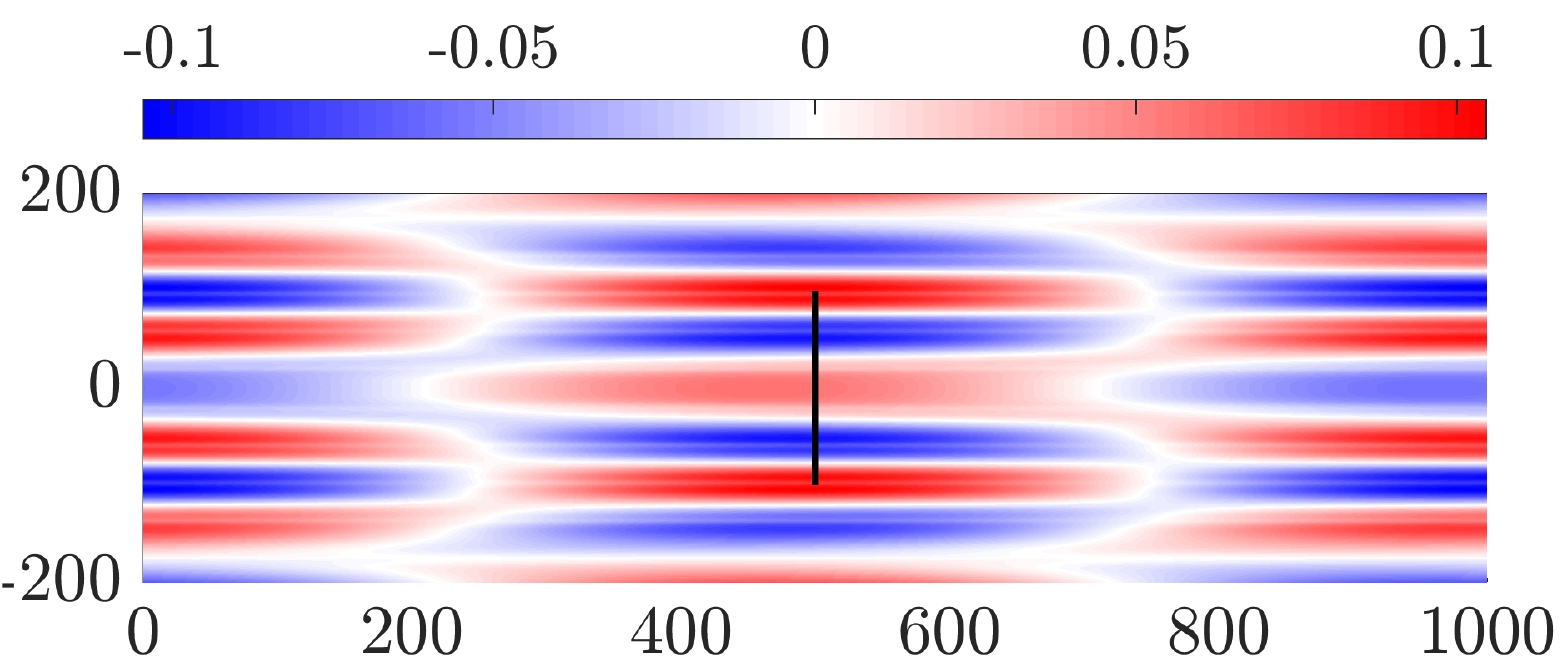}
       \\ $x^+$
       \end{tabular}
       &\hspace{.2cm}
       \begin{tabular}{c}
        \vspace{.3cm}
        \normalsize{\rotatebox{90}{$y^+$}}
       \end{tabular}
       &\hspace{-.3cm}
    \begin{tabular}{c}
       \includegraphics[width=0.4\textwidth]{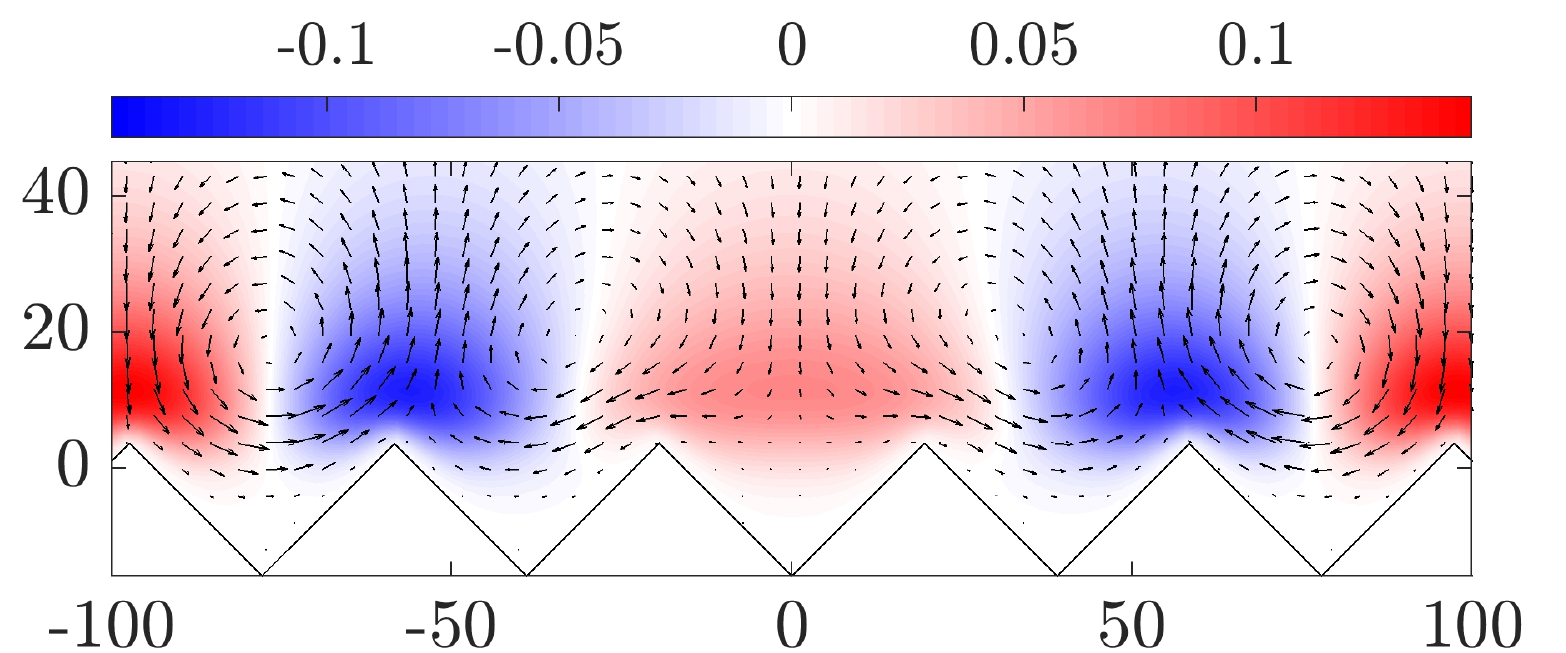}
       \\ $z^+$
       \end{tabular}
       \end{tabular}
       \end{center}
        \caption{Dominant flow structures in the vicinity of the lower wall of a turbulent channel flow with $Re_\tau=186$ over triangular riblets with $\alpha=90\degree$ and {(a)(b)} $\omega_z = 160$; {(c)(d)} $\omega_z = 50$; and {(e)(f)} $\omega_z = 30$. These flow structures {correspond to $(\lambda_x^+, \lambda_z^+) = (1000,100)$, typical scales of the near-wall cycle, and are extracted from the dominant eigenmode pair of the covariance matrix $\Phi_\theta(k_x)$.} Left column: $x-z$ slice of the streamwise velocity $u$ at $y^+ \approx 6$; Right column: $y-z$ slice of $u$ along with the vector field $(v, w)$ at $x^+ = 500$, which corresponds to the thick black lines in the left column. Color plots are used for the streamwise velocity fluctuation $u$ and vector fields identify streamwise vortices.}
        \label{fig.NWcycle}
\end{figure}

For different sizes of riblets, figure~\ref{fig.NWcycle} compares the flow structures that correspond to the near-wall cycle in a turbulent channel flow with $Re_\tau=186$. Flow patterns resulting from the combination of streaks and vortices can be clearly observed. In particular, it is evident that the quasi-streamwise vortices and regions of high and low streamwise velocity are pushed above the riblet tips creating a region of limited turbulence in the riblet grooves, and effectively impeding the transfer of mean momentum toward the lower wall. {The first two rows of figure~\ref{fig.NWcycle}} illustrate the flow structures over small- and optimal-sized riblets, respectively. The dominance of streamwise elongated structures that follow the length-scales of the near-wall cycle is evident in these two scenarios. {However,} as shown in {figures~\ref{fig.d90omz30NWcyclexz} and~\ref{fig.d90omz30NWcycleyz}}, the flow over larger riblets is contaminated by multiple energetically relevant spanwise length-scales. This indicates energy distribution across multiple Fourier modes beyond the ones that are relevant in the near-wall cycle. This distribution of energy is caused by the interaction of near-wall turbulence with the spanwise-periodic surface, which leads to the generation of secondary flow structures that follow the spatial frequency of riblets close to their tips~\citep{goltua98}; see {figure~\ref{fig.d90omz30NWcycleyz}}. 

The cross-plane views in figure~\ref{fig.NWcycle} also show that, for larger riblets, secondary flow structures begin to penetrate into the grooves. This induces high-momentum flow into the viscous flow regime, which is reflected by an increase in the amplitude of velocity fluctuations at $y^+ \approx 6$ and the breakdown stage of the viscous regime which precedes the deterioration of drag reduction. Moreover, the kinetic energy corresponding to the length-scale of the near-wall cycle varies from $0.0498$, for a channel flow with smooth walls, to $0.0468$, $0.0441$, and $0.0609$ (given by $\bar{E}(k_x,\theta)$ defined in~\eqref{eq.Ebar}), for small, optimal, and large riblets, respectively. This shows that  {small and optimally sized} riblets suppress the energy of near-wall cycle flow structures while larger ones can promote their energy, which is consistent with the trend observed in~\S~\ref{sec.energyspectra}.

\subsection{Spanwise rollers resembling Kelvin-Helmholtz vortices}
\label{sec.KHinstability}

\begin{figure}
        \begin{tabular}{ccccc}
        \hspace{-.4cm}
        \begin{tabular}{c}
		\vspace{.4cm}
                \rotatebox{90}{\normalsize $y^+$}
	    \end{tabular}
	    &
        \hspace{-.5cm}
        \begin{tabular}{c}
                DNS\\
                \includegraphics[width=0.242\textwidth]{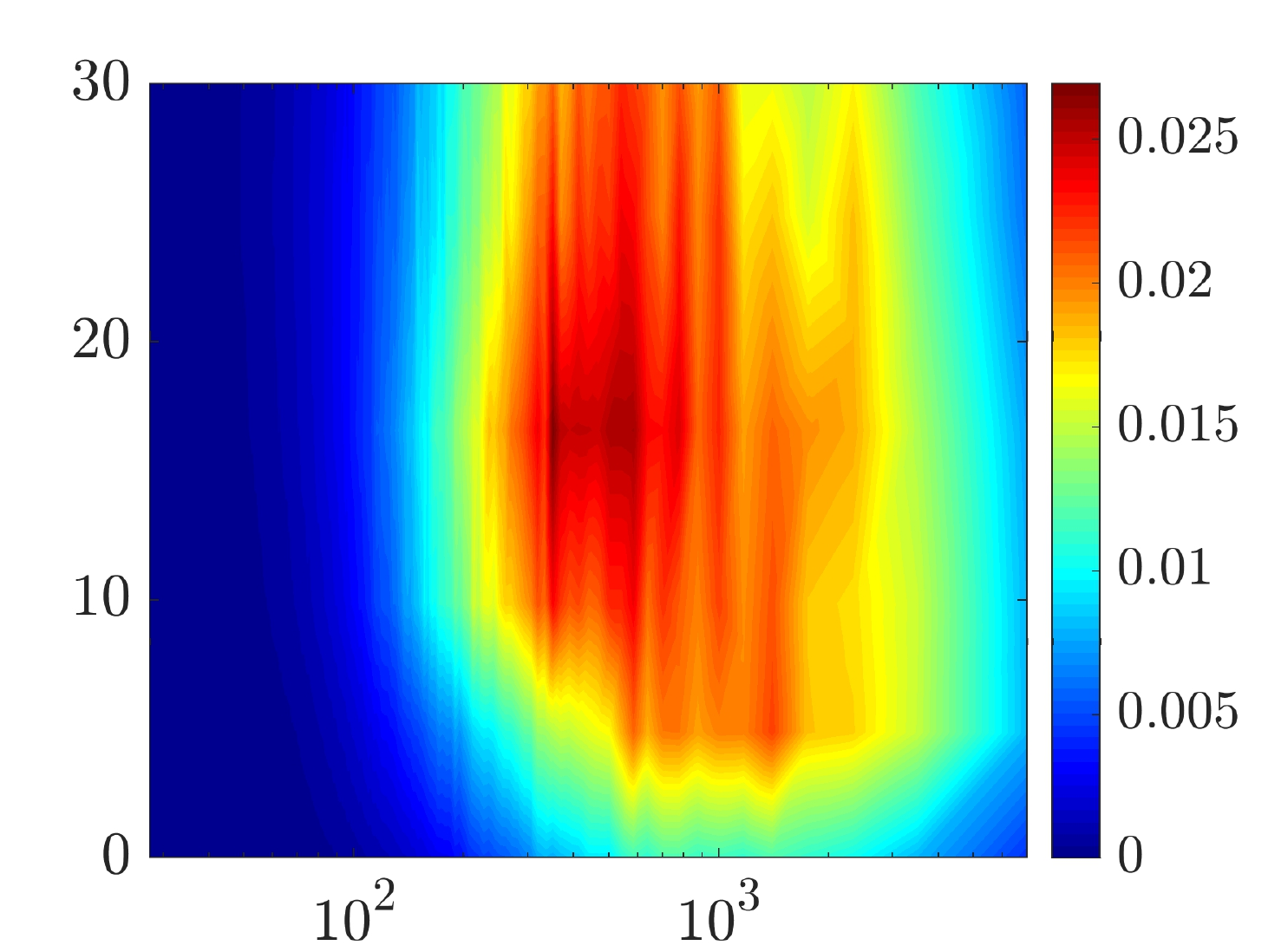}
        \end{tabular}
        &
        \hspace{-.5cm}
        \begin{tabular}{c}
                $\omega_z=160$\\
                \includegraphics[width=0.24\textwidth]{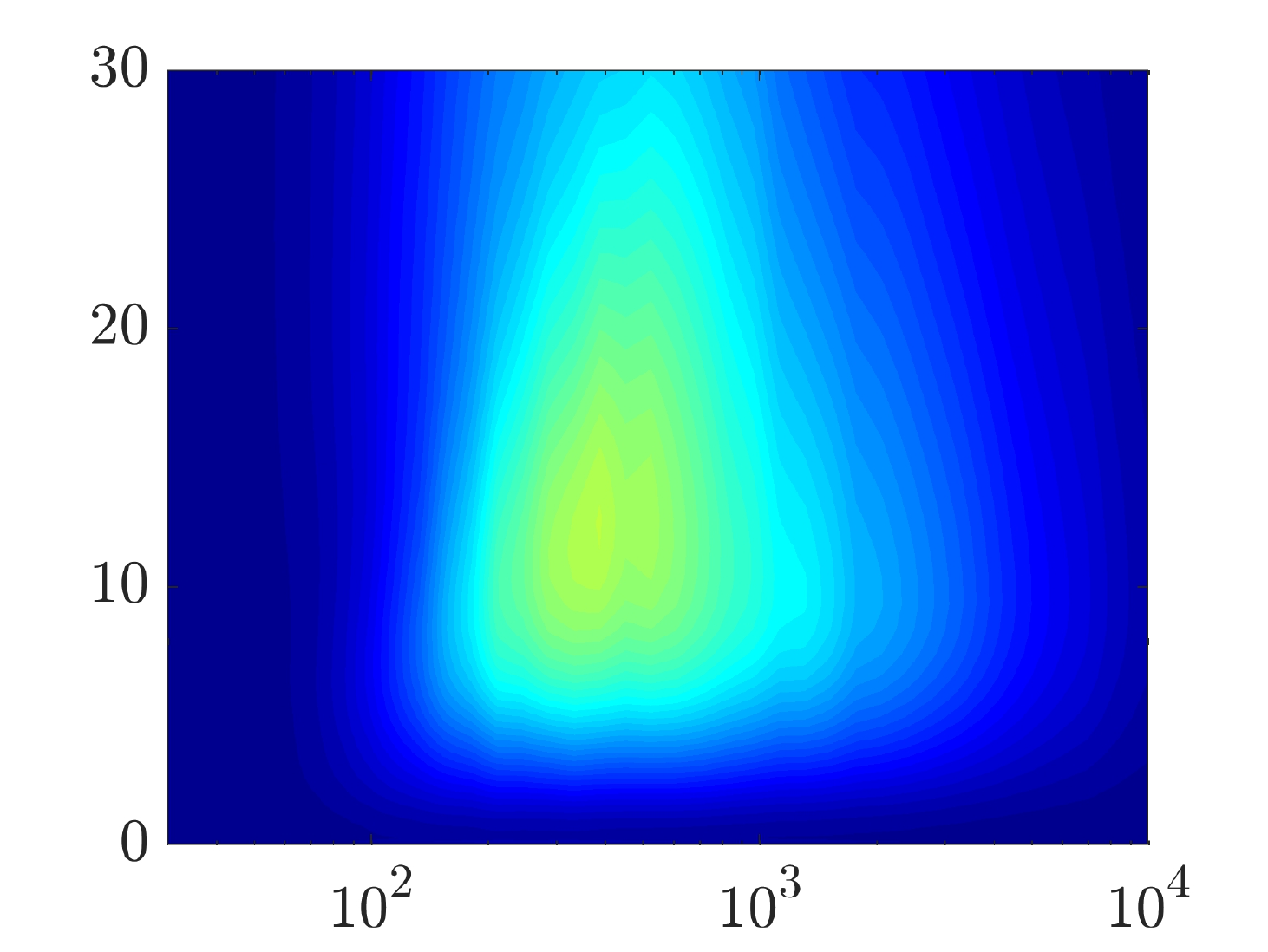}
        \end{tabular}
        &\hspace{-.5cm}
        \begin{tabular}{c}
                $\omega_z=50$\\
                \includegraphics[width=0.24\textwidth]{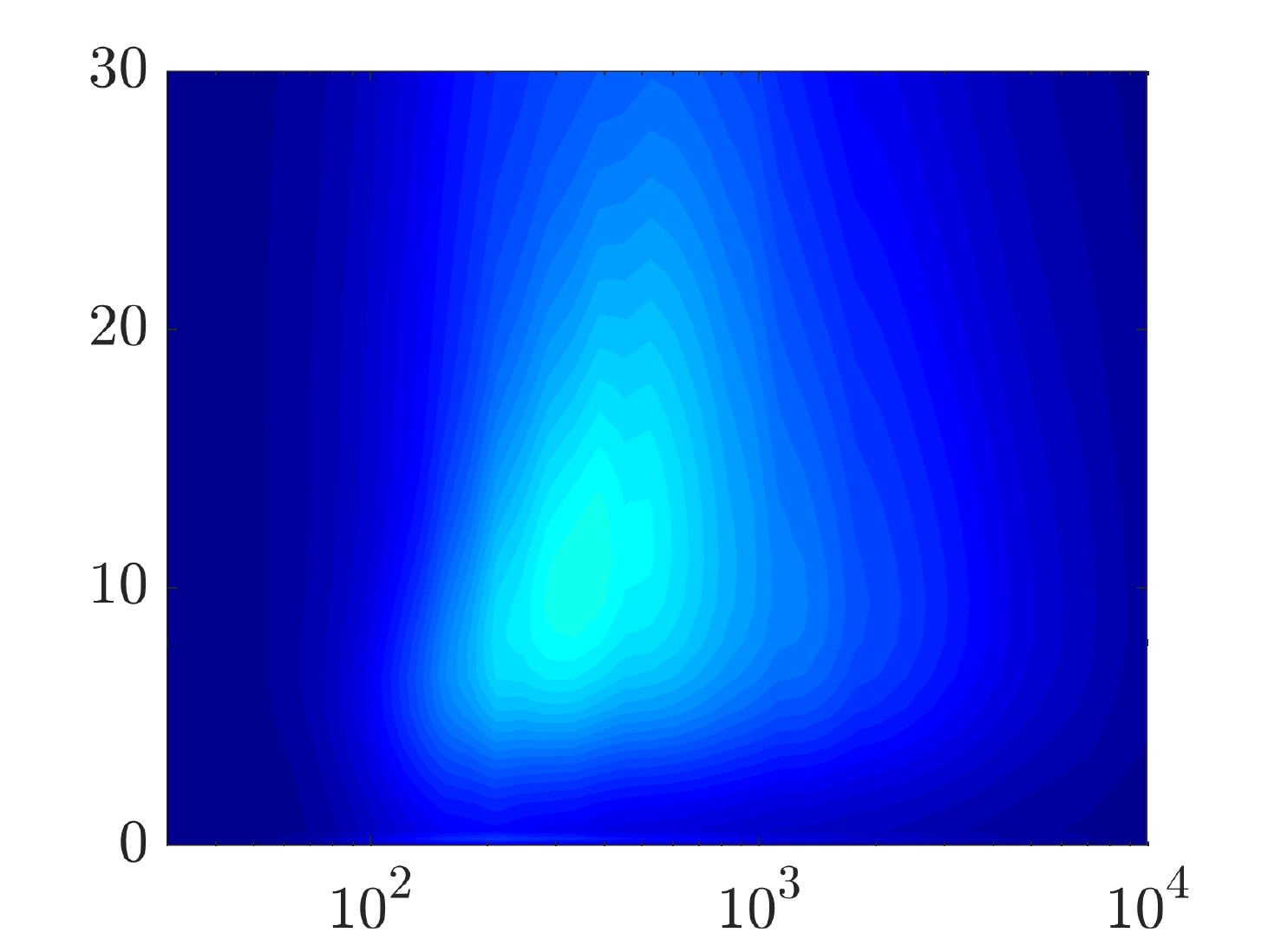}
        \end{tabular}
        &\hspace{-.5cm}
        \begin{tabular}{c}
                $\omega_z=30$\\
                \includegraphics[width=0.242\textwidth]{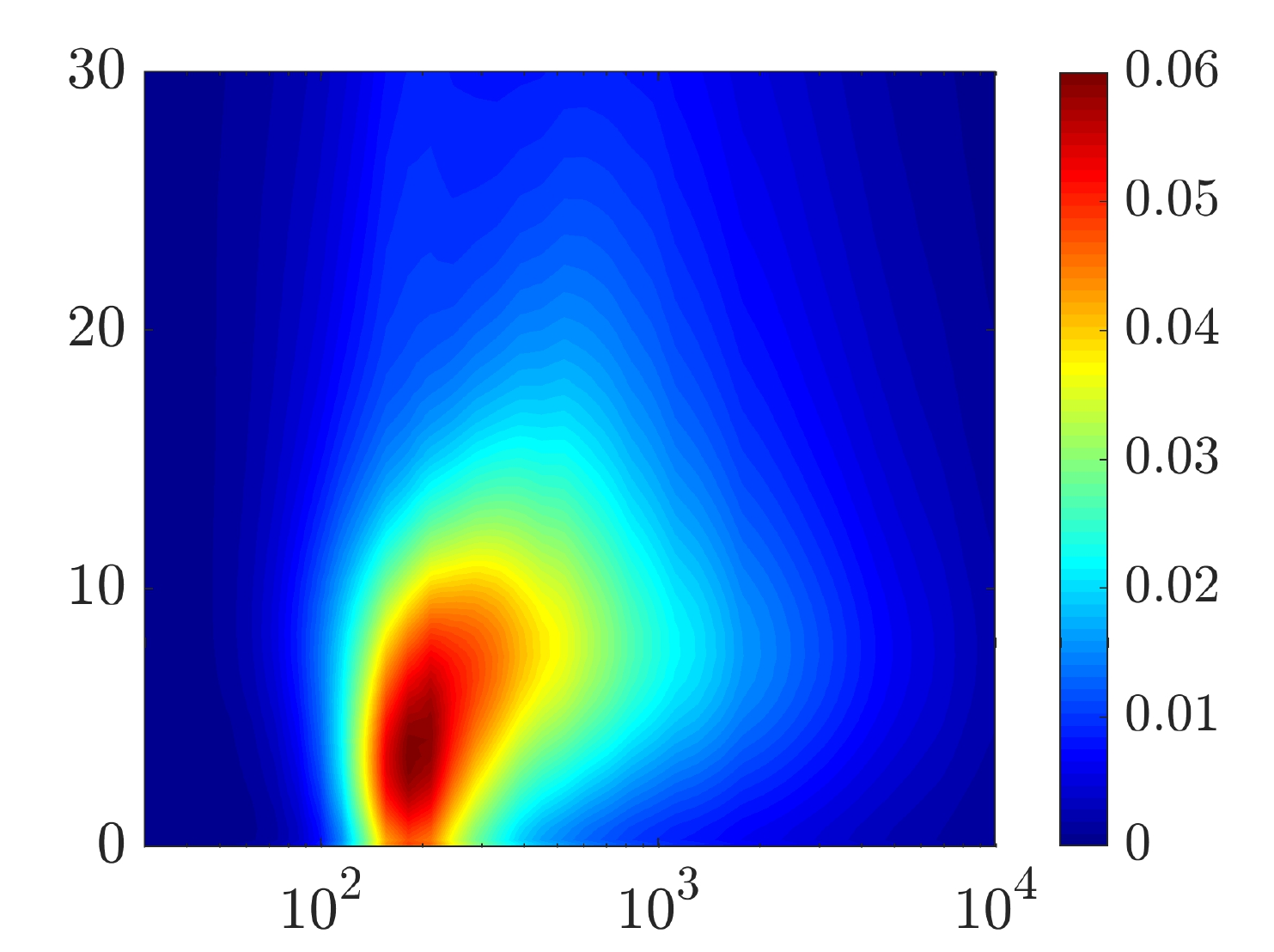}
        \end{tabular}
        \\[-.1cm]
        \hspace{-.4cm}
        \begin{tabular}{c}
		\vspace{.4cm}
                \rotatebox{90}{\normalsize $y^+$}
	    \end{tabular}
	    &
        \hspace{-.5cm}
        \begin{tabular}{c}
                \includegraphics[width=0.242\textwidth]{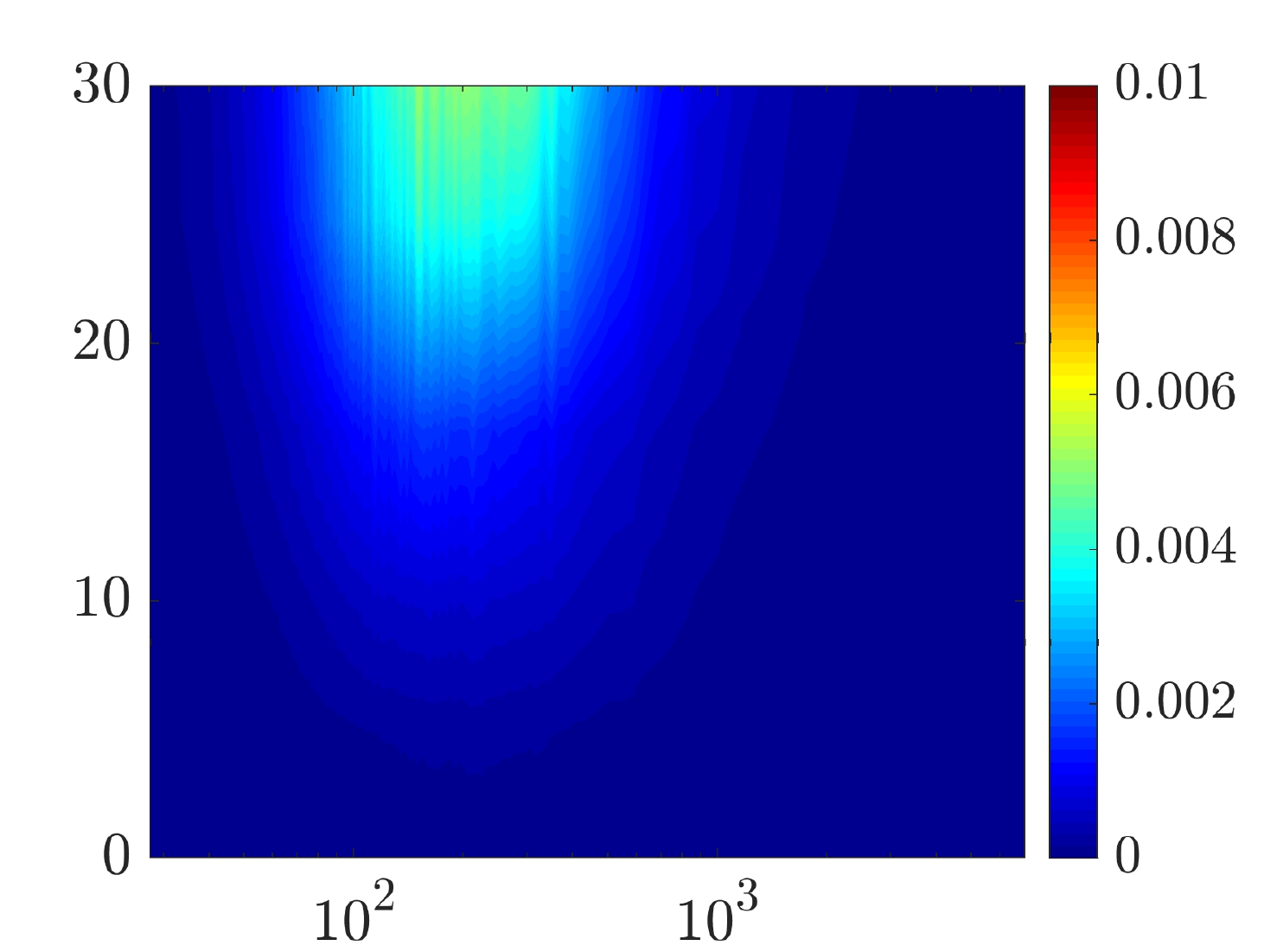}
        \end{tabular}
        &
        \hspace{-.5cm}
        \begin{tabular}{c}
                \includegraphics[width=0.24\textwidth]{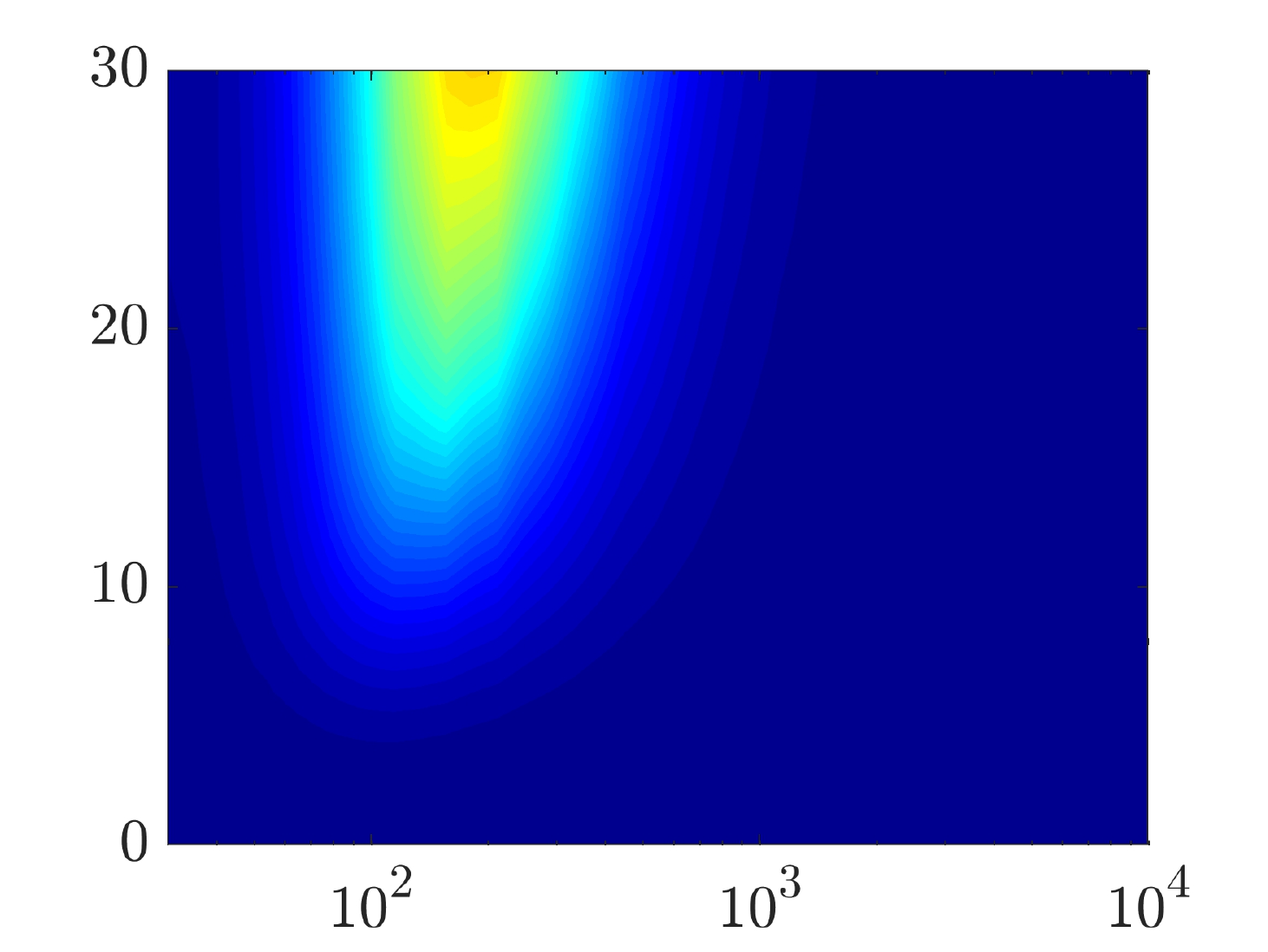}
        \end{tabular}
        &\hspace{-.5cm}
        \begin{tabular}{c}
                \includegraphics[width=0.24\textwidth]{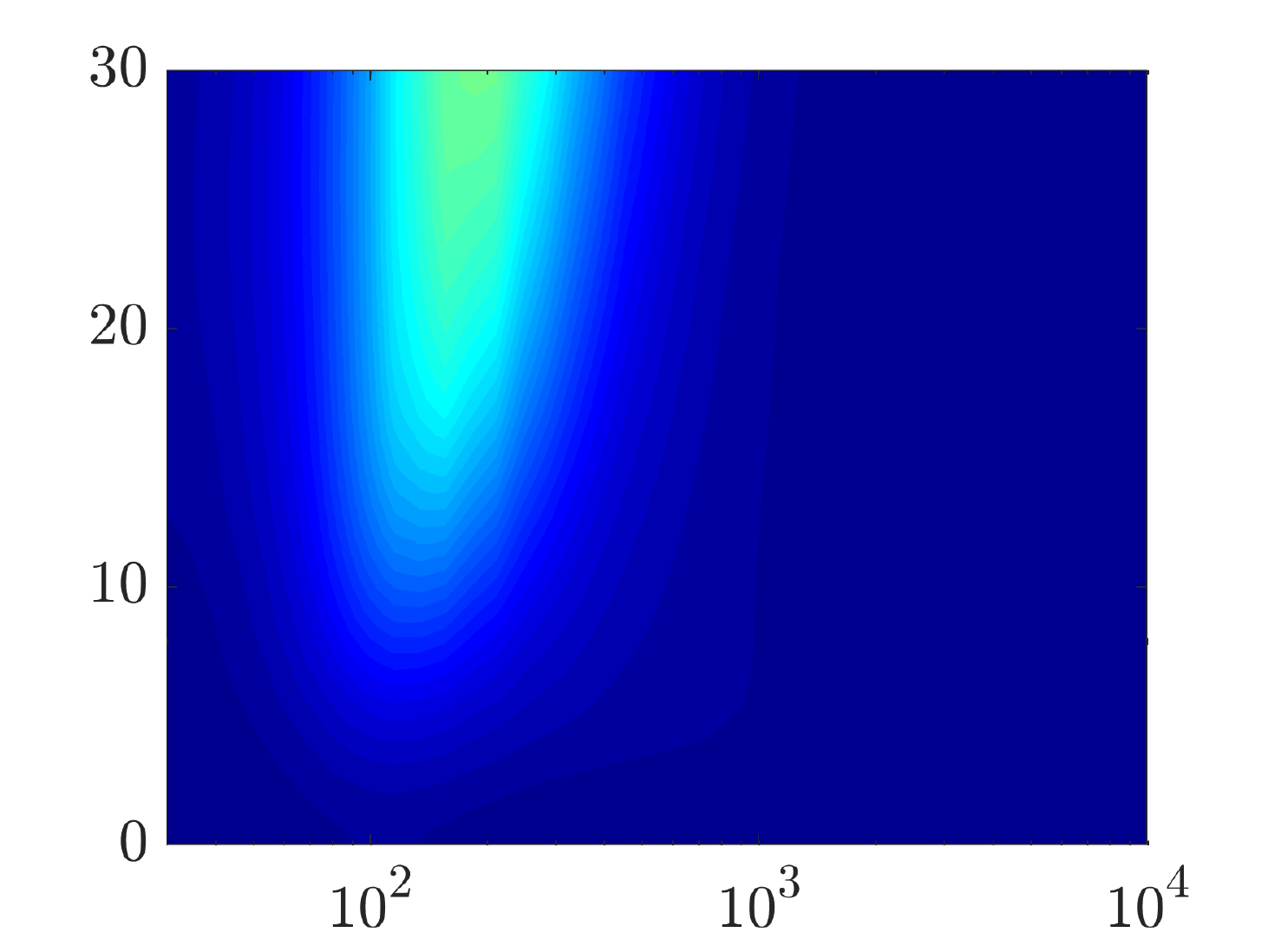}
        \end{tabular}
        &\hspace{-.5cm}
        \begin{tabular}{c}
                \includegraphics[width=0.242\textwidth]{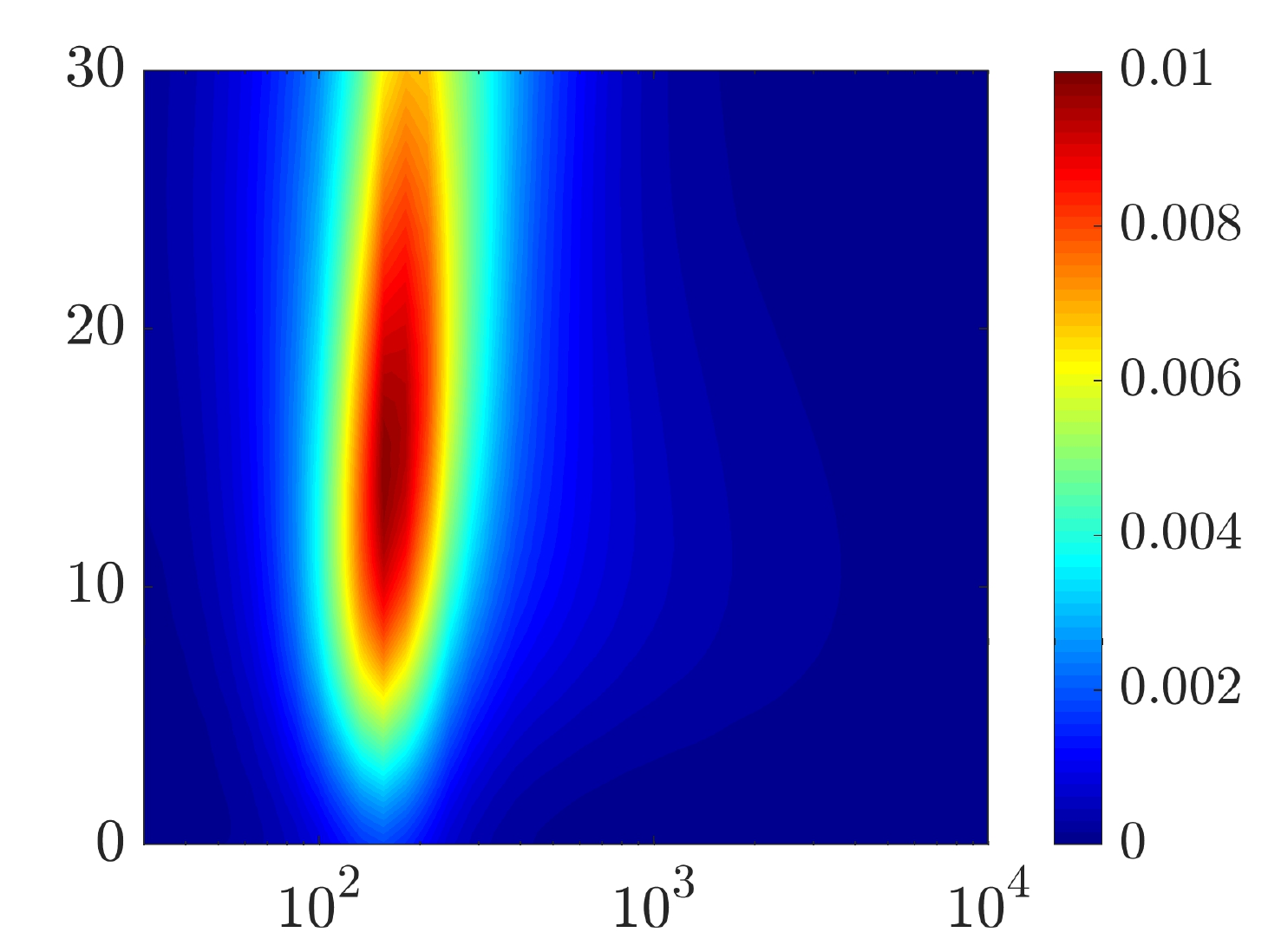}
        \end{tabular}
        \\[-.1cm]
        \hspace{-.4cm}
        \begin{tabular}{c}
		\vspace{.4cm}
                \rotatebox{90}{\normalsize $y^+$}
	    \end{tabular}
	    &
        \hspace{-.5cm}
        \begin{tabular}{c}
                \includegraphics[width=0.242\textwidth]{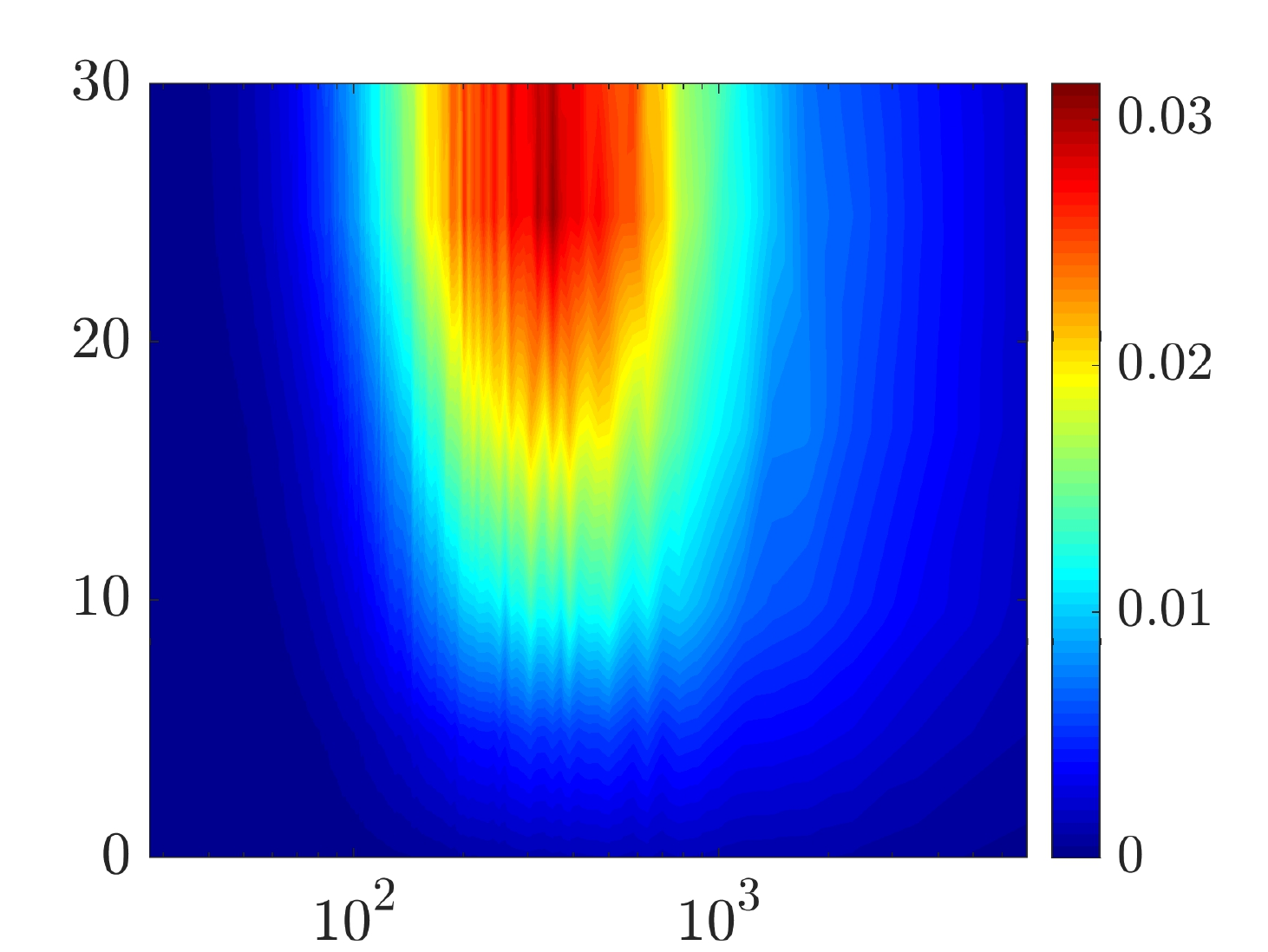}
        \end{tabular}
        &
        \hspace{-.5cm}
        \begin{tabular}{c}
                \includegraphics[width=0.24\textwidth]{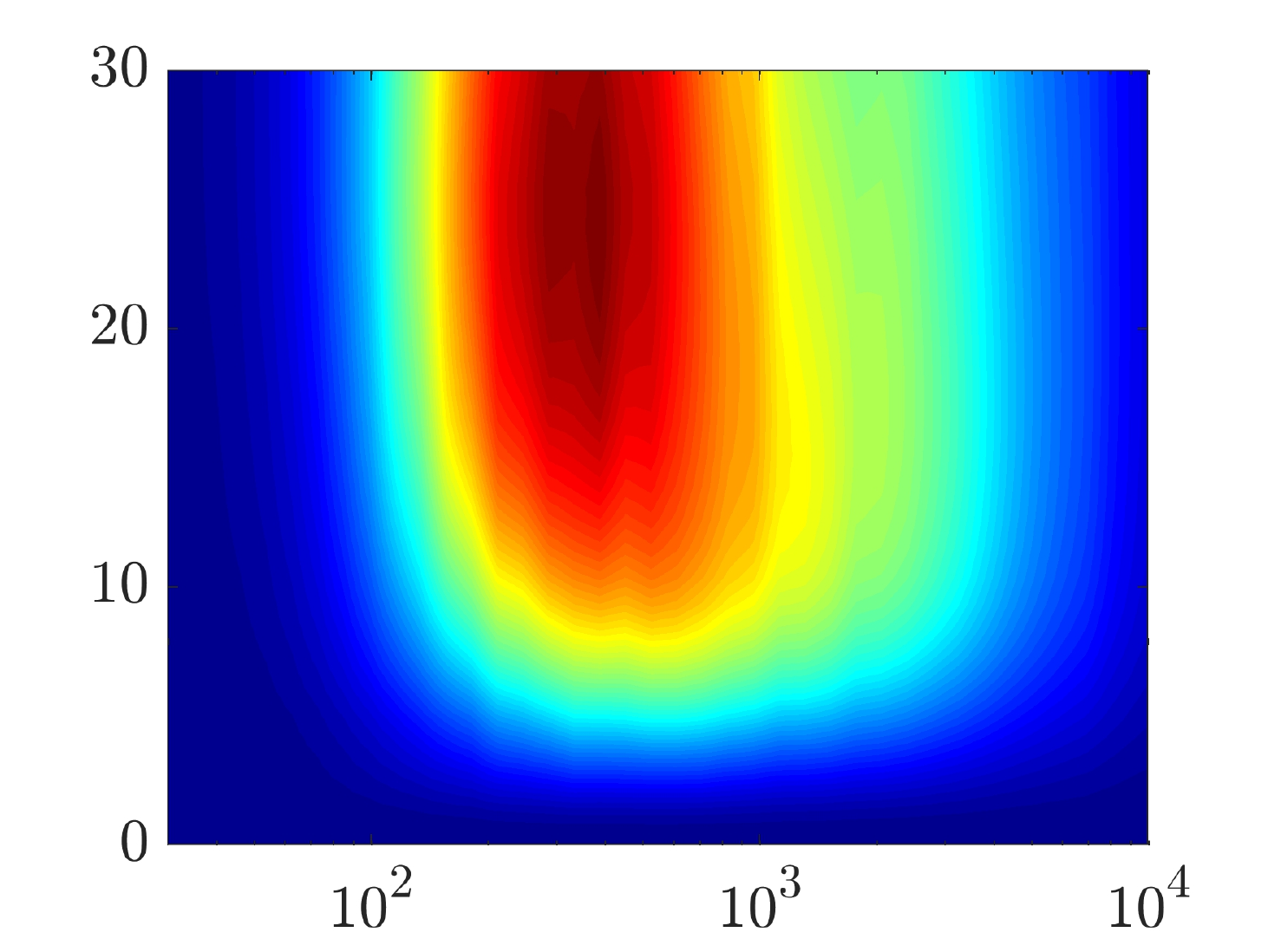}
        \end{tabular}
        &\hspace{-.5cm}
        \begin{tabular}{c}
                \includegraphics[width=0.24\textwidth]{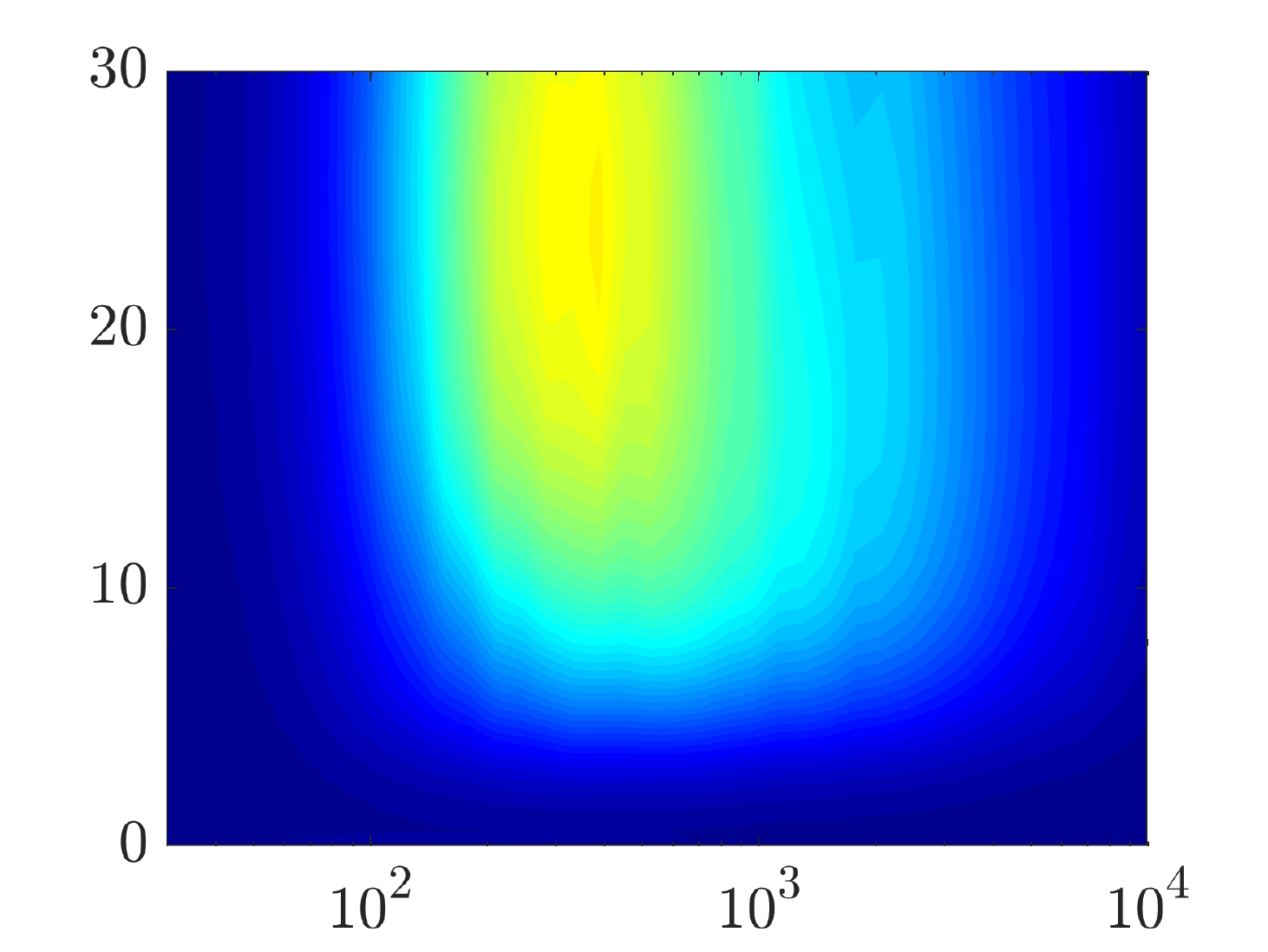}
        \end{tabular}
        &\hspace{-.5cm}
        \begin{tabular}{c}
                \includegraphics[width=0.242\textwidth]{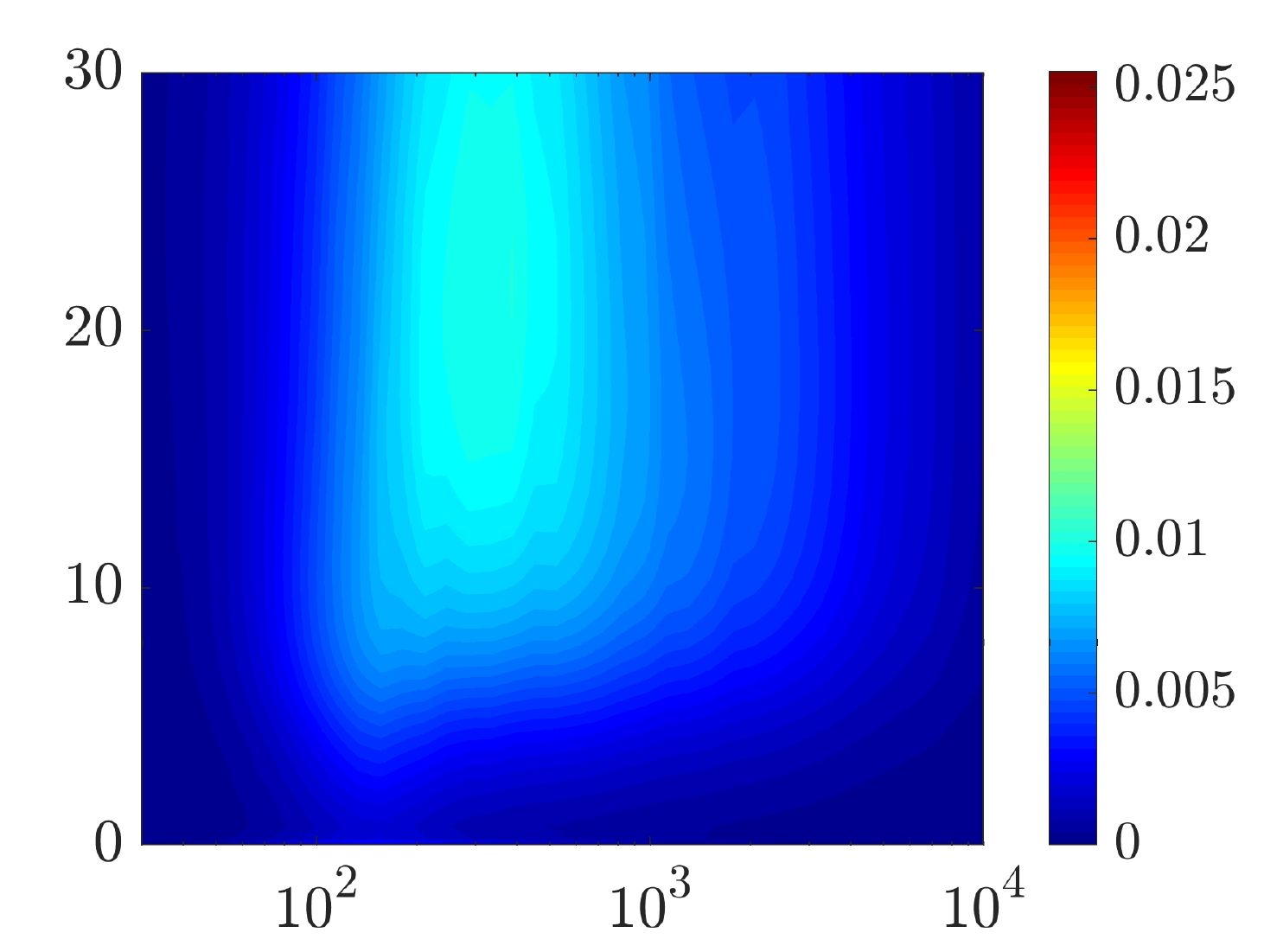}
        \end{tabular}
        \\[-.1cm]
        \hspace{-.4cm}
        \begin{tabular}{c}
		\vspace{.4cm}
                \rotatebox{90}{\normalsize $y^+$}
	    \end{tabular}
	    &
        \hspace{-.5cm}
        \begin{tabular}{c}
                \includegraphics[width=0.242\textwidth]{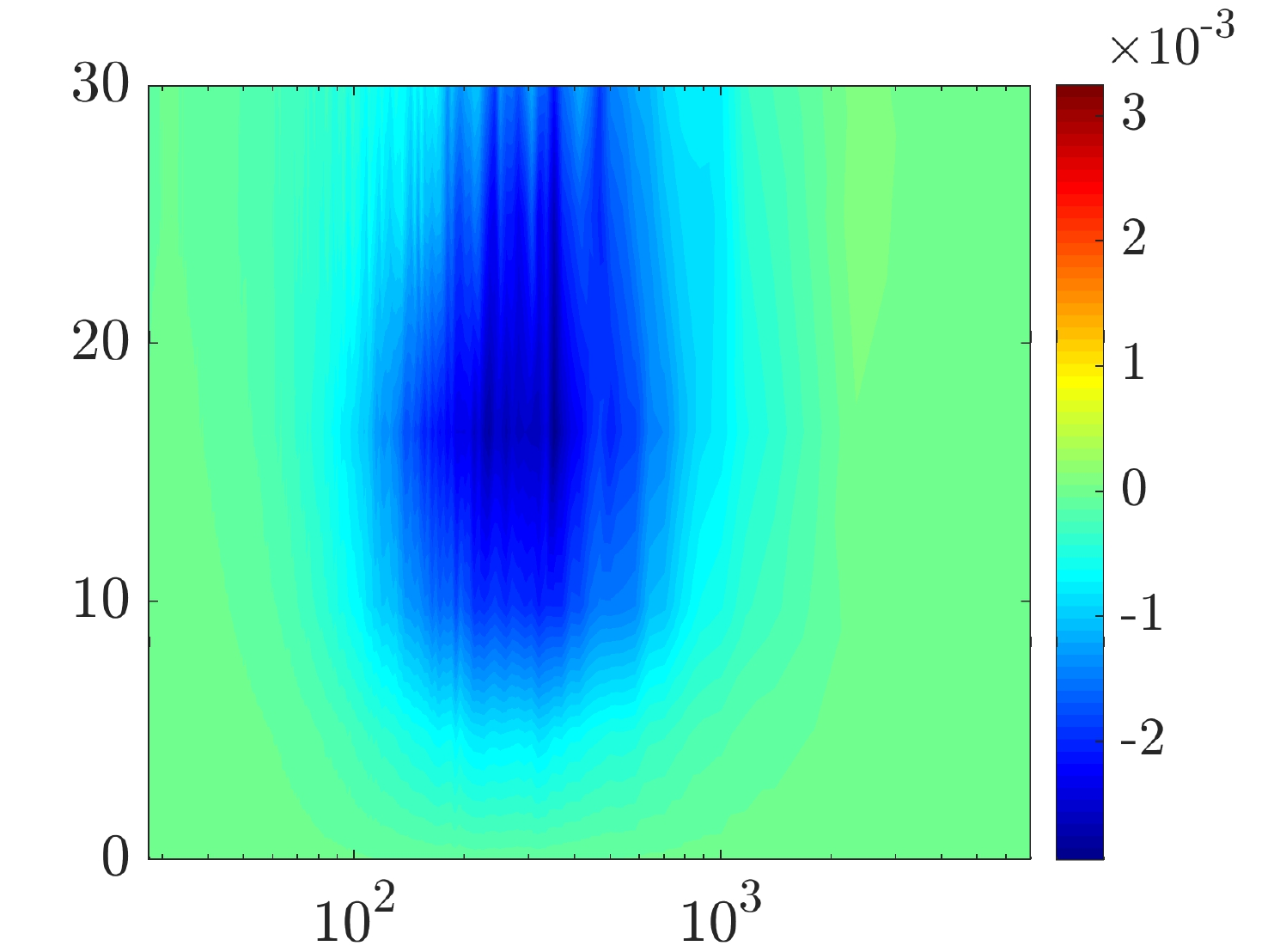}
        \end{tabular}
        &
        \hspace{-.5cm}
        \begin{tabular}{c}
                \includegraphics[width=0.24\textwidth]{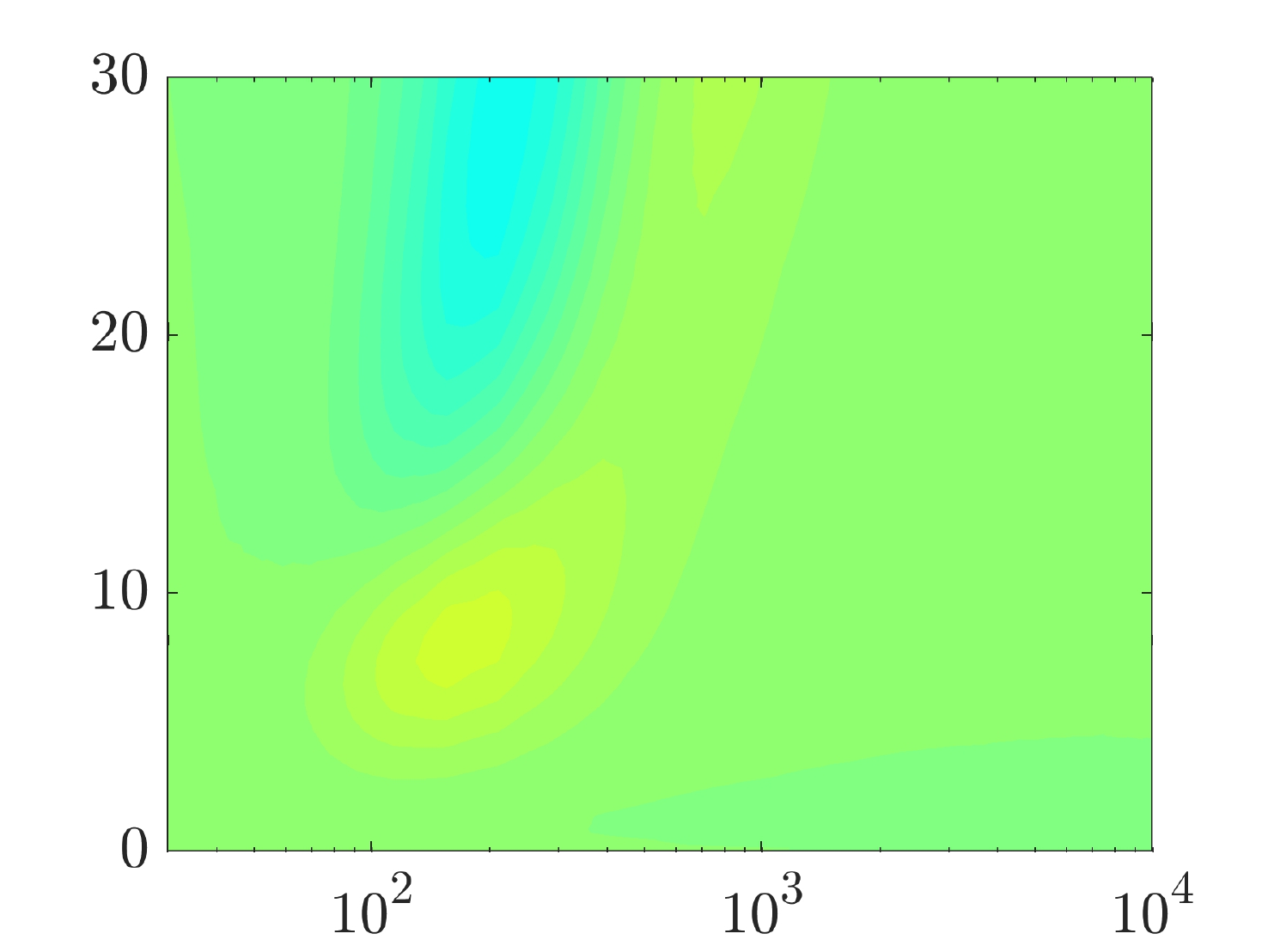}
        \end{tabular}
        &\hspace{-.5cm}
        \begin{tabular}{c}
                \includegraphics[width=0.24\textwidth]{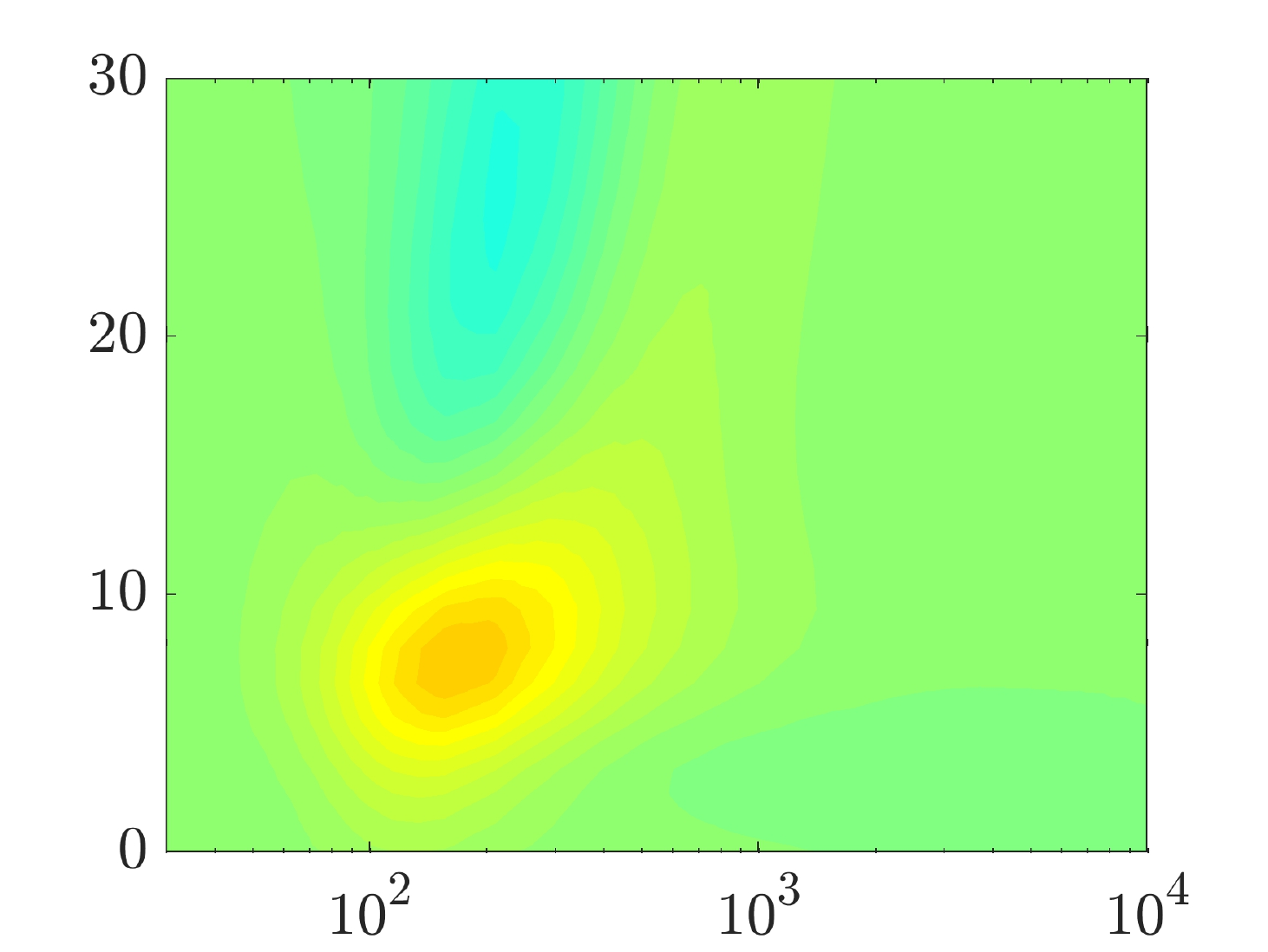}
        \end{tabular}
        &\hspace{-.5cm}
        \begin{tabular}{c}
                \includegraphics[width=0.242\textwidth]{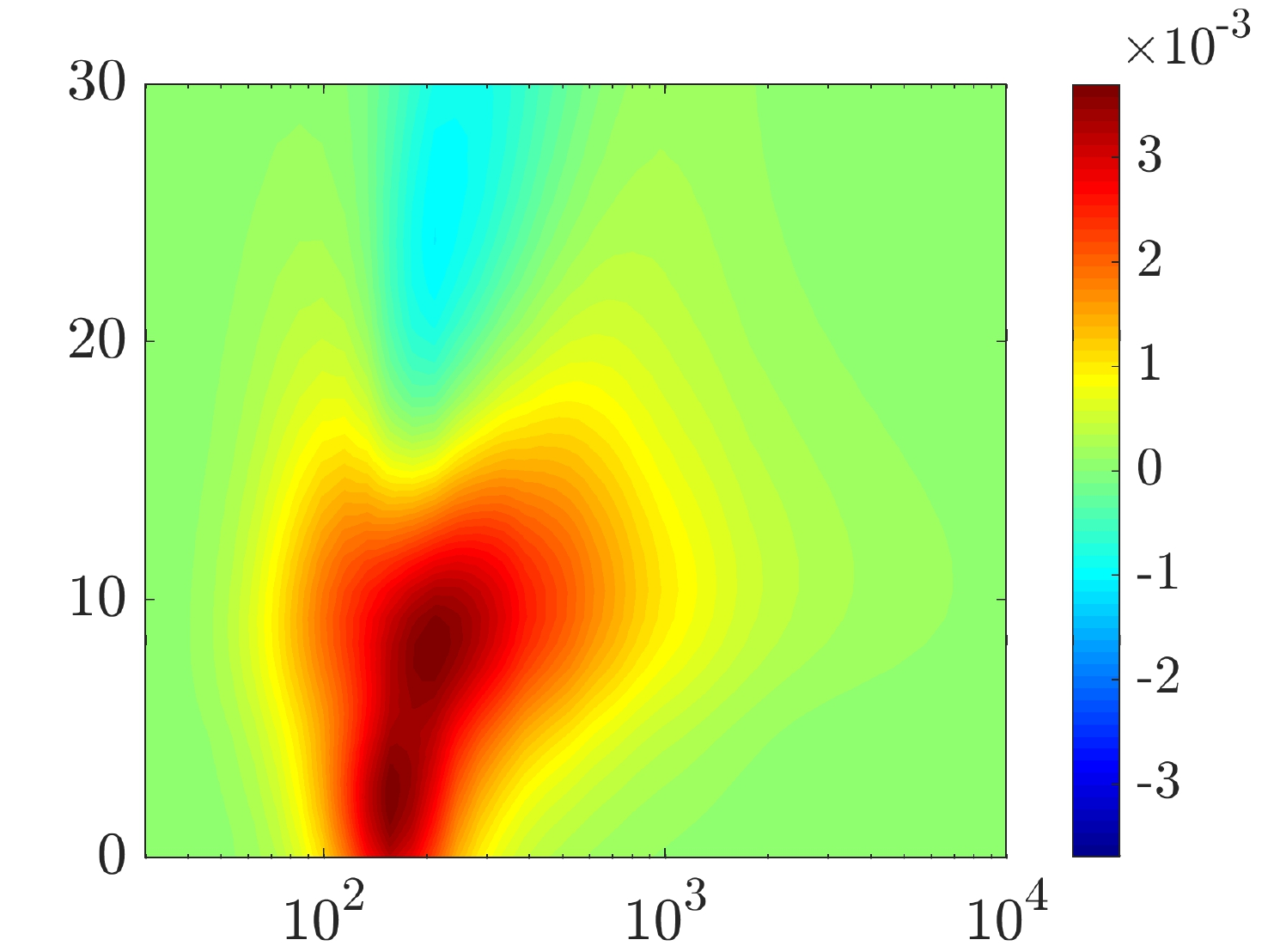}
        \end{tabular}
        \\[-.1cm]
        &
	\hspace{-.5cm}
         {\normalsize $\lambda^+_x$}
        &
        \hspace{-.5cm}
        {\normalsize $\lambda^+_x$}
        &
        \hspace{-.5cm}
        {\normalsize $\lambda^+_x$}
        &
        \hspace{-.5cm}
        {\normalsize $\lambda^+_x$}
        \end{tabular}
        \caption{Premultiplied streamwise co-spectra at $\theta=0$ for a turbulent channel flow with $Re_\tau=186$ over triangular riblets of tip angle $\alpha=90\degree$. The four rows correspond to {$k_xE_{uu}$}, {$k_xE_{vv}$}, {$k_xE_{ww}$}, and {$-k_xE_{uv}$}; the four columns represent DNS data for the uncontrolled channel flow~\citep{deljim03,deljimzanmos04} and data resulting from our computations for the flow with riblets of spatial frequency $\omega_z=160$, $50$, and $30$, respectively.}
        \label{fig.cospectrakz0}
\end{figure}

In addition to the influence of secondary flow structures around the tip of riblets, the breakdown of the viscous regime and decrease in drag reduction can also result from the amplification of spanwise rollers that are induced by a two-dimensional \mbox{K-H} instability~\citep{garjim11b}. The amplification of long spanwise scales for large riblets is evident from the energy spectra shown in figure~\ref{fig.spectrums}. Figure~\ref{fig.cospectrakz0} shows the DNS-based premultiplied streamwise co-spectra of various Reynolds stresses for infinitely wide scales ($\theta = 0$) in a turbulent channel flow with $Re_\tau=186$~\citep{deljim03,deljimzanmos04}, as well as the corresponding co-spectra resulting from our analysis of the flow over triangular riblets of different size. For larger riblets, the amplification of the co-spectra corresponding to streamwise and wall-normal intensities become stronger and occur closer to the wall. However, this trend is not observed for the co-spectrum corresponding to the spanwise turbulence intensity, which shows smaller amplification of channel-wide scales for riblets of larger size. Nevertheless, as the size of riblets increases, the co-spectrum corresponding to the wall-shear stress, {$-k_xE_{uv}$}, starts to show signs of suppression in the vicinity of the wall; the penetration of negative shear stress into the riblet grooves is evident from the co-spectrum associated with spatial frequency $\omega_z=30$. The co-spectrum {$-k_xE_{uv}$} shows the largest suppression of shear stress for streamwise wavelengths $\lambda_x^+\approx 200$. Our results demonstrate that large riblets result in the suppression of shear stress within the grooves, which is consistent with our earlier findings that showed a degradation of drag reduction for such sizes (cf.\ figure~\ref{fig.DRml}). We note that our computations illustrate that the trends observed for the energy spectra do not vary for different shapes of riblets (i.e., different values of $\alpha$).

\begin{figure}
        \begin{tabular}{cccccc}
        \hspace{-.4cm}
        \subfigure[]{\label{fig.90domz160yp5spectrum}}
        &&
        \hspace{-.65cm}
        \subfigure[]{\label{fig.90domz50yp5spectrum}}
        &&
        \hspace{-.65cm}
        \subfigure[]{\label{fig.90domz30yp5spectrum}}
        &
        \\[-.2cm]
        \hspace{-.4cm}
        \begin{tabular}{c}
		\vspace{.4cm}
                \rotatebox{90}{\normalsize $\lambda^+_z$}
	    \end{tabular}
        &
        \hspace{-.5cm}
        \begin{tabular}{c}
                \includegraphics[width=0.315\textwidth]{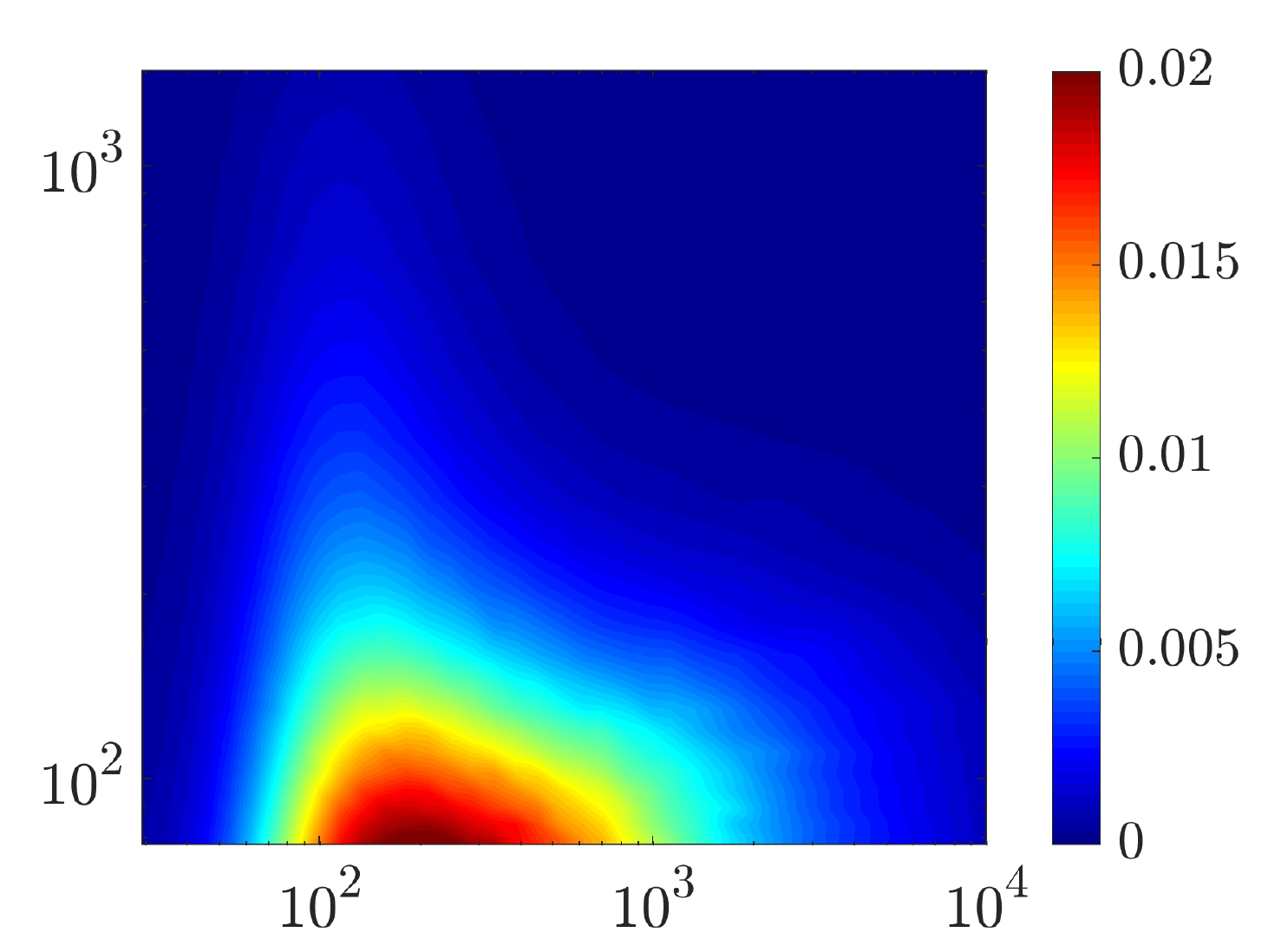}
        \end{tabular}
        &&
        \hspace{-.5cm}
        \begin{tabular}{c}
                \includegraphics[width=0.315\textwidth]{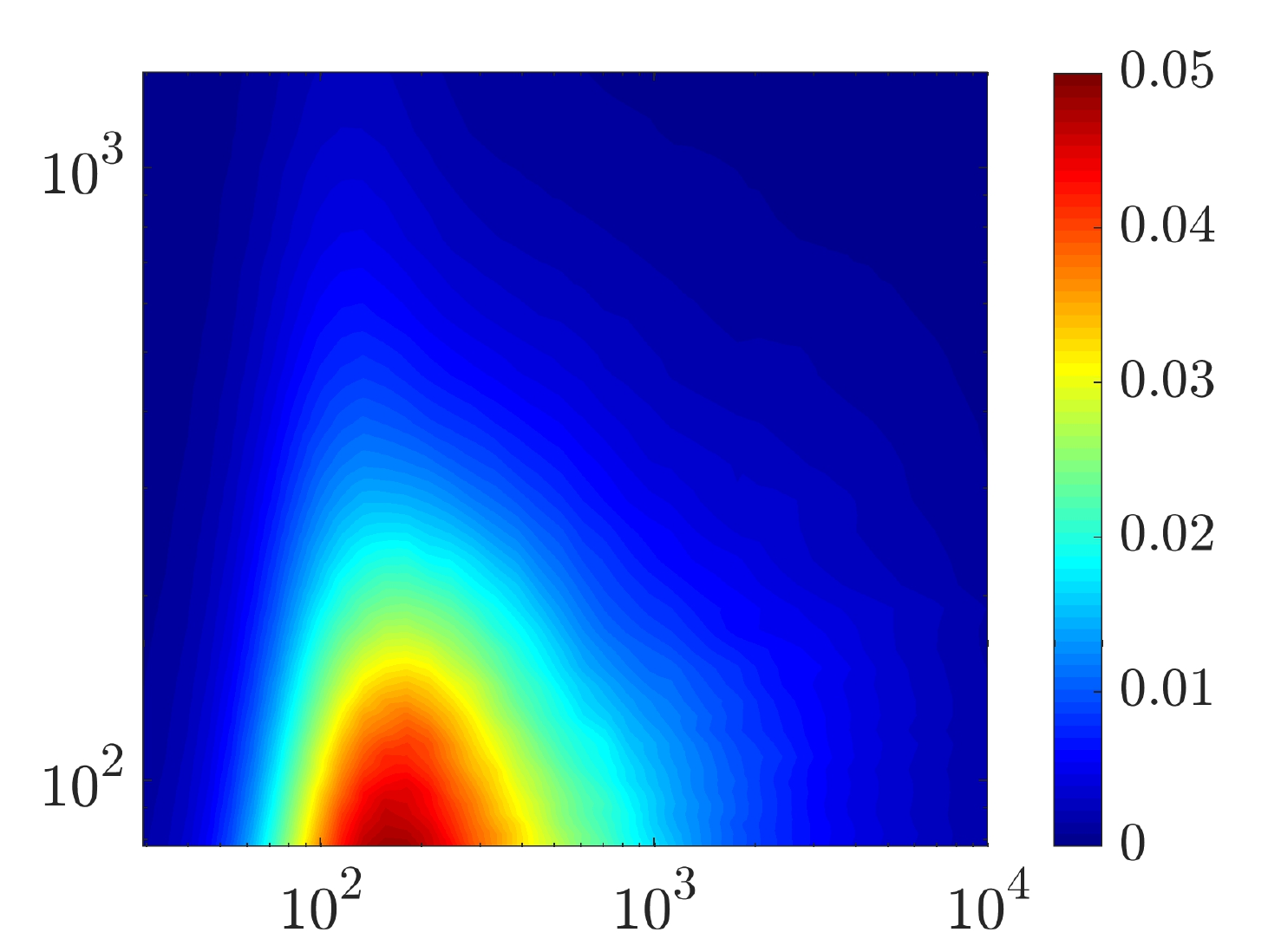}
        \end{tabular}
        &&\hspace{-.5cm}
        \begin{tabular}{c}
                \includegraphics[width=0.32\textwidth]{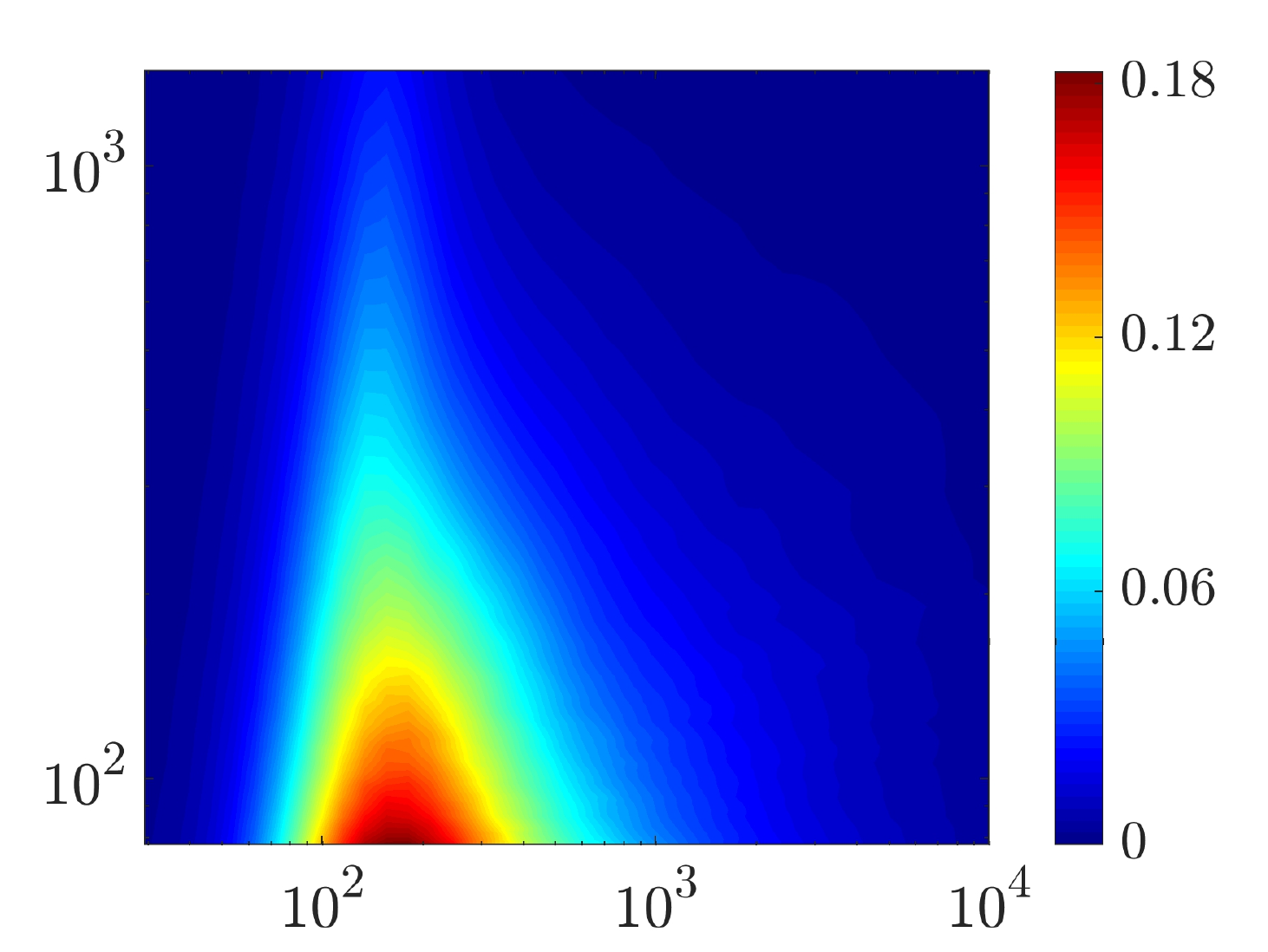}
        \end{tabular}
        \\[-.1cm]
        &\hspace{-.5cm}
         {\normalsize $\lambda^+_x$}
        &&
        \hspace{-.5cm}
        {\normalsize $\lambda^+_x$}
        &&
        \hspace{-.6cm}
        {\normalsize $\lambda^+_x$}
        \end{tabular}
        \caption{Premultiplied energy spectra of wall-normal velocity, $ {k_x k_z} E_{vv}$, at $y^+=5$ in turbulent channel flow with $Re_\tau=186$ over triangular riblets of tip angle $\alpha=90\degree$ and (a) $\omega_z=160$; (b) $\omega_z=50$; and (c) $\omega_z=30$.}
        \label{fig.yp5spectrum}
\end{figure}

For largest riblets ($\omega_z=30$), figure~\ref{fig.cospectrakz0} illustrates that the streamwise ($k_x E_{uu}$) and wall-normal ($k_x E_{vv}$) contributions to the energy spectra are significantly larger than the spanwise ($k_x E_{ww}$) contribution. Furthermore, the wall-normal spectrum shows significant amplification of wall-separated flow structures. Figure~\ref{fig.yp5spectrum} shows the premultiplied  {energy spectrum of wall-normal velocity}, $k_x k_z E_{vv}$, at $y^+=5$ for different  sizes of riblets. We note that energy amplification becomes larger as the size of riblets increases. For the riblets with $\omega_z=30$, figure~\ref{fig.90domz30yp5spectrum} shows that the band of streamwise and spanwise scales corresponding to $\lambda^+_x \in [100,300]$ and $\lambda_z^+>300$, is significantly more amplified than the other two cases, which is \mbox{consistent with the DNS-result of~\citet{garjim11b}.}

\begin{figure}
	\begin{centering}
	\begin{tabular}{lc}
	\subfigure[]{\label{fig.omz30Zrolleravg}}&
	 \\[-.3cm]
	        &
            \hspace{-0.5cm}
            \begin{tabular}{rc}
                \begin{tabular}{r}
                \vspace{.4cm}
                \rotatebox{90}{\normalsize $y^+$}
            \end{tabular}
            &\hspace{-0.4cm}
            \begin{tabular}{c}
            \includegraphics[width=.92\textwidth]{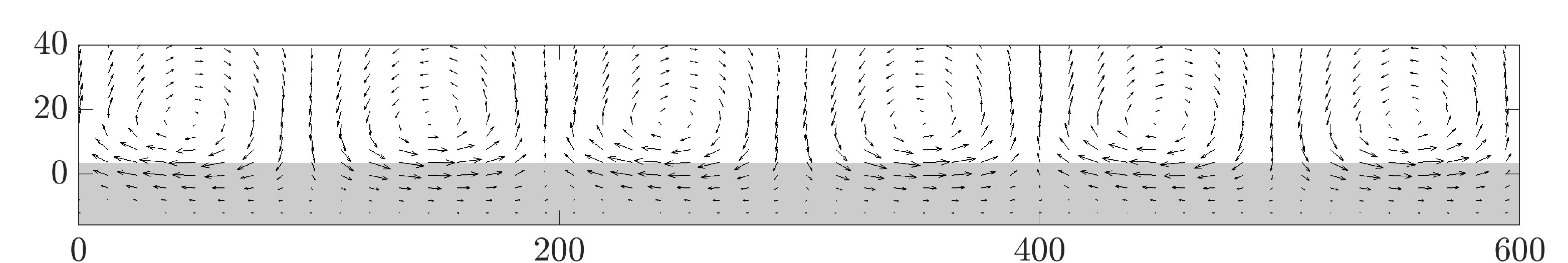}
            \\
            {\normalsize $x^+$}
            \end{tabular}
            \end{tabular}
	\\[-.2cm]
	\subfigure[]{\label{fig.omz30Zrollerxz}}&
	\\[-.6cm]
            &
            \hspace{-0.3cm}
            \begin{tabular}{rc}
                \begin{tabular}{r}
                \vspace{.4cm}
                \rotatebox{90}{\normalsize $z^+$}
            \end{tabular}
            &\hspace{-0.3cm}
            \begin{tabular}{c}
            \includegraphics[width=.8\textwidth]{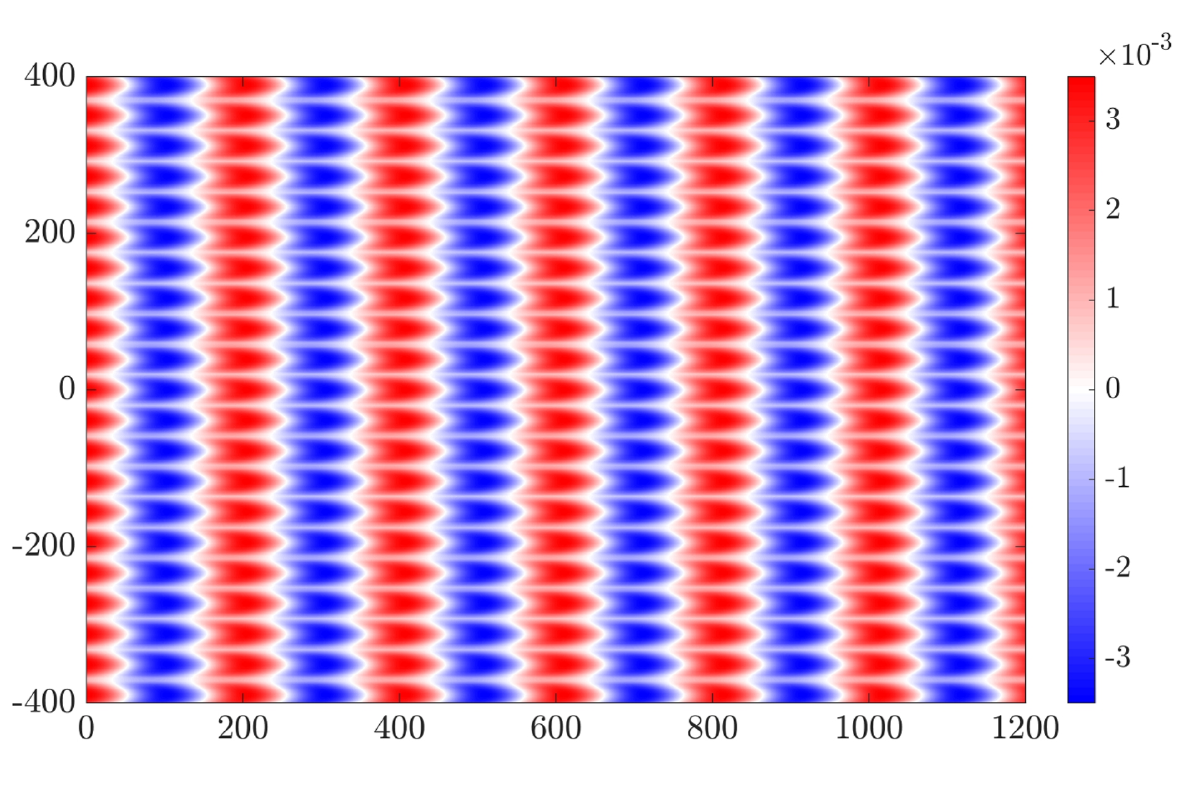}
            \\[-.15cm]
            {\normalsize $x^+$}
            \end{tabular}
            \end{tabular}
	\end{tabular}
	\caption{Turbulent flow structures corresponding to spanwise {elongated rollers with $\lambda_x^+\approx200$ that are extracted from the dominant eigenmode of the covariance matrix $\Phi_\theta(k_x)$} for a turbulent channel flow with $Re_\tau = 186$ over triangular riblets of $\alpha=90\degree$ and $\omega_z = 30$. (a) Vector field denotes the in-plane velocities $(u, v)$; and (b) $x-z$ slice of the wall-normal velocity $v$ at $y^+ = 5$.}
	\label{fig.spanwiseroller}
	\end{centering}
\end{figure}

Figure~\ref{fig.spanwiseroller} shows the flow structures that are extracted from our model for $(k_x,\theta)=(5.76,0)$ which correspond to spanwise-averaged infinitely long scales in $z$ and the peak in the shear stress co-spectrum for riblets with $\omega_z=30$. These flow structures indicate that the dominant eigenmode of the covariance matrix~\eqref{eq.output-covariance} resembles a spanwise-constant roller centered {at} $y^+ = 17.7$, which penetrates well into the grooves, thereby causing the breakdown of the viscous regime (figure~\ref{fig.omz30Zrolleravg}).  {The streamwise wavelength of these flow structures extracted from the  {premultiplied} co-spectrum $-k_xE_{uv}$ is $\lambda_x^+ \approx 200$. The wall-normal velocity $v$ at $y^+=5$ corresponding to the same horizontal wavenumbers is plotted in figure~\ref{fig.omz30Zrollerxz}. These flow structures are reminiscent of the spanwise rollers identified using DNS of turbulent channel flow over square riblets; see figure 14 in~\cite{garjim11b}.}

\begin{figure}
	\begin{center}
        \begin{tabular}{cc}
        \begin{tabular}{c}
                \vspace{.4cm}
                \rotatebox{90}{\normalsize $y^+$}
        \end{tabular}
        &\hspace{-.5cm}
        \begin{tabular}{c}
                \includegraphics[width=.5\textwidth]{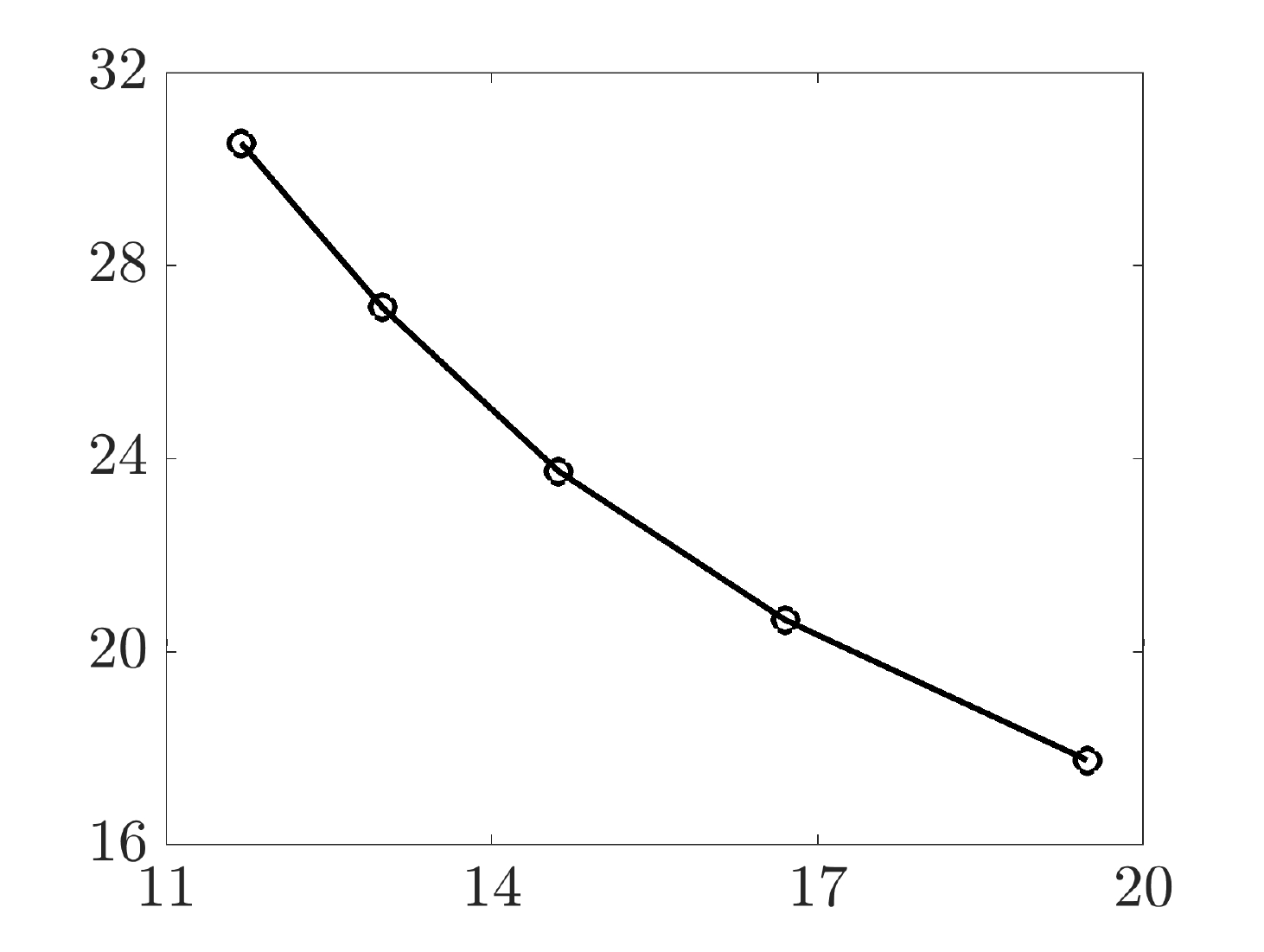}
                \\[-.1cm]
                {\normalsize $l_g^+$}
        \end{tabular}
        \end{tabular}
        \end{center}
        \caption{Wall-normal locations of the core of spanwise {elongated rollers with $\lambda_x^+\approx200$} in a turbulent channel flow with $Re_\tau=186$ over triangular riblets with $\alpha=90^\degree$ and sizes specified by $l_g^+$.}
        \label{fig.vortexcoreypvslg}
\end{figure}

Figure~\ref{fig.vortexcoreypvslg} shows the wall-normal location corresponding to the center of the spanwise rollers that appear above riblets with $\alpha=90\degree$ and larger size relative to the optimal value $l_g^+=11.7$. For these riblets, the spanwise rollers have similar dominant streamwise length scales ($\lambda_x^+ \approx 200$). For larger riblets, the core of the spanwise rollers moves down toward the riblets. Thus, as the size of riblets increases and the grooves become deeper, the dominant turbulent flow structures penetrate further down into the viscous region in the grooves.

	\vspace*{-2ex}

\subsection{Very large scale motions in turbulent channel flow with $Re_\tau=547$}
\label{sec.VLSM}

Apart from the dominant flow structures associated with the near-wall cycle that are centered at $y^+\approx10$, other flow structures within the logarithmic and wake region become significant in wall-bounded shear flows with  {high} Reynolds numbers. An important class of streamwise elongated flow structures that reside in the logarithmic region, i.e., $3\sqrt{Re_\tau} < y^+ < 0.15 Re_\tau$, determine very large scale motions (VLSM)~\citep{hutmar07,monstewilcho07,marmonhulsmi13}. Relative to the near-wall cycle, VLSMs are centered farther away from the wall. Yet, they exhibit a wall-normal reach that can influence the energy transfer at the wall~\citep{marmckmonnagsmisre10}. It is thus relevant to explore the effect of riblets on the energy and locality of these flow structures.

\begin{figure}
	\begin{centering}
	\begin{tabular}{cc}
	\hspace{-.4cm}\subfigure[]{\label{fig.Re547VLSMomz360}}& 
	 \\[-.6cm]	 
	 	\begin{tabular}{c}
        \vspace{.2cm}
        \small{\rotatebox{90}{$y^+$}}
       \end{tabular}
       &\hspace{-.5cm}
	\begin{tabular}{c}
       \includegraphics[width=0.9\textwidth]{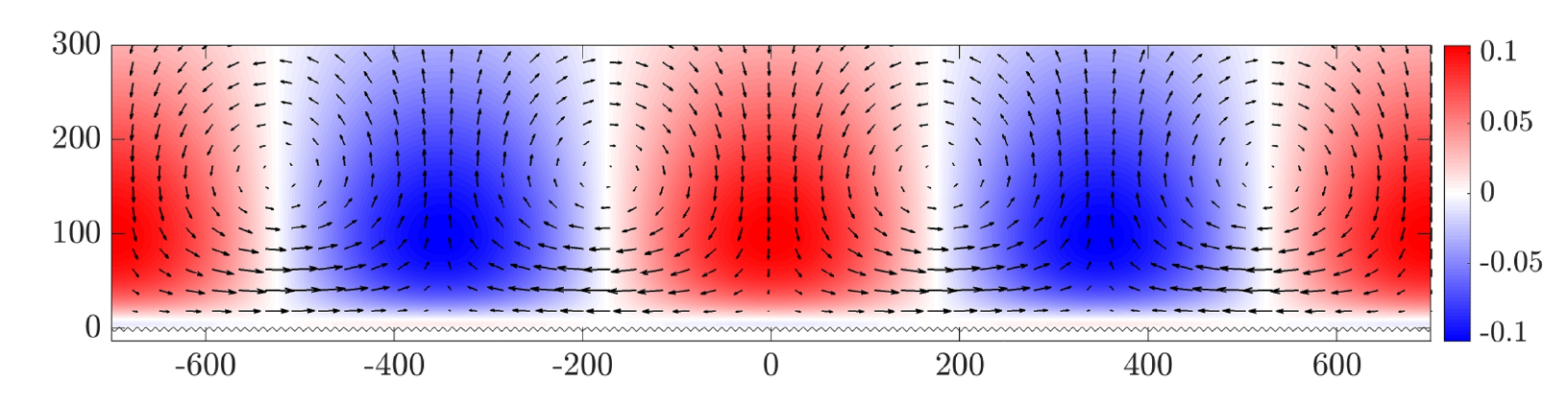}
       \end{tabular}
	\\[-.2cm]
	\hspace{-.41cm}\subfigure[]{\label{fig.Re547VLSMomz175}}& 
	\\[-.6cm]	
	       \begin{tabular}{c}
        \vspace{.2cm}
        \small{\rotatebox{90}{$y^+$}}
       \end{tabular}
       &\hspace{-.5cm}
	\begin{tabular}{c}
       \includegraphics[width=0.9\textwidth]{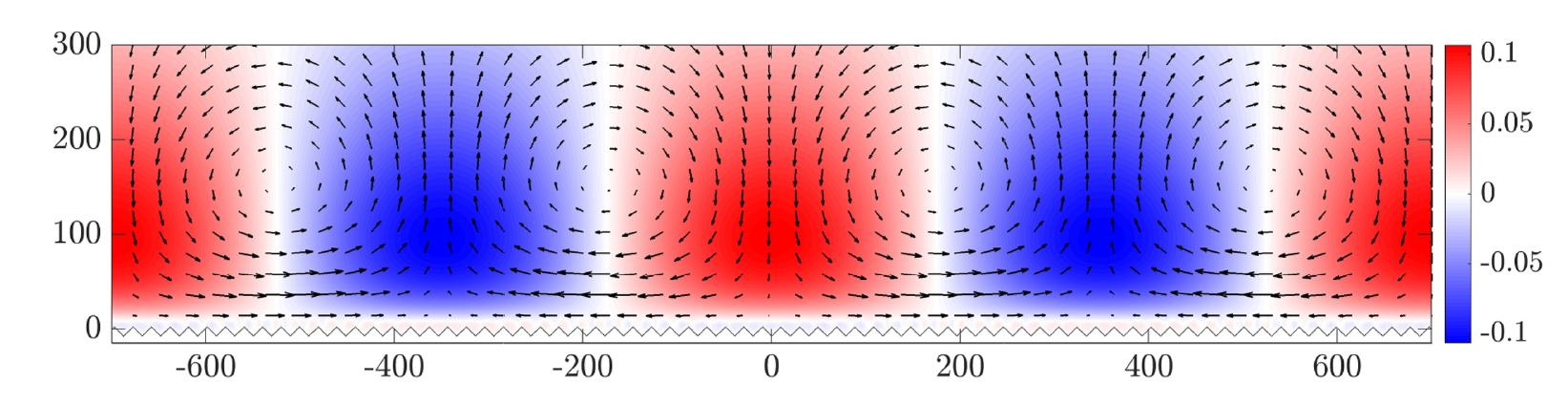}
       \end{tabular}
    \\[-.2cm]
	\hspace{-.4cm}\subfigure[]{\label{fig.Re547VLSMomz90}}& 
	\\[-.6cm]	
	       \begin{tabular}{c}
        \vspace{.5cm}
        \small{\rotatebox{90}{$y^+$}}
       \end{tabular}
       &\hspace{-.5cm}
	\begin{tabular}{c}
       \includegraphics[width=0.9\textwidth]{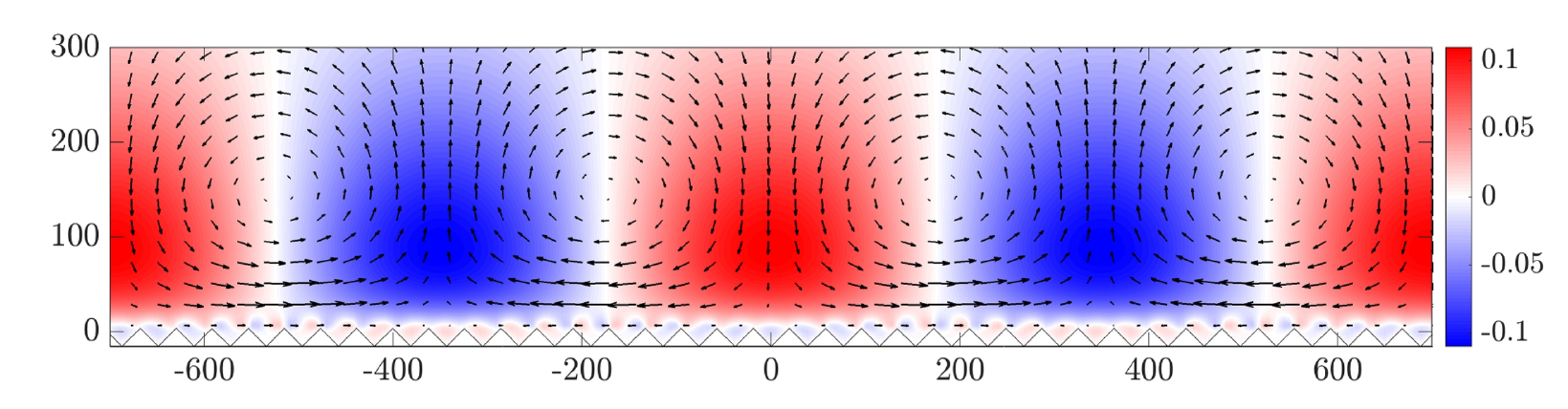}
       \\[-0.cm] {\small $z^+$}
       \end{tabular}
	\end{tabular}
	\caption{{Spatial structure of the streamwise velocity (red and blue colors denoting regions of high and low velocity) and the $(v,w)$ vector field (quiver lines) corresponding to VLSMs with $(\lambda_x^+,\, \lambda_z^+) = (1400,\, 700)$ in a turbulent channel flow with $Re_\tau=547$ over triangular riblets with $\alpha=90\degree$ and (a) $\omega_z = 360$; (b) $\omega_z = 175$; and (c) $\omega_z = 90$.}}
	\label{fig.VLSMstructures}
	\end{centering}
\end{figure}

For a turbulent channel flow with $Re_\tau=547$, VLSMs are characterized by wall-parallel wavelengths $(\lambda_x^+,\, \lambda_z^+) \approx (1400,\, 700)$ that can be extracted from the premultiplied one-dimensional energy spectrum generated from DNS data~\citep{deljim03,deljimzanmos04}. For various sizes of riblets with $\alpha = 90\degree$, figure~\ref{fig.VLSMstructures} shows the spatial structure of the principal eigenvector of the steady-state covariance matrix $\Phi_\theta(k_x)$ (equation~ {\eqref{eq.output-covariance}}) corresponding to such VLSMs. In this figure, riblets with $\omega_z = 175$ (figure~\ref{fig.Re547VLSMomz175}) provide the maximum drag reduction; cf.~figure~\ref{fig.DRml}. Figure~\ref{fig.VLSMstructures} demonstrates that small- and optimal-size riblets have little influence on the shape of the VLSMs. In contrast, large riblets distort the shape of such flow structures close to the wall.  {Moreover, the core of flow structures (i.e., the wall-normal location of the maximum streamwise velocity) shifts from $y^+\approx 85$ in the flow over smooth walls to $y^+\approx 90.2$, $95.7$, and $86.0$ for small, optimal, and large riblets, respectively. Thus, larger size riblets allow the VLSMs that are otherwise pushed away to settle into the riblet valleys; see figure~\ref{fig.Re547VLSMomz90}. This is consistent with the downward shift of near-wall flow structures predicted in \S~\ref{sec.NWcycle}, which results in an increase in skin-friction drag.}

 {In a turbulent channel flow with smooth walls at $Re_\tau=547$, the kinetic energy associated with $(\lambda_x^+,\, \lambda_z^+) = (1400,\, 700)$ is $0.0606$. In the presence of small, optimal, and large riblets the kinetic energy for this wavelength pair shifts to {$0.0602$, $0.0599$, and $0.0611$ (given by $\bar{E}(k_x,\theta)$ defined in~\eqref{eq.Ebar})}, respectively. This is consistent with the experimental study of~\cite{leelee01} where small changes to the turbulent kinetic energy and velocity fluctuations in the outer layer of turbulent channel flow over drag-reducing surfaces with semi-circular grooves were observed. The same reference also reports an increase in the turbulent kinetic energy that arises from a drag-increasing surface with larger grooves.}

	\vspace*{-2ex}
\section{Concluding remarks}
\label{sec.conclution}

We have developed a model-based framework for evaluating the effect of surface corrugation on skin-friction drag and kinetic energy in turbulent channel flows. The influence of the corrugated surface is captured via a volume penalization technique that enters as a feedback term into the governing equations. Our simulation-free approach utilizes eddy-viscosity-enhanced NS equations and it consists of two steps: (i) we use the turbulent viscosity of the turbulent channel flow with smooth walls to capture the effect of the corrugated surface on the turbulent base velocity; and (ii) we use second-order statistics of stochastically forced equations linearized around this base velocity profile to assess the role of velocity fluctuations and correct the turbulent viscosity model. This correction perturbs the turbulent base velocity profile obtained in the first step and refines our prediction of skin-friction drag.

For a turbulent channel flow with streamwise-aligned spanwise-periodic triangular riblets on the lower wall, we demonstrate that the base flow computed in the first step of our approach does not capture drag-reducing trends reported in experiments and simulations. Incorporating the influence of fluctuations on the turbulent viscosity significantly improves our predictions. Our results demonstrate good agreement with experimental and numerical results both in capturing drag-reducing trends and in identifying optimal shapes and sizes of riblets for the largest drag reduction. We also investigate the dependence of the turbulent kinetic energy of fluctuations on the size of riblets and demonstrate similar trends to what we observe for drag reduction. Building on this similarity and data obtained through a parametric study for riblets of various shapes and sizes, we extract a linear regression model and show that energy can be used as a surrogate for predicting the effect of riblets on skin-friction drag in the viscous regime.

The steady-state covariance matrices that we compute also allow us to examine the impact of riblets on dominant turbulent flow structures. We show that small-size triangular riblets limit the wall-normal transfer of momentum associated with the near-wall cycle and the generation of secondary flow structures around the tips. Our model captures the penetration of secondary vortices into riblet grooves and predicts that drag reduction reduces for large-size riblets. We also investigate the amplification of spanwise rollers that resemble \mbox{K-H} vortices and show {good} agreement between our predictions of their streamwise length-scale and core location with previous numerical studies. Finally, for turbulent channel flow with $Re_\tau=547$, we study the influence of riblet size on the wall-normal reach of VLSMs and demonstrate that  {drag-reducing riblets have little effect on the energy of these flow structures, but large riblets can increase their strength.}

 {Our long term objective is the development of a framework for the low complexity modeling of turbulent flows over corrugated surfaces that can bypass the need for costly numerical simulations and experiments and guide the optimal design of drag-reducing surfaces. The present work represents a step in this direction in that it provides a method for improving predictions to the turbulent viscosity and the mean velocity in channel flows over spanwise periodic surfaces.} We anticipate that incorporation of data from numerical simulations and experiments can further improve the predictive capability of  {the framework based on} stochastically forced linearized NS \mbox{equations~\citep{zarchejovgeoTAC17,zarjovgeoJFM17,zargeojovARC20}.}

\section*{Acknowledgments}
We thank Mr.\ A.\ Chavarin and Prof.\ M.\ Luhar for insightful discussions. Financial support from the Office of Naval Research under Award N00014-17-1-2308 and the Air Force Office of Scientific Research under Awards FA9550-16-1-0009 and FA9550-18-1-0422 is gratefully acknowledged. The Center for High-Performance Computing at the University of Southern California is acknowledged for providing computing resources.

\section*{A declaration of interest}

The authors report no conflict of interest.

\appendix

\section{Operators $A_\theta$, $B_\theta$, and $C_\theta$ in equations~\eqref{eq.evolutionform}}
\label{sec.A-C}

The dynamical generator $\bA_\theta$ in equations~\eqref{eq.evolutionform} has a bi-infinite structure shown in equation~\eqref{eq.Aform}, in which elements $\bA_{n,m}$ contain four operators,
\begin{align*}
	\bA_{n,m}
	\;=\;
	\tbt{\bA_{n,m,1,1}}{\bA_{n,m,1,2}}{\bA_{n,m,2,1}}{\bA_{n,m,2,2}}.
\end{align*}
For the operators on the main diagonal, $\bA_{n,0}$, we have
{
\begin{align*}
        \ba{rcl}
        \bA_{n,0,1,1}
        &\!\!\!=\!\!\!&
        \Delta_n^{-1}\left[(1\,+\,\nu_{T})\Delta_n^2 +\, \nu''_{T}(\partial_y^2 \,+\,k_n^2) \,+\,2\nu'_{T}\Delta_n \right]/Re \,+\; \Gamma_{n,0,1,1}
        \\[0.2cm]
        \bA_{n,0,1,2}
        &\!\!\!=\!\!\!&
        \Gamma_{n,0,1,2}
        \\[0.2cm]
        \bA_{n,0,2,1}
        &\!\!\!=\!\!\!&
        \Gamma_{n,0,2,1}
        \\[0.2cm]
        \bA_{n,0,2,2}
        &\!\!\!=\!\!\!& \left[(1 \,+\,\nu_{T})\Delta_n \,+\,\nu'_{T} \right]/Re \,+\;\Gamma_{n,0,2,2}
        \ea
\end{align*}}
and for the off-diagonal ones, $A_{n,m}$ with $m \neq 0$, we have
\begin{align*}
        \ba{rclrcl}
        \bA_{n,m,1,1}
        &\!\!\!=\!\!\!&
        \Gamma_{n,m,1,1},&\;
        \bA_{n,m,1,2}
        &\!\!\!=\!\!\!&
        \Gamma_{n,m,1,2},
        \\[0.2cm]
        \bA_{n,m,2,1}
        &\!\!\!=\!\!\!&
        \Gamma_{n,m,2,1},&\;
        \bA_{n,m,2,2}
        &\!\!\!=\!\!\!&
        \Gamma_{n,m,2,2}
        \ea
\end{align*}
where
\begin{align*}
        \ba{rcl}
        \Gamma_{n,m,1,1}
        &=&
        \Delta_n^{-1} \!\! \left[2\, \mri m\, k_x\, \omega_z\dfrac{\theta_{n+m}}{k_{n+m}^2}\left({\bar{U}}'_{-m}\partial_y +\, {\bar{U}}_{-m}\partial_{yy}\right)\right.
        +\, \mri k_x \left({\bar{U}}''_{-m} \,-\, {\bar{U}}_{-m}\Delta_{n+m}\right)
        \\[0.5cm]
        &&
        +\,\mri k_x(m\, \omega_z)^2{\bar{U}}_{-m} - 2\, m\, k_x \omega_z \theta_{n+m}{\bar{U}}_{-m}
        +\, m\, \omega_z (m\omega_z \,-\, 2\, \theta_{n+m})a_{-m}
        \\[0.2cm]
        &&
         -\, a_{-m}\Delta_{n+m} -\, a'_{-m}\partial_y + \left. m\, \omega_z\dfrac{\theta_{n+m}}{k_{n+m}^2}\left(a'_{-m}\partial_y\,+\,a_{-m}\partial_{yy}\right)\right]
	\ea
\end{align*}

\begin{align*}
        \ba{rcl}
        \Gamma_{n,m,1,2}
        &=&
        \Delta_n^{-1}\left[2\,\dfrac{\mri m\, k_x^2\,\omega_z}{k_{n+m}^2}\left({\bar{U}}'_{-m}\,+\,{\bar{U}}_{-m}\partial_y\right)\right.
         +\,\left.\dfrac{m\, k_x\,\omega_z}{k_{n+m}^2}\left(a'_{-m} +\, a_{-m}\partial_y\right)\right]
       \\[0.5cm]
        \Gamma_{n,m,2,1}
        &=&
        \mri m\, \omega_z \left({\bar{U}}'_{-m} \,-\, {\bar{U}}_{-m}\partial_y\right) \,-\, \mri \theta_{n+m}{\bar{U}}'_{-m}
        \\[0.2cm]
        && + \left[\mri(m\,\omega_z)^2\dfrac{\theta_{n+m}}{k_{n+m}^2}\,{\bar{U}}'_{-m} - \dfrac{m\, k_x\, \omega_z}{k_{n+m}^2}\,a_{-m} \right]\partial_y
        \\[0.5cm]
        \Gamma_{n,m,2,2}
        &=&
        -\mri k_x {\bar{U}}_{-m} \,-\, a_{-m}
         +\, \dfrac{\mri\, k_x(m\,\omega_z)^2}{k_{n+m}^2}\,{\bar{U}}_{-m} \,+\, m\, \omega_z\dfrac{\theta_{n+m}}{k_{n+m}^2}\, a_{-m}
        \ea
\end{align*}
Here, $\theta_{n+m} = (n\,+\,m)\omega_z \,+\, \theta$, $k_{n+m}^2 = k_x^2 \,+\, \theta_{n+m}^2$, and $\Delta_{n+m}=\partial_{yy} \,-\, k_{n+m}^2$. 

The input operator $\bB_{\theta}$ takes the form
	$
	\bB_{\theta}
	=
	\diag \left\{\, \ldots, \bB_{n-1}, \bB_n, \bB_{n+1}, \ldots \,\right\}
	$
where
\begin{align}
\label{eq.Bn}
        \bB_n
        \;=\;
        \begin{bmatrix}
        \bB_{v}
        \\[0.1cm]
        \bB_{\eta}
        \end{bmatrix}
        \;=\;
        \begin{bmatrix}
        \;-\mri k_x \Delta_n^{-1} \partial_y& -\mri k_n^2\Delta_n^{-1} & -\mri \theta_n \Delta_n^{-1} \partial_y\;
        \\[0.1cm]
        \;\mri \theta_n I & 0 & -\mri k_x I \;
        \end{bmatrix}.
\end{align}
Similarly, the output operator $\bC_{\theta}$ is given by
	$
	\bC_{\theta}
	=
	\diag \left\{\, \ldots, \bC_{n-1}, \bC_n, \bC_{n+1}, \ldots \,\right\}
	$
where
\begin{align}
\label{eq.Cn}
        \bC_n
        \;=\;
        \begin{bmatrix}
        \bC_{u}
        \\[0.1cm]
        \bC_{v}
        \\[0.1cm]
        \bC_{w}
        \end{bmatrix}
        \;=\;
        \begin{bmatrix}
        \;( \mri k_x /k_n^2 )\partial_y& -(\mri \theta_n/k_n^2) I \;
        \\[0.1cm]
        \;I & 0 \;
        \\[0.1cm]
        \; (\mri  \theta_n/k_n^2 ) \partial_y & ( \mri k_x/k_n^2) I \;
        \end{bmatrix}.
\end{align}

\section{Computing corrections ($k_c,\eps_c$) to ($k,\eps$)}
\label{sec.kecorrection}

{Following~\eqref{eq.kandepsilon},} we show that the effect of fluctuations around the mean velocity on corrections $k_c$ and $\epsilon_c$ can be obtained from the correction {$X_{\theta,c}(k_x)$} to the steady-state covariance:
\begin{align*}
\ba{rcl}
k_c(y)
&\!\!=\!\!&
\ds{\int_{0}^{\infty}\int_{0}^{{\tfrac{\omega_z}{2}}}\sum_{n \, \in \, \bbZ}K_k(y,k_x,\theta_n)\,\mathrm{d}\theta\,\mathrm{d}k_x},
\\[0.4cm]
\eps_c(y)
&\!\!=\!\!&
\ds{\int_{0}^{\infty}\int_{0}^{{\tfrac{\omega_z}{2}}}\sum_{n \, \in \, \bbZ}K_\eps(y,k_x,\theta_n)\,\mathrm{d}\theta\,\mathrm{d}k_x}
\ea
\end{align*}
where $K_k(y,k_x,\theta_n)$ and $K_\eps(y,k_x,\theta_n)$ are obtained by taking the diagonal components of matrices $N_k$ and $N_\eps$, respectively:
\begin{align*}
\ba{rcl}
\!\!{N_k(y,k_x,\theta_n)}
&\!\!\!=\!\!\!&
\dfrac{1}{2} \left(C_uX_cC_u^* \;+\; C_vX_cC_v^* \;+\; C_wX_cC_w^* \right),
\\[0.4cm]
\!\!{N_\eps(y,k_x,\theta_n)}
&\!\!\!=\!\!\!&
2\left(k_x^2C_uX_cC_u^* \,+\, D_yC_vX_cC_v^*D_y^* \,+\, \theta_n^2C_wX_cC_w^* \,-\,\mri k_xD_yC_uX_cC_v^* \right.
\\[0.2cm]
&&
\!\left.+\, k_x\theta_nC_uX_cC_w^* \,+\,\mri \theta_nD_yC_vX_cC_w^*D_y^*\right) \,+\, D_yC_uX_cC_u^*D_y^* \,+\, k_x^2C_vX_cC_v^*
\\[0.2cm]
&&
+\, D_yC_wX_cC_w^*D_y^*  \,+\, \theta_n^2C_uX_cC_u^* \,+\, \theta_n^2C_vX_cC_v^* \,+\, k_x^2C_wX_cC_w^*.
\ea
\end{align*}
Here, {the terms on the right-hand-side of the equations are at the wavenumber pair $(k_x,\,\theta_n)$;} $D_y$ denotes the finite-dimensional representation of $\partial_y$ and $C_u$, $C_v$, and $C_w$ are finite-dimensional approximations of the output operators in~\eqref{eq.Cn} {and the covariance matrix $X_c(k_x,\theta_n)$ can be obtained as
\begin{align*}
	X_c(k_x,\theta_n)
	\;=\;
	X_d(k_x,\theta_n)
	\,-\,
	X_s(k_x,\theta_n),
\end{align*}
where $X_s(k_x,\theta_n)$ and $X_d(k_x,\theta_n)$ represent the steady-state covariance matrix in channel flow over smooth walls and the main diagonal blocks of the solution to Lyapunov equation~\eqref{eq.lyap}, ${X}_\theta(k_x)$, respectively. {Note that these matrices have been confined to the wall-normal range $y\in[-1, 1]$ to provide appropriate comparison between channel flows over smooth and corrugated surfaces.}

	%\newpage
%\bibliographystyle{jfm}
%% Note the spaces between the initials
%\bibliography{../bib/reference,../bib/couette,../bib/mj-complete-bib,../bib/periodic,../bib/covariance,../bib/control-pde,../bib/channel,../bib/ref-added-rm,../bib/low_rank_bib,../bib/tg,../bib/ref-added-az,../bib/mj-unrefereed-bib,../bib/PSE,../bib/ref-added-wr}

\end{document}